\documentclass{article}
\usepackage{graphicx} %
\usepackage{braket}
\usepackage{amsfonts}
\usepackage{amsmath, amsthm}
\usepackage{pagecolor}
\usepackage{xcolor}
\usepackage{amssymb}
\usepackage{tikzit}
\usepackage{tikz, circuitikz}
\usepackage[a4paper, total={6in, 8in}]{geometry}
\usepackage[outline]{contour}

\usepackage{xcolor}
\usepackage{hyperref}

\tikzstyle{env}=[copoint,regular polygon rotate=0,minimum width=0.2cm, fill=black]

\tikzstyle{probs}=[shape=semicircle,fill=white,draw=black,shape border rotate=180,minimum width=1.2cm]

\tikzstyle{nudge}=[yshift=0.6mm]

\tikzstyle{every picture}=[baseline=-0.25em,scale=0.5]
\tikzstyle{dotpic}=[] 
\tikzstyle{diredges}=[every to/.style={diredge}]
\tikzstyle{math matrix}=[matrix of math nodes,left delimiter=(,right delimiter=),inner sep=2pt,column sep=1em,row sep=0.5em,nodes={inner sep=0pt},text height=1.5ex, text depth=0.25ex]

\tikzstyle{inline text}=[text height=1.5ex, text depth=0.25ex,yshift=0.5mm]
\tikzstyle{label}=[font=\footnotesize,text height=1.5ex, text depth=0.25ex]
\tikzstyle{left label}=[label,anchor=east,xshift=2mm]
\tikzstyle{right label}=[label,anchor=west,xshift=-2mm]

\tikzstyle{braceedge}=[decorate,decoration={brace,amplitude=2mm,raise=-1mm}]
\tikzstyle{small braceedge}=[decorate,decoration={brace,amplitude=1mm,raise=-1mm}]

\tikzstyle{doubled}=[line width=1.6pt] 
\tikzstyle{boldedge}=[doubled,shorten <=-0.17mm,shorten >=-0.17mm]
\tikzstyle{boldedgegray}=[doubled,gray,shorten <=-0.17mm,shorten >=-0.17mm]
\tikzstyle{singleedgegray}=[gray]

\tikzstyle{semidoubled}=[line width=1.4pt] 
\tikzstyle{semiboldedgegray}=[semidoubled,gray,shorten <=-0.17mm,shorten >=-0.17mm]

\tikzstyle{boxedge}=[semiboldedgegray]

\tikzstyle{boldedgedashed}=[very thick,dashed,shorten <=-0.17mm,shorten >=-0.17mm]
\tikzstyle{vboldedgedashed}=[doubled,dashed,shorten <=-0.17mm,shorten >=-0.17mm]
\tikzstyle{left hook arrow}=[left hook-latex]
\tikzstyle{right hook arrow}=[right hook-latex]
\tikzstyle{sembracket}=[line width=0.5pt,shorten <=-0.07mm,shorten >=-0.07mm]

\tikzstyle{causal edge}=[->,thick,gray]
\tikzstyle{causal nondir}=[thick,gray]
\tikzstyle{timeline}=[thick,gray, dashed]

\tikzstyle{cedge}=[<->,thick,gray!70!white]

\tikzstyle{empty diagram}=[draw=gray!40!white,dashed,shape=rectangle,minimum width=1cm,minimum height=1cm]
\tikzstyle{empty diagram small}=[draw=gray!50!white,dashed,shape=rectangle,minimum width=0.6cm,minimum height=0.5cm]

\tikzstyle{dot}=[inner sep=0mm,minimum width=2mm,minimum height=2mm,draw,shape=circle]  
\tikzstyle{Wsquare}=[white dot, shape=regular polygon, rounded corners=0.8 mm, minimum size=3.3 mm, regular polygon sides=3, outer sep=-0.2mm]
\tikzstyle{Wsquareadj}=[white dot, shape=regular polygon, rounded corners=0.8 mm, minimum size=3.3 mm, regular polygon sides=3, outer sep=-0.2mm, regular polygon rotate=180]
\tikzstyle{ddot}=[inner sep=0mm, doubled, minimum width=2.5mm,minimum height=2.5mm,draw,shape=circle]

\tikzstyle{black dot}=[dot,fill=black]
\tikzstyle{white dot}=[dot,fill=white,,text depth=-0.2mm]
\tikzstyle{white Wsquare}=[Wsquare,fill=white,,text depth=-0.2mm]
\tikzstyle{white Wsquareadj}=[Wsquareadj,fill=white,,text depth=-0.2mm]
\tikzstyle{green dot}=[white dot] 
\tikzstyle{gray dot}=[dot,fill=gray!40!white,,text depth=-0.2mm]
\tikzstyle{red dot}=[gray dot] 

\tikzstyle{black ddot}=[ddot,fill=black]
\tikzstyle{white ddot}=[ddot,fill=white]
\tikzstyle{gray ddot}=[ddot,fill=gray!40!white]

\tikzstyle{gray edge}=[gray!60!white]

\tikzstyle{small dot}=[inner sep=0.5mm,minimum width=0pt,minimum height=0pt,draw,shape=circle]

\tikzstyle{small black dot}=[small dot,fill=black]
\tikzstyle{small white dot}=[small dot,fill=white]
\tikzstyle{small gray dot}=[small dot,fill=gray!40!white]

\tikzstyle{very small dot}=[inner sep=0.3mm,minimum width=0pt,minimum height=0pt,draw,shape=circle]

\tikzstyle{very small black dot}=[very small dot,fill=black]
\tikzstyle{very small white dot}=[small dot,fill=white]
\tikzstyle{very small gray dot}=[small dot,fill=gray!40!white]

\tikzstyle{causal dot}=[inner sep=0.4mm,minimum width=0pt,minimum height=0pt,draw=white,shape=circle,fill=gray!40!white]

\tikzstyle{phase dimensions}=[minimum size=5mm,font=\footnotesize,rectangle,rounded corners=2.5mm,inner sep=0.2mm,outer sep=-2mm]
\tikzstyle{dphase dimensions}=[minimum size=5mm,font=\footnotesize,rectangle,rounded corners=2.5mm,inner sep=0.2mm,outer sep=-2mm]

\tikzstyle{white phase dot}=[dot,fill=white,phase dimensions]
\tikzstyle{white phase ddot}=[ddot,fill=white,dphase dimensions]

\tikzstyle{white rect ddot}=[draw=black,fill=white,doubled,minimum size=5mm,font=\footnotesize,rectangle,rounded corners=2.5mm,inner sep=0.2mm]
\tikzstyle{gray rect ddot}=[draw=black,fill=gray!40!white,doubled,minimum size=6mm,font=\footnotesize,rectangle,rounded corners=3mm]

\tikzstyle{gray phase dot}=[dot,fill=gray!40!white,phase dimensions]
\tikzstyle{gray phase ddot}=[ddot,fill=gray!40!white,dphase dimensions]
\tikzstyle{grey phase dot}=[gray phase dot]
\tikzstyle{grey phase ddot}=[gray phase ddot]

\tikzstyle{small phase dimensions}=[minimum size=4mm,font=\tiny,rectangle,rounded corners=2mm,inner sep=0.2mm,outer sep=-2mm]
\tikzstyle{small dphase dimensions}=[minimum size=4mm,font=\tiny,rectangle,rounded corners=2mm,inner sep=0.2mm,outer sep=-2mm]

\tikzstyle{small gray phase dot}=[dot,fill=gray!40!white,small phase dimensions]
\tikzstyle{small gray phase ddot}=[ddot,fill=gray!40!white,small dphase dimensions]

\tikzstyle{small map}=[draw,shape=rectangle,minimum height=4mm,minimum width=4mm,fill=white]

\tikzstyle{cnot}=[fill=white,shape=circle,inner sep=-1.4pt]

\tikzstyle{asym hadamard}=[fill=white,draw,shape=NEbox,inner sep=0.6mm,font=\footnotesize,minimum height=4mm]
\tikzstyle{asym hadamard conj}=[fill=white,draw,shape=NWbox,inner sep=0.6mm,font=\footnotesize,minimum height=4mm]
\tikzstyle{asym hadamard dag}=[fill=white,draw,shape=SEbox,inner sep=0.6mm,font=\footnotesize,minimum height=4mm]

\tikzstyle{hadamard}=[fill=white,draw,inner sep=0.6mm,font=\footnotesize,minimum height=4mm,minimum width=4mm]
\tikzstyle{small hadamard}=[fill=white,draw,inner sep=0.6mm,minimum height=1.5mm,minimum width=1.5mm]
\tikzstyle{small hadamard rotate}=[small hadamard,rotate=45]
\tikzstyle{dhadamard}=[hadamard,doubled]
\tikzstyle{small dhadamard}=[small hadamard,doubled]
\tikzstyle{small dhadamard rotate}=[small hadamard rotate,doubled]
\tikzstyle{antipode}=[white dot,inner sep=0.3mm,font=\footnotesize]

\tikzstyle{scalar}=[diamond,draw,inner sep=0.5pt,font=\small]
\tikzstyle{dscalar}=[diamond,doubled, draw,inner sep=0.5pt,font=\small]

\tikzstyle{small box}=[rectangle,inline text,fill=white,draw,minimum height=5mm,yshift=-0.5mm,minimum width=5mm,font=\small]
\tikzstyle{small gray box}=[small box,fill=gray!30]
\tikzstyle{medium box}=[rectangle,inline text,fill=white,draw,minimum height=5mm,yshift=-0.5mm,minimum width=8mm,font=\small]
\tikzstyle{square box}=[small box] 
\tikzstyle{medium gray box}=[small box,fill=gray!30]
\tikzstyle{semilarge box}=[rectangle,inline text,fill=white,draw,minimum height=5mm,yshift=-0.5mm,minimum width=12.5mm,font=\small]
\tikzstyle{large box}=[rectangle,inline text,fill=white,draw,minimum height=5mm,yshift=-0.5mm,minimum width=15mm,font=\small]
\tikzstyle{large gray box}=[small box,fill=gray!30]

\tikzstyle{Bayes box}=[rectangle,fill=black,draw, minimum height=3mm, minimum width=3mm]

\tikzstyle{gray square point}=[small box,fill=gray!50]

\tikzstyle{dphase box white}=[dhadamard]
\tikzstyle{dphase box gray}=[dhadamard,fill=gray!50!white]
\tikzstyle{phase box white}=[hadamard]
\tikzstyle{phase box gray}=[hadamard,fill=gray!50!white]

\tikzstyle{point}=[regular polygon,regular polygon sides=3,draw,scale=0.75,inner sep=-0.5pt,minimum width=9mm,fill=white,regular polygon rotate=180]
\tikzstyle{copoint}=[regular polygon,regular polygon sides=3,draw,scale=0.75,inner sep=-0.5pt,minimum width=9mm,fill=white]
\tikzstyle{dpoint}=[point,doubled]
\tikzstyle{dcopoint}=[copoint,doubled]

\tikzstyle{wide copoint}=[fill=white,draw,shape=isosceles triangle,shape border rotate=90,isosceles triangle stretches=true,inner sep=0pt,minimum width=1.5cm,minimum height=6.12mm]
\tikzstyle{wide point}=[fill=white,draw,shape=isosceles triangle,shape border rotate=-90,isosceles triangle stretches=true,inner sep=0pt,minimum width=1.5cm,minimum height=6.12mm,yshift=-0.0mm]
\tikzstyle{wide point plus}=[fill=white,draw,shape=isosceles triangle,shape border rotate=-90,isosceles triangle stretches=true,inner sep=0pt,minimum width=1.74cm,minimum height=7mm,yshift=-0.0mm]

\tikzstyle{wide dpoint}=[fill=white,doubled,draw,shape=isosceles triangle,shape border rotate=-90,isosceles triangle stretches=true,inner sep=0pt,minimum width=1.5cm,minimum height=6.12mm,yshift=-0.0mm]

\tikzstyle{tinypoint}=[regular polygon,regular polygon sides=3,draw,scale=0.55,inner sep=-0.15pt,minimum width=6mm,fill=white,regular polygon rotate=180] 

\tikzstyle{white point}=[point]
\tikzstyle{white dpoint}=[dpoint]
\tikzstyle{green point}=[white point] 
\tikzstyle{white copoint}=[copoint]
\tikzstyle{gray point}=[point,fill=gray!40!white]
\tikzstyle{gray dpoint}=[gray point,doubled]
\tikzstyle{red point}=[gray point] 
\tikzstyle{gray copoint}=[copoint,fill=gray!40!white]
\tikzstyle{gray dcopoint}=[gray copoint,doubled]

\tikzstyle{white point guide}=[regular polygon,regular polygon sides=3,font=\scriptsize,draw,scale=0.65,inner sep=-0.5pt,minimum width=9mm,fill=white,regular polygon rotate=180]

\tikzstyle{black point}=[point,fill=black,font=\color{white}]
\tikzstyle{black copoint}=[copoint,fill=black,font=\color{white}]

\tikzstyle{tiny gray point}=[tinypoint,fill=gray!40!white]

\tikzstyle{diredge}=[->]
\tikzstyle{ddiredge}=[<->]
\tikzstyle{rdiredge}=[<-]
\tikzstyle{thickdiredge}=[->, very thick]
\tikzstyle{pointer edge}=[->,very thick,gray]
\tikzstyle{pointer edge part}=[very thick,gray]
\tikzstyle{dashed edge}=[dashed]
\tikzstyle{thick dashed edge}=[very thick,dashed]
\tikzstyle{thick gray dashed edge}=[thick dashed edge,gray!40]
\tikzstyle{thick map edge}=[very thick,|->]

\makeatletter
\newcommand{\boxshape}[3]{%
\pgfdeclareshape{#1}{
\inheritsavedanchors[from=rectangle] 
\inheritanchorborder[from=rectangle]
\inheritanchor[from=rectangle]{center}
\inheritanchor[from=rectangle]{north}
\inheritanchor[from=rectangle]{south}
\inheritanchor[from=rectangle]{west}
\inheritanchor[from=rectangle]{east}
\backgroundpath{
\southwest \pgf@xa=\pgf@x \pgf@ya=\pgf@y
\northeast \pgf@xb=\pgf@x \pgf@yb=\pgf@y

\@tempdima=#2
\@tempdimb=#3

\pgfpathmoveto{\pgfpoint{\pgf@xa - 5pt + \@tempdima}{\pgf@ya}}
\pgfpathlineto{\pgfpoint{\pgf@xa - 5pt - \@tempdima}{\pgf@yb}}
\pgfpathlineto{\pgfpoint{\pgf@xb + 5pt + \@tempdimb}{\pgf@yb}}
\pgfpathlineto{\pgfpoint{\pgf@xb + 5pt - \@tempdimb}{\pgf@ya}}
\pgfpathlineto{\pgfpoint{\pgf@xa - 5pt + \@tempdima}{\pgf@ya}}
\pgfpathclose
}
}}

\boxshape{NEbox}{0pt}{0pt}
\boxshape{SEbox}{0pt}{-5pt}
\boxshape{NWbox}{5pt}{0pt}
\boxshape{SWbox}{-5pt}{0pt}
\boxshape{EBox}{-3pt}{3pt}
\boxshape{WBox}{3pt}{-3pt}
\makeatother

\tikzstyle{cloud}=[shape=cloud,draw,minimum width=1.5cm,minimum height=1.5cm]

\tikzstyle{map}=[draw,shape=NEbox,inner sep=2pt,minimum height=6mm,fill=white]
\tikzstyle{dashedmap}=[draw,dashed,gray,shape=NEbox,inner sep=2pt,minimum height=6mm,fill=white]
\tikzstyle{medium dashedmap}=[draw,dashed,gray,shape=NEbox,inner sep=2pt,minimum height=6mm,fill=white,minimum width=7mm]
\tikzstyle{semilarge dashedmap}=[draw,dashed,gray,shape=NEbox,inner sep=2pt,minimum height=6mm,fill=white,minimum width=9.5mm]
\tikzstyle{large dashedmap}=[draw,dashed,gray,shape=NEbox,inner sep=2pt,minimum height=6mm,fill=white,minimum width=12mm]
\tikzstyle{very large dashedmap}=[draw,dashed,gray,shape=NEbox,inner sep=2pt,minimum height=6mm,fill=white,minimum width=17mm]

\tikzstyle{dashed map}=[fill=white, draw=gray, shape=rectangle, style=map, dashed]

\tikzstyle{mapdag}=[draw,shape=SEbox,inner sep=2pt,minimum height=6mm,fill=white]
\tikzstyle{mapadj}=[draw,shape=SEbox,inner sep=2pt,minimum height=6mm,fill=white]
\tikzstyle{maptrans}=[draw,shape=SWbox,inner sep=2pt,minimum height=6mm,fill=white]
\tikzstyle{mapconj}=[draw,shape=NWbox,inner sep=2pt,minimum height=6mm,fill=white]

\tikzstyle{medium map}=[draw,shape=NEbox,inner sep=2pt,minimum height=6mm,fill=white,minimum width=7mm]
\tikzstyle{medium map dag}=[draw,shape=SEbox,inner sep=2pt,minimum height=6mm,fill=white,minimum width=7mm]
\tikzstyle{medium map adj}=[draw,shape=SEbox,inner sep=2pt,minimum height=6mm,fill=white,minimum width=7mm]
\tikzstyle{medium map trans}=[draw,shape=SWbox,inner sep=2pt,minimum height=6mm,fill=white,minimum width=7mm]
\tikzstyle{medium map conj}=[draw,shape=NWbox,inner sep=2pt,minimum height=6mm,fill=white,minimum width=7mm]
\tikzstyle{semilarge map}=[draw,shape=NEbox,inner sep=2pt,minimum height=6mm,fill=white,minimum width=9.5mm]
\tikzstyle{semilarge map trans}=[draw,shape=SWbox,inner sep=2pt,minimum height=6mm,fill=white,minimum width=9.5mm]
\tikzstyle{semilarge map adj}=[draw,shape=SEbox,inner sep=2pt,minimum height=6mm,fill=white,minimum width=9.5mm]
\tikzstyle{semilarge map dag}=[draw,shape=SEbox,inner sep=2pt,minimum height=6mm,fill=white,minimum width=9.5mm]
\tikzstyle{semilarge map conj}=[draw,shape=NWbox,inner sep=2pt,minimum height=6mm,fill=white,minimum width=9.5mm]
\tikzstyle{large map}=[draw,shape=NEbox,inner sep=2pt,minimum height=6mm,fill=white,minimum width=12mm]
\tikzstyle{large map conj}=[draw,shape=NWbox,inner sep=2pt,minimum height=6mm,fill=white,minimum width=12mm]
\tikzstyle{very large map}=[draw,shape=NEbox,inner sep=2pt,minimum height=6mm,fill=white,minimum width=17mm]
\tikzstyle{very very large map}=[draw,shape=NEbox,inner sep=2pt,minimum height=6mm,fill=white,minimum width=50mm]
\tikzstyle{large map dag}=[draw,shape=SEbox,inner sep=2pt,minimum height=6mm,fill=white,minimum width=12mm]

\tikzstyle{medium dmap}=[draw,doubled,shape=NEbox,inner sep=2pt,minimum height=6mm,fill=white,minimum width=7mm]
\tikzstyle{medium dmap dag}=[draw,doubled,shape=SEbox,inner sep=2pt,minimum height=6mm,fill=white,minimum width=7mm]
\tikzstyle{medium dmap adj}=[draw,doubled,shape=SEbox,inner sep=2pt,minimum height=6mm,fill=white,minimum width=7mm]
\tikzstyle{medium dmap trans}=[draw,doubled,shape=SWbox,inner sep=2pt,minimum height=6mm,fill=white,minimum width=7mm]
\tikzstyle{medium dmap conj}=[draw,doubled,shape=NWbox,inner sep=2pt,minimum height=6mm,fill=white,minimum width=7mm]
\tikzstyle{semilarge dmap}=[draw,doubled,shape=NEbox,inner sep=2pt,minimum height=6mm,fill=white,minimum width=9.5mm]
\tikzstyle{semilarge dmap trans}=[draw,doubled,shape=SWbox,inner sep=2pt,minimum height=6mm,fill=white,minimum width=9.5mm]
\tikzstyle{semilarge dmap adj}=[draw,doubled,shape=SEbox,inner sep=2pt,minimum height=6mm,fill=white,minimum width=9.5mm]
\tikzstyle{semilarge dmap dag}=[draw,doubled,shape=SEbox,inner sep=2pt,minimum height=6mm,fill=white,minimum width=9.5mm]
\tikzstyle{semilarge dmap conj}=[draw,doubled,shape=NWbox,inner sep=2pt,minimum height=6mm,fill=white,minimum width=9.5mm]
\tikzstyle{large dmap}=[draw,doubled,shape=NEbox,inner sep=2pt,minimum height=6mm,fill=white,minimum width=12mm]
\tikzstyle{large dmap conj}=[draw,doubled,shape=NWbox,inner sep=2pt,minimum height=6mm,fill=white,minimum width=12mm]
\tikzstyle{large dmap trans}=[draw,doubled,shape=SWbox,inner sep=2pt,minimum height=6mm,fill=white,minimum width=12mm]
\tikzstyle{large dmap adj}=[draw,doubled,shape=SEbox,inner sep=2pt,minimum height=6mm,fill=white,minimum width=12mm]
\tikzstyle{large dmap dag}=[draw,doubled,shape=SEbox,inner sep=2pt,minimum height=6mm,fill=white,minimum width=12mm]
\tikzstyle{very large dmap}=[draw,doubled,shape=NEbox,inner sep=2pt,minimum height=6mm,fill=white,minimum width=19.5mm]

\tikzstyle{muxbox}=[draw,shape=rectangle,minimum height=3mm,minimum width=3mm,fill=white]
\tikzstyle{dmuxbox}=[muxbox,doubled]

\tikzstyle{box}=[draw,shape=rectangle,inner sep=2pt,minimum height=6mm,minimum width=6mm,fill=white]
\tikzstyle{dbox}=[draw,doubled,shape=rectangle,inner sep=2pt,minimum height=6mm,minimum width=6mm,fill=white]
\tikzstyle{dmap}=[draw,doubled,shape=NEbox,inner sep=2pt,minimum height=6mm,fill=white]
\tikzstyle{dmapdag}=[draw,doubled,shape=SEbox,inner sep=2pt,minimum height=6mm,fill=white]
\tikzstyle{dmapadj}=[draw,doubled,shape=SEbox,inner sep=2pt,minimum height=6mm,fill=white]
\tikzstyle{dmaptrans}=[draw,doubled,shape=SWbox,inner sep=2pt,minimum height=6mm,fill=white]
\tikzstyle{dmapconj}=[draw,doubled,shape=NWbox,inner sep=2pt,minimum height=6mm,fill=white]

\tikzstyle{ddmap}=[draw,doubled,dashed,shape=NEbox,inner sep=2pt,minimum height=6mm,fill=white]
\tikzstyle{ddmapdag}=[draw,doubled,dashed,shape=SEbox,inner sep=2pt,minimum height=6mm,fill=white]
\tikzstyle{ddmapadj}=[draw,doubled,dashed,shape=SEbox,inner sep=2pt,minimum height=6mm,fill=white]
\tikzstyle{ddmaptrans}=[draw,doubled,dashed,shape=SWbox,inner sep=2pt,minimum height=6mm,fill=white]
\tikzstyle{ddmapconj}=[draw,doubled,dashed,shape=NWbox,inner sep=2pt,minimum height=6mm,fill=white]

\boxshape{sNEbox}{0pt}{3pt}
\boxshape{sSEbox}{0pt}{-3pt}
\boxshape{sNWbox}{3pt}{0pt}
\boxshape{sSWbox}{-3pt}{0pt}
\tikzstyle{smap}=[draw,shape=sNEbox,fill=white]
\tikzstyle{smapdag}=[draw,shape=sSEbox,fill=white]
\tikzstyle{smapadj}=[draw,shape=sSEbox,fill=white]
\tikzstyle{smaptrans}=[draw,shape=sSWbox,fill=white]
\tikzstyle{smapconj}=[draw,shape=sNWbox,fill=white]

\tikzstyle{dsmap}=[draw,dashed,shape=sNEbox,fill=white]
\tikzstyle{dsmapdag}=[draw,dashed,shape=sSEbox,fill=white]
\tikzstyle{dsmaptrans}=[draw,dashed,shape=sSWbox,fill=white]
\tikzstyle{dsmapconj}=[draw,dashed,shape=sNWbox,fill=white]

\boxshape{mNEbox}{0pt}{10pt}
\boxshape{mSEbox}{0pt}{-10pt}
\boxshape{mNWbox}{10pt}{0pt}
\boxshape{mSWbox}{-10pt}{0pt}
\tikzstyle{mmap}=[draw,shape=mNEbox]
\tikzstyle{mmapdag}=[draw,shape=mSEbox]
\tikzstyle{mmaptrans}=[draw,shape=mSWbox]
\tikzstyle{mmapconj}=[draw,shape=mNWbox]

\tikzstyle{mmapgray}=[draw,fill=gray!40!white,shape=mNEbox]
\tikzstyle{smapgray}=[draw,fill=gray!40!white,shape=sNEbox]

\makeatletter

\pgfdeclareshape{cornerpoint}{
\inheritsavedanchors[from=rectangle] 
\inheritanchorborder[from=rectangle]
\inheritanchor[from=rectangle]{center}
\inheritanchor[from=rectangle]{north}
\inheritanchor[from=rectangle]{south}
\inheritanchor[from=rectangle]{west}
\inheritanchor[from=rectangle]{east}
\backgroundpath{
\southwest \pgf@xa=\pgf@x \pgf@ya=\pgf@y
\northeast \pgf@xb=\pgf@x \pgf@yb=\pgf@y

\pgfmathsetmacro{\pgf@shorten@left}{\pgfkeysvalueof{/tikz/shorten left}}
\pgfmathsetmacro{\pgf@shorten@right}{\pgfkeysvalueof{/tikz/shorten right}}

\pgfpathmoveto{\pgfpoint{0.5 * (\pgf@xa + \pgf@xb)}{\pgf@ya - 5pt}}
\pgfpathlineto{\pgfpoint{\pgf@xa - 8pt + \pgf@shorten@left}{\pgf@yb - 1.5 * \pgf@shorten@left}}
\pgfpathlineto{\pgfpoint{\pgf@xa - 8pt + \pgf@shorten@left}{\pgf@yb}}
\pgfpathlineto{\pgfpoint{\pgf@xb + 8pt - \pgf@shorten@right}{\pgf@yb}}
\pgfpathlineto{\pgfpoint{\pgf@xb + 8pt - \pgf@shorten@right}{\pgf@yb - 1.5 * \pgf@shorten@right}}
\pgfpathclose
}
}

\pgfdeclareshape{cornercopoint}{
\inheritsavedanchors[from=rectangle] 
\inheritanchorborder[from=rectangle]
\inheritanchor[from=rectangle]{center}
\inheritanchor[from=rectangle]{north}
\inheritanchor[from=rectangle]{south}
\inheritanchor[from=rectangle]{west}
\inheritanchor[from=rectangle]{east}
\backgroundpath{
\southwest \pgf@xa=\pgf@x \pgf@ya=\pgf@y
\northeast \pgf@xb=\pgf@x \pgf@yb=\pgf@y

\pgfmathsetmacro{\pgf@shorten@left}{\pgfkeysvalueof{/tikz/shorten left}}
\pgfmathsetmacro{\pgf@shorten@right}{\pgfkeysvalueof{/tikz/shorten right}}

\pgfpathmoveto{\pgfpoint{0.5 * (\pgf@xa + \pgf@xb)}{\pgf@yb + 5pt}}
\pgfpathlineto{\pgfpoint{\pgf@xa - 8pt + \pgf@shorten@left}{\pgf@ya + 1.5 * \pgf@shorten@left}}
\pgfpathlineto{\pgfpoint{\pgf@xa - 8pt + \pgf@shorten@left}{\pgf@ya}}
\pgfpathlineto{\pgfpoint{\pgf@xb + 8pt - \pgf@shorten@right}{\pgf@ya}}
\pgfpathlineto{\pgfpoint{\pgf@xb + 8pt - \pgf@shorten@right}{\pgf@ya + 1.5 * \pgf@shorten@right}}
\pgfpathclose
}
}

\makeatother

\pgfkeyssetvalue{/tikz/shorten left}{0pt}
\pgfkeyssetvalue{/tikz/shorten right}{0pt}

\tikzstyle{kpoint common}=[draw,fill=white,inner sep=1pt,minimum height=4mm]
\tikzstyle{kpoint sc}=[shape=cornerpoint,kpoint common]
\tikzstyle{kpoint adjoint sc}=[shape=cornercopoint,kpoint common]
\tikzstyle{kpoint}=[shape=cornerpoint,shorten left=5pt,kpoint common]
\tikzstyle{kpoint adjoint}=[shape=cornercopoint,shorten left=5pt,kpoint common]
\tikzstyle{kpoint conjugate}=[shape=cornerpoint,shorten right=5pt,kpoint common]
\tikzstyle{kpoint transpose}=[shape=cornercopoint,shorten right=5pt,kpoint common]
\tikzstyle{kpoint symm}=[shape=cornerpoint,shorten left=5pt,shorten right=5pt,kpoint common]

\tikzstyle{black kpoint}=[shape=cornerpoint,shorten left=5pt,kpoint common,fill=black,font=\color{white}]
\tikzstyle{black kpoint adjoint}=[shape=cornercopoint,shorten left=5pt,kpoint common,fill=black,font=\color{white}]
\tikzstyle{black kpointadj}=[shape=cornercopoint,shorten left=5pt,kpoint common,fill=black,font=\color{white}]

\tikzstyle{black dkpoint}=[shape=cornerpoint,shorten left=5pt,kpoint common,fill=black, doubled,font=\color{white}]
\tikzstyle{black dkpoint adjoint}=[shape=cornercopoint,shorten left=5pt,kpoint common,fill=black, doubled,font=\color{white}]
\tikzstyle{black dkpointadj}=[shape=cornercopoint,shorten left=5pt,kpoint common,fill=black, doubled,font=\color{white}] 

\tikzstyle{kpointdag}=[kpoint adjoint]
\tikzstyle{kpointadj}=[kpoint adjoint]
\tikzstyle{kpointconj}=[kpoint conjugate]
\tikzstyle{kpointtrans}=[kpoint transpose]

\tikzstyle{big kpoint}=[kpoint, minimum width=1.2 cm, minimum height=8mm, inner sep=4pt, text depth=3mm]

\tikzstyle{wide kpoint}=[kpoint, minimum width=1 cm, inner sep=2pt]
\tikzstyle{wide kpointdag}=[kpointdag, minimum width=1 cm, inner sep=2pt]
\tikzstyle{wide kpointconj}=[kpointconj, minimum width=1 cm, inner sep=2pt]
\tikzstyle{wide kpointtrans}=[kpointtrans, minimum width=1 cm, inner sep=2pt]

\tikzstyle{gray kpoint}=[kpoint,fill=gray!50!white]
\tikzstyle{gray kpointdag}=[kpointdag,fill=gray!50!white]
\tikzstyle{gray kpointadj}=[kpointadj,fill=gray!50!white]
\tikzstyle{gray kpointconj}=[kpointconj,fill=gray!50!white]
\tikzstyle{gray kpointtrans}=[kpointtrans,fill=gray!50!white]

\tikzstyle{gray dkpoint}=[kpoint,fill=gray!50!white,doubled]
\tikzstyle{gray dkpointdag}=[kpointdag,fill=gray!50!white,doubled]
\tikzstyle{gray dkpointadj}=[kpointadj,fill=gray!50!white,doubled]
\tikzstyle{gray dkpointconj}=[kpointconj,fill=gray!50!white,doubled]
\tikzstyle{gray dkpointtrans}=[kpointtrans,fill=gray!50!white,doubled]

\tikzstyle{white label}=[draw,fill=white,rectangle,inner sep=0.7 mm]
\tikzstyle{gray label}=[draw,fill=gray!50!white,rectangle,inner sep=0.7 mm]
\tikzstyle{black label}=[draw,fill=black,rectangle,inner sep=0.7 mm]

\tikzstyle{dkpoint}=[kpoint,doubled]
\tikzstyle{wide dkpoint}=[wide kpoint,doubled]
\tikzstyle{dkpointdag}=[kpoint adjoint,doubled]
\tikzstyle{wide dkpointdag}=[wide kpointdag,doubled]
\tikzstyle{dkcopoint}=[kpoint adjoint,doubled]
\tikzstyle{dkpointadj}=[kpoint adjoint,doubled]
\tikzstyle{dkpointconj}=[kpoint conjugate,doubled]
\tikzstyle{dkpointtrans}=[kpoint transpose,doubled]

\tikzstyle{kscalar}=[kpoint common, shape=EBox, inner xsep=-1pt, inner ysep=3pt,font=\small]
\tikzstyle{kscalarconj}=[kpoint common, shape=WBox, inner xsep=-1pt, inner ysep=3pt,font=\small]

\tikzstyle{spekpoint}=[kpoint sc,minimum height=5mm,inner sep=3pt]
\tikzstyle{spekcopoint}=[kpoint adjoint sc,minimum height=5mm,inner sep=3pt]

\tikzstyle{dspekpoint}=[spekpoint,doubled]
\tikzstyle{dspekcopoint}=[spekcopoint,doubled]

 \tikzstyle{discard}=[ground,rotate=180,scale=1.5,inner sep=-2mm]
  \tikzstyle{upground}=[ground,rotate=180,scale=1,inner sep=-2mm]
 \tikzstyle{downground}=[circuit ee IEC,thick,ground,rotate=-90,scale=1.5,inner sep=-2mm]

\tikzstyle{maxmix}=[regular polygon,regular polygon sides=3,draw=black,xscale=0.4,yscale=0.3,inner sep=-0.5pt,minimum width=10mm,fill=gray,regular polygon rotate=180]

 \tikzstyle{bigground}=[regular polygon,regular polygon sides=3,draw=gray,scale=0.50,inner sep=-0.5pt,minimum width=10mm,fill=gray]

\tikzstyle{arrs}=[-latex,font=\small,auto]
\tikzstyle{arrow plain}=[arrs]
\tikzstyle{arrow dashed}=[dashed,arrs]
\tikzstyle{arrow bold}=[very thick,arrs]
\tikzstyle{arrow hide}=[draw=white!0,-]
\tikzstyle{arrow reverse}=[latex-]
\tikzstyle{cdnode}=[]

\tikzstyle{green dashed arrow}=[green, arrow dashed]
\tikzstyle{dashed blue}=[blue, dashed]
\tikzstyle{red dashed arrow}=[red, arrow dashed]
\tikzstyle{orange arrow}=[orange, arrs]
\tikzstyle{blue arrow}=[blue, arrs]
\tikzstyle{magenta arrow}=[magenta, arrs]

\tikzstyle{dotted line}=[-, style=dotted, tikzit draw=brown]
\tikzstyle{dashed line}=[-, style=dashed, tikzit draw=cyan]
\tikzstyle{green fill line}=[-, fill={green!90!black}, tikzit draw=green]
\tikzstyle{blue fill}=[-, fill=blue, tikzit fill=blue, tikzit draw={rgb,255: red,102; green,117; blue,255}]
\tikzstyle{red}=[-, draw=red, tikzit draw=red]
\tikzstyle{blue}=[-, draw=blue, tikzit draw=blue]
\tikzstyle{thick black}=[-, draw=black, tikzit draw=black, line width=1pt]
\tikzstyle{dotted red}=[-, draw=red, style=dotted, tikzit draw=red]
\tikzstyle{dotted blue}=[-, draw=blue, tikzit draw=blue, style=dotted]
\tikzstyle{dashed thick blue}=[-, draw={rgb,255: red,28; green,176; blue,255}, tikzit draw={rgb,255: red,83; green,19; blue,156}, line width=1pt, style=dashed]
\tikzstyle{dashed thick red}=[-, draw=red, tikzit draw={rgb,255: red,255; green,100; blue,10}, line width=1pt, style=dashed]
\tikzstyle{green}=[-, draw=green, tikzit draw=green]
\tikzstyle{dotted green}=[-, draw=green, tikzit draw=green, style=dotted]
\tikzstyle{arrow}=[->]
\tikzstyle{arrow green dashed}=[draw=green, ->, tikzit draw=green, style=dashed]
\tikzstyle{arrow dashed red}=[draw=red, ->, style=dashed, tikzit draw=red]
\tikzstyle{dashed green}=[-, tikzit draw=green, draw=green, style=dashed]

\newcommand{\ca}{\mathcal A}
\newcommand{\cb}{\mathcal B}
\newcommand{\cc}{\mathcal C}
\newcommand{\cd}{\mathcal D}

\newcommand{\ch}{\mathcal H}

\newcommand{\cl}{\mathcal L}

\newcommand{\co}{\mathcal O}
\newcommand{\calp}{\mathcal P}

\newcommand{\cs}{\mathcal S}

\newcommand{\cu}{\mathcal U}
\newcommand{\cv}{\mathcal V}
\newcommand{\cw}{\mathcal W}
\newcommand{\cx}{\mathcal X}

\newtheorem{theorem}{Theorem}
\newtheorem{lemma}{Lemma}
\newtheorem{definition}{Definition}
\newtheorem{corollary}{Corollary}
\begin{document}

\title{Quantum influences and event relativity}
    \author{Nick Ormrod\footnote{nicholas.ormrod@cs.ox.ac.uk} \ and Jonathan Barrett\footnote{jonathan.barrett@cs.ox.ac.uk} \\
Quantum Group, Department of Computer Science, University of Oxford}

\date{}
\maketitle

\begin{abstract}
    We develop a new interpretation of quantum theory by combining insights from extended Wigner's friend scenarios and quantum causal modelling. In this interpretation, which synthesizes ideas from relational quantum mechanics and consistent histories, events obtain relative to a set of systems, and correspond to projectors that are picked out by causal structure. We articulate these ideas using a precise mathematical formalism. Using this formalism, we show through specific examples and general constructions how quantum phenomena can be modelled and paradoxes avoided; how different scenarios may be classified and the framework of quantum causal models extended; and how one can approach decoherence and emergent classicality without relying on quantum states.
\end{abstract}

\section{Introduction}

Nearly a century after the core ideas of quantum theory were first stitched together, there remains little consensus over whether they paint us any clear, observer-independent picture of reality. It wouldn't be unreasonable to conclude that the task of interpreting quantum theory is too difficult; that a human being attempting to achieve a realistic understanding of the formalism is like a dog trying to figure out how a television works. Nevertheless, there are reasons for hope.

Two recent developments in particular suggest there is considerable progress yet to be made on quantum interpretation. Firstly, the articulation of \cite{wiseman2015causarum, bong2020strong}, and the no-go theorems for \cite{bong2020strong, haddara2022possibilistic, leegwater2022greenberger, healey2018quantum, ormrod2022no, wiseman2022thoughtful, ormrod2023theories}, the \textit{absoluteness of observed events}. At a minimum, absoluteness assumes that there are unique and nonrelational facts about what is observed. So, according to absoluteness, if Alice sees a pointer indicate ``up'', then there can be no sense in which she also sees it indicate ``down''. But if absoluteness is denied, it might be claimed that Alice sees the pointer indicate ``down'' in another world, or relative to another ``perspective'', or something similar.

Although the absoluteness assumption was presumably once thought to be self-evident, it is becoming increasingly clear that it is in tension with the universality of unitary quantum theory. For example, the local friendliness theorem \cite{bong2020strong, haddara2022possibilistic} shows that the only way of reconciling unitary quantum theory with absoluteness is through nonlocal causal influences, superdeterminism, or retrocausality. Other results \cite{leegwater2022greenberger, healey2018quantum, ormrod2022no} show that nonabsoluteness is inevitable if unitary quantum theory can be applied to calculate joint probability distributions for arbitrary spacelike separated measurements. And a recent paper \cite{ormrod2023theories} shows that even when unitary quantum theory is not assumed, some very natural properties of a general theory always lead to a similar no-go result for absoluteness.

So perhaps a good interpretation of quantum theory should deny absoluteness. Unfortunately, that is easier said than done. That is, while it is easy to state the absoluteness assumption positively, it is much harder to articulate a satisfactory sense in which absoluteness might fail. For example, the Everettian brand of nonabsoluteness involves a multiplicity of worlds, many or most of which do not follow the Born rule frequencies. At least prima facie, this would appear to undermine our usual conviction that observing Born rule frequencies gives us a reason to believe in quantum theory, while observing other frequencies would give us a reason to reject it \cite{albert2010probability, adlam2014problem} (although see e.g.\ \cite{wallace2012emergent, greaves2010everett} for Everettian responses). Or, take the sort of nonabsoluteness that arises from the consistent histories  \cite{griffiths1984consistent} formalism.\footnote{In particular, we refer to the sort of nonabsoluteness from the ``many histories'' interpretation of the formalism \cite{dowker1996consistent}.} As Dowker and Kent \cite{dowker1996consistent} have argued extensively, this version of nonabsoluteness leads to issues with predicting and explaining the persistently approximately classical character of our experiences.

As a final example, consider Rovelli's relational quantum mechanics (RQM) \cite{rovelli1996relational}. This approach doesn't appear to suffer the same objections as Everett and many histories just outlined, but it does face an even more fundamental problem: vagueness. Historically, physics has achieved conceptual clarity and quantitative precision by formalizing its claims mathematically where possible. But RQM's most important and least intuitive claims are often stately almost entirely in the English language. For example,
\begin{quote}
    ``Events happen in interactions between any two systems and can be described as the actualisation of the value of a variable of one system relative to the other.'' \cite{di2022relational}
\end{quote}
As will become clear from our approach, we think this statement contains some important insights, especially regarding the connection between events and interaction. But without a clear formalism in which to anchor one's understanding of ambiguous notions such as ``interaction''  and ``relative'', it isn't clear exactly what is being said. 

But that's where the second recent development comes in; the second ``reason for hope''. We are speaking of the discovery of a natural, elegant, and useful theory of causation in quantum mechanics, in the form of quantum causal models \cite{Allen_2017, barrett2020quantum, barrett2021cyclic, ormrod2023causal}. At the heart of this framework is the assumption that causal influences should be defined quantum-theoretically. In this paper, we will show that such \textit{quantum influences} provide a natural understanding of \textit{event relativity}. Quantum influences allow one to pin down the idea that events emerge out of the interactions between a given subset of systems, and events are naturally relational on such a view because there are many different subsets. The absoluteness assumption fails because Alice might see ``up'' relative to one set of interacting systems, and ``down'' relative to another. Ultimately, this will lead us to a precise, observer-independent, and relational interpretation of the quantum theory of finite-dimensional unitary circuits, which aims to combine the best parts of Everett, consistent histories, and relational quantum mechanics (RQM).


In a little more detail, at the heart of this interpretation lies the observation that the causal structure of a set of unitarily interacting systems singles out certain families of projectors onto those systems -- families that are decoherent relative to the other systems under consideration. Remarkably, the families of projectors that get selected are \textit{special} enough such that they always generate a consistent set of histories, meaning they naturally give rise to the idea of events stochastically distributed according to the (generalized) Born rule, and, at the same time, they are \textit{general} enough to model any phenomena from standard operational quantum theory, plus (extended) Wigner's friend scenarios. The interpretation claims that reality is described by a unitary circuit, and, relative to every subset of its systems (i.e.\ its wires), exactly one consistent history gets realized.

First, we will lay out in greater detail the problem we are aiming to solve, outlining the major pitfalls of ``standard quantum theory'', as well as the merits and shortcomings of the attempt to overcome them using consistent histories (Section \ref{sec:interpretations_ch}).
After that, we will gradually introduce the core ideas of our interpretation (Sections \ref{sec:influences}, \ref{sec:events}), then axiomatize it, apply it in several scenarios, and show that it can reproduce any operational phenomena from the standard theory (Section \ref{sec:theory_models}). We then introduce a classification scheme for various quantum phenomena using a central concept from the interpretation (Section \ref{sec:classications}). As we shall argue, this result appears as if it may form the basis for an extension to quantum causal modelling in which (roughly speaking) different proofs of nonclassicality are shown to be equivalent to different fine-grained quantum causal structures. Finally, we discuss whether the interpretation can be viewed as describing fundamental physics (Section \ref{sec:status}), before we conclude by discussing possible applications of the formalism to emergent classicality and quantum gravity (Section \ref{sec:discussion}).

\section{Interpretation and consistent histories} \label{sec:interpretations_ch}

We start by explaining exactly what problem we are aiming to address by introducing an interpretation of quantum theory. This will lead us to a discussion of the consistent histories formalism \cite{griffiths1984consistent}, which aims to address a similar problem, and which is the appropriate starting point for the interpretation we will introduce. \\

\textbf{Interpretation.} Why did we suggest that the standard quantum theory is not ``clear'', or ``observer-independent''? The problem lies in its vague and dualistic approach to dynamics. The theory tells us that there are two types of evolution: the reversible and linear unitary time evolution, and the irreversible and nonlinear ``collapse of the wavefunction''. But we are not then given an underlying dynamical rule that synthesizes the two evolutions, nor even a precise answer to when (or why) one evolution takes over from the other. Of course, it is usually said that nonlinearity takes over when a ``measurement'' takes place. But then we are not given a precise definition of a measurement!\footnote{We are told what happens when a measurement takes place (e.g.\ the quantum state collapses onto an eigenstate of the measured observable), but we are not told when the measurement does take place.} As  memorably put in \cite{maudlin1995three}, ``[t]his is not
much better than saying that the evolution is linear
except when it is cloudy, and saying no more about how
many, or what kind, of clouds precipitate this radical
shift in the operation of fundamental physical law''. In his essay \textit{Against ``measurement''}, John Bell also laments the vagueness:
\begin{quote}
What exactly qualifies some physical systems to play the role of `measurer’? Was the wavefunction of the world waiting to jump for thousands of millions of years until a single-celled living creature appeared? Or did it have to wait a little longer, for
some better qualified system... with a PhD?
 \cite{bell1990against}
\end{quote}

One might object that this vagueness isn't so problematic if we regard physics as only a tool for predicting what we shall observe. Don't we know well enough in practice what a measurement is, and isn't quantum theory good enough at predicting probabilities for measurement outcomes, despite the vagueness? Even if we ignore the fact that there are conceivable experiments for which the standard theory fails to make clear predictions (e.g.\ (extended) Wigner's friend scenarios), this attitude has the pitfall that it makes it harder to address a number of important physical questions. For example, cosmology is generally considered a legitimate field of physics, but no observer measures the universe as a whole. Or, it is generally regarded as important to understand how an approximately classical world can emerge from an underlying quantum reality. But such an account will not satisfy many if it has to fall back on the phrase ``because we measure it'', without spelling out exactly what that means in physical terms. And it is hard not to suspect that the vagueness of quantum theory itself is a significant part of why we struggle to construct an adequate quantum theory of gravity. It therefore seems there is much to be gained from a more precise and less anthropocentric formulation of quantum theory, independently of one's preferred philosophy of physics.

Coming up with such a formulation is what we mean when we talk about ``interpreting'' quantum theory. Note, then, that for us, providing an interpretation does \textit{not} mean offering a series of principles formulated in the English language that aim to explain the formalism of standard quantum theory. Rather, one needs a better formalism, in which imprecise or anthropocentric notions such as measurement no longer play a fundamental role. Since the problem is vagueness, the solution must be clarity. 

We therefore say that an interpretation of quantum theory is really just a more precise and less anthropocentric theory, complete with a mathematical formalism that makes its physical claims clear. On top of these basic requirements, we assume that the following are all desirable features in an interpretation:
\begin{enumerate}
    \item it maintains that all time evolution is unitary,
    \item it nevertheless avoids describing the universe as a unitarily evolving quantum state,\footnote{Part of our motivation for (ii) comes from the argument from \cite{brown2005solving} that interpretations like Bohm theory that postulate a unitarily evolving quantum state \textit{plus other stuff} are simply committed to the Everettian hypothesis that the world is a branching multiverse, \textit{plus other stuff}. If this is right, and if, as we suspect, the Everett interpretation cannot satisfactorily recover the Born rule, then it would seem that any interpretation that posits a unitarily evolving quantum state faces a similar problem.}
    \item it adds very little additional structure to the standard quantum formalism, and 
    \item it nevertheless introduces considerable explanatory power.
\end{enumerate}

The goal of this paper is to develop an interpretation of quantum theory that achieves these desiderata by combining quantum influences with event relativity. \\

\textbf{Consistent histories.} The interpretation that we will develop can be seen as a refinement of the consistent histories formalism \cite{griffiths1984consistent}, which already achieves some of these desiderata. The formalism is based on the insight that standard quantum theory can be purged of dynamical dualism simply by restricting its use. 

To explain how, it is useful to first describe in detail why the dynamical dualism is necessary in the standard theory. In the standard theory, a probability distribution for the outcomes $o_k$ of $N$ projector-valued measurements (PVMs) $\{P_k^{o_k}\}_{o_k}$ performed in sequence on an initial quantum state $\rho$ is given by
\begin{equation} \label{eqLch_prob}
\begin{split}
        p(o_1, \ldots, o_N) &= {\rm Tr}(P_N^{o_n}U_N \ldots U_2P_1^{o_1}U_1 \rho U_1^\dag P_1^{o_1} U_2^\dag \ldots U_N^\dag P_N^{o_N}) 
        \\ &={\rm Tr}(\tilde{P}_N^{o_N}\ldots \tilde{P}_1^{o_1} \rho \tilde{P}_1^{o_1}\ldots \tilde{P}_N^{o_N}),
\end{split}
\end{equation}
where we have assumed unitary evolution in between the measurements, and tildes denote Heisenberg representations of projectors $\tilde{P}_k^{o_k}:=U_1^\dag \ldots U_k^\dag P_k^{o_k} U_k \ldots U_1$.
But what if one then wants to calculate the joint probabilities for a different experiment in which, for example, the second PVM was omitted? In general, they will \textit{not} be given by marginalizing over $o_2$ in the expression above. That is, the probabilities for the remaining $N-1$ measurements depend on whether or not the second measurement was performed. To explain this, we are pushed towards the idea that measurements cause a nonlinear ``disturbance'' to the otherwise linear unitary evolution. But this leads to the vagueness and anthropocentricity that was lamented a moment ago.

To avoid the dualism, the consistent histories formalism simply suggests that we restrict our attention to the (very) special cases where $p(o_1, \ldots, o_N)$ happens to be linear in each projector (due to the particular choice of $\rho$ and PVMs). In these special cases, the decision of whether or not to perform a measurement does not affect the probabilities for the remaining measurements,\footnote{Explicitly: the linearity of $p(o_1, \ldots, o_N)$ in e.g.\ $P_2^{o_2}$ is equivalent to the statement that Re(Tr$(P_N^{o_N}\ldots P_2^{o_2}P_1^{o_1} \rho P_1^{o_1}P_2^{o_2'}\ldots P_N^{o_N}))=0$ for all $o_2 \neq o_2'$, but this means that $\sum_{o_2} p(o_1, \ldots, o_N)= \sum_{o_2o_2'} {\rm Tr}((P_N^{o_N}\ldots P_2^{o_2}P_1^{o_1} \rho P_1^{o_1}P_2^{o_2'}\ldots P_N^{o_N})={\rm Tr}(P_N^{o_N}\ldots P_3^{o_3}P_1^{o_1}\rho P_1^{o_1}P_3^{o_3} \ldots P_N^{o_N})$, which is precisely the formula we would use if we omitted the second PVM. Thus performing the second measurement doesn't ``disturb'' the probabilities for the other measurement outcomes.}  so one does not need to posit measurement disturbance, or any interruption to the unitary dynamics. 

Naively, one might expect that when one restricts the use of the standard formalism in this way, certain phenomena from standard quantum theory can no longer be modelled. But fortunately, this turns out not to be the case: by representing measurements explicitly as unitary interactions, and considering projectors onto the measurement devices rather than onto the measured system itself, one can always take a model in which (\ref{eqLch_prob}) is nonlinear and transform it to one in which it is linear, and yet in which the same statistics are reproduced. Thus one can argue that the notions of measurement disturbance, collapse of the wavefunction, and nonlinear evolution are not necessary for quantum theory, after all. Rather, they only appear necessary when one fails to take into account all of the systems whose interaction is required for a measurement to take place.

Because a consistent historian no longer needs to invoke the idea of measurement disturbance, they also no longer need to think of the $o_k$ as ``measurement outcomes'' at all. Instead, they might be thought of as physical \textit{events}, which may or may not be involved in some measurement. We shall reflect this with a change of notation, writing $e_k$ instead of $o_k$ from now on. Similarly, there is no longer any need to think of the $\{P_k^{e_k}\}_{e_k}$ as projector-valued ``measurements''. Instead, from now on we adopt the more neutral terminology of \textit{projective decompositions} (which, like PVMs, are mathematically defined as a family of orthogonal projects that sum to the identity operator). A list $(e_1, \ldots, e_N)$ shall be called a history, and the complete set of such histories associated with some linear probability function is called a \textit{consistent set} of histories.

Using the consistent histories approach, one can maintain universal unitarity (desideratum 1) without adding much if anything to the formalism (desideratum 3). But the status of the quantum state on this approach remains not entirely clear \cite{wallace2008philosophy}. Also, it is very important to note that there are very many different and incompatible consistent sets of histories -- in general, an uncountably infinite number, corresponding to the continuum of different sets of projective decompositions that lead to a linear probability function. Should one then say that each consistent set has an equal physical significance? That is, given some sequence of unitary transformations $\{U_i\}_{i=1}^N$, should one say that a consistent history gets realized relative to every definable set of consistent histories?

On the one hand, the absoluteness no-go theorems \cite{bong2020strong, haddara2022possibilistic, leegwater2022greenberger, healey2018quantum, ormrod2022no, wiseman2022thoughtful, ormrod2023theories} might serve as a justification for saying ``yes''. But on the other hand, it isn't clear that this very radically relational approach has much predictive power. For example, one obviously wants to be able to unambiguously predict that the sun will rise tomorrow morning, but most consistent sets of histories that have described the sun rising in the past will not describe it rising in the future (because they don't involve the relevant projectors). Should we therefore be surprised tomorrow at dawn?

One might conclude that, while the no-go theorems for absolute events motivate \textit{some} notion of event relativity, the sort of event relativity that most obviously arises from consistent histories is too extreme. By appealing to quantum influences, this paper shall articulate a less extreme conception of event relativity. We shall show how, relative to a given subset of a set of unitarily interacting subsystems, a \textit{unique} consistent set of histories is privileged by causal structure. It will then be possible to postulate that relative to any set of systems, precisely one history is realized with a probability given by (\ref{eqLch_prob}). On this view, events and histories fail to be absolute precisely because they are relative to sets of systems. And so, relative to any set of systems, one can make clear predictions about whether the sun will rise again.

\section{Interference influences} \label{sec:influences}

This section will introduce the causal concepts that will subsequently be deployed to articulate a more satisfactory conception of event relativity, and, ultimately, to interpret quantum theory.\\

\textbf{Background.} Throughout this paper, we understand causal influences as \textit{dynamical dependencies}. In the classical case, an influence from one variable to another would mean that the second variable depends nontrivially on the first in a function that describes the dynamics. 
In the quantum case, an influence from one system to another means that the second system depends nontrivially on the first in a unitary channel that describes the dynamics.\footnote{We note that on this approach causal influences can be no more essentially connected with agents and signalling than dynamics are. In particular, it would be wrong to conflate causation with signalling, since there might be a dynamical connection between systems that agents cannot exploit because they lack knowledge of some relevant parameters.} Let us now make this statement more precise.  Given a unitary channel $\cu: A \otimes B \rightarrow C \otimes D$, we say that there is \textbf{no} quantum causal influence from $A$ to $D$, written $A \not\rightarrow D$, if and only if the channel obtained from tracing out $C$ is equivalent to a channel that traces out $A$. That is, 

\begin{equation} \label{eq:ni_def_no_diagram}
  A \not\rightarrow D \quad \Longleftrightarrow \quad \exists \cd: B \rightarrow D \ \  {\rm such \ that} \  \   {\rm Tr}_C\cu(\cdot) = \cd( {\rm Tr}_A(\cdot)).
\end{equation}
Or, writing the exact same expression diagrammatically:
\begin{equation} \label{eq:ni_def_diagram}
   A \not\rightarrow D \quad \Longleftrightarrow \quad \exists \cd \  {\rm such \ that} \  \begin{tikzpicture}
	\begin{pgfonlayer}{nodelayer}
		\node [style=large map] (0) at (1.5, 0) {$\cu$};
		\node [style=none] (1) at (0.5, -2) {};
		\node [style=none] (3) at (2.5, -2) {};
		\node [style=none] (4) at (2.5, 2) {};
		\node [style=none] (6) at (4, 0) {$=$};
		\node [style=none] (7) at (5.5, -2) {};
		\node [style=upground] (9) at (5.5, -1.5) {};
		\node [style=map] (10) at (7.5, 0) {$\cd$};
		\node [style=none] (11) at (7.5, -2) {};
		\node [style=none] (12) at (7.5, 2) {};
		\node [style=none] (13) at (0, -1.5) {$A$};
		\node [style=none] (14) at (3, -1.5) {$B$};
		\node [style=none] (15) at (3, 1.5) {$D$};
		\node [style=none] (16) at (-0.25, 1.5) {$C$};
		\node [style=upground] (17) at (0.5, 1.1) {};
		\node [style=none] (18) at (5, -1.5) {$A$};
		\node [style=none] (19) at (8, -1.5) {$B$};
		\node [style=none] (20) at (8, 1.5) {$D$};
	\end{pgfonlayer}
	\begin{pgfonlayer}{edgelayer}
		\draw (3.center) to (4.center);
		\draw (11.center) to (12.center);
		\draw (7.center) to (9);
		\draw (1.center) to (17);
	\end{pgfonlayer}
\end{tikzpicture}    
\end{equation}

This turns out to be equivalent to many other definitions of a quantum influence, some of which express very different intuitions \cite{schumacher2005locality, ormrod2023causal}. A particularly salient one is the stipulation that all operators on the systems commute in the Heisenberg picture:
\begin{equation} \label{eq:quantuminfluence_comm}
    A \not\rightarrow D \quad \Longleftrightarrow \quad [M_A \otimes I_B, \ \cu^\dag(I_C \otimes N_D)] = 0 \ \forall M_A, N_D
\end{equation}
This makes it easy to see 
some attractive properties \cite{barrett2020quantum, barrett2021cyclic} of quantum influences. One of these is a \textit{time-symmetry} property, that $A$ influences $D$ through $\cu$ if and only if $D$ influences $A$ through $\cu^\dag$. Another is \textit{causal atomicity}, that influences among composite systems are uniquely fixed by influences among the most elementary subsystems. For instance, $A$ influences the composite system $D_1 \otimes D_2$ if and only if $A$ influences $D_1$ or $A$ influences $D_2$. In a theory known for its nonseparable state space, this is quite remarkable.

Quantum influences permit causal models of Bell inequality violations without superluminal influences; something that is not possible using classical causal models without retrocausality or superdeterminism. Moreover, they permit causal models of the violations that do not require fine-tuning; something that is entirely impossible with classical causal models \cite{Wood_2015}. 

In summary, quantum influences are naturally defined using the standard formalism (good from the point of view of desideratum (3)), they have nice properties, and they help us to respond to Bell inequality violations. In light of all of this, it seems natural to also look to quantum influences for guidance in the quest for a satisfactory conception of event relativity. \\

\textbf{Interference influences.} We can find such guidance, but it requires a more fine-grained conception of a quantum influence -- one that holds between particular projective decompositions $\{P_A^i\}$ and $\{P_D^j\}$ rather than between $A$ and $D$ themselves. (\ref{eq:quantuminfluence_comm}) suggests that we could say such an influence holds if some of the Heisenberg projectors, defined by
    \begin{equation}
    \begin{split}
        \tilde{P}_A^i :&= P_A^i \otimes I_B \\
        \tilde{P}_D^j :&= \cu^\dag(I_C \otimes P_D^j),
    \end{split}
    \end{equation}
    do not commute. And it turns out that this idea is indeed equivalent to a particular sort of dynamical dependence. 
\begin{theorem} \label{thm:int_equivalence}
    Consider a unitary channel $\cu: A \otimes B \rightarrow C \otimes D$, the projective decomposition $\{P_A^i\}$ on $A$, and the projective decomposition $\{P_D^j\}$ on $D$.
    Then 
    \begin{equation} \label{eq:int_inf_op}
    \begin{split}
        [\tilde{P}_A^i, \tilde{P}_D^j] = 0 \ \ \forall i, j \\
        \Longleftrightarrow {\rm Tr} \big((I_C \otimes P_D^j)\cu(V_{\vec{\phi}}(\cdot)V_{\vec{\phi}}^\dag)\big)={\rm Tr} \big((I_C \otimes P_D^j)\cu(\cdot)\big)  \ \ \forall j, \ \forall V_{\vec{\phi}}
    \end{split}
    \end{equation}
    where  $V_{\vec{\phi}}$ is any unitary of the form $V_{\vec{\phi}} = \sum_i e^{i\phi_i}P_A^i \otimes I_B$.
\end{theorem}

Theorem \ref{thm:int_equivalence} is proven in Appendix \ref{app:interference_equivalence}. Thinking in terms of standard quantum theory for a moment, we see that the Heisenberg projectors fail to commute if and only if a message can be sent by shifting the relative phases between the $P_A^i$ and then received by performing the PVM defined by $\{P_D^j\}$. Again, noncommutation obtains if and only if the selection of a unique $P_D^j$ is sensitive to the \textit{interference} between the different $P_A^i$ -- if $\{P_A^i\}$ is not \textit{decoherent} from the point of view of $\{P_D^j\}$. We therefore call this influence relation an \textit{interference influence}.

\begin{definition} \label{def:int}
   Given a unitary channel $\cu: A \otimes B \rightarrow C \otimes D$, there is \textbf{no} interference influence from $\{P_A^i\}$ to $\{P_D^j\}$, written $\{P_A^i\} \not\rightarrow  \{P_D^j\}$, if and only if $[\tilde{P}_A^i, \tilde{P}_D^j] = 0 \ \ \forall i, j$.
\end{definition}

It is evident from their definition that interference influences have a time symmetry property. It is also clear that they satisfy a sort of causal atomicity, that there is an influence from $A$ to $D$ if and only if there is an interference influence between at least one pair $(\{P_A^i\}, \{P_D^j\})$ of associated projective decompositions:
\begin{equation} \label{eq:fine_graining}
    A \rightarrow D \quad \Longleftrightarrow \quad \exists \{P_A^i\}, \ \{P_D^j\}: \ \{P_A^i\} \rightarrow \{P_D^j\} 
\end{equation}
This property means that the list of all the interference influences between projective decompositions on the input and output subsystems of a unitary channel contains strictly more information than the list of all the quantum influences between the subsystems themselves. In this sense, interference influences are a fine-graining of the quantum influences on which the causal models of \cite{barrett2020quantum, barrett2021cyclic} are based.

The lack of an interference influence provides a notion of decoherence that is time-symmetric and defined entirely in terms of dynamics rather than states. So it is natural to suspect that interference influences might be of use in describing the emergence of events from dynamics.

\section{Event relativity} \label{sec:events}

And indeed they are. As in consistent histories \cite{griffiths1984consistent}, we wish to understand an event $e$ as the (stochastic) selection of a unique projector $P^e \in \mathbb{D}$ from a projective decomposition $\mathbb{D}$. But which projective decomposition? This section will show that certain projective decompositions are privileged by causal structure, and give rise to consistent sets of histories. And this will allow us to conceive of events as emerging out of causation.

RQM hints at a similar idea with its suggestion that events arise from interactions \cite{rovelli1996relational, di2022relational}. But, as Brukner points out \cite{brukner2021qubits}, if an interaction is conceived in terms of some entangled state that it produces, say $\ket{\Phi^+}=\frac{1}{\sqrt{2}}(\ket{0}\ket{0}+\ket{1}\ket{1})$, it doesn't generally select any preferred decomposition. The response from Adlam and Rovelli \cite{adlam2022information} is to let a thousand flowers bloom: perhaps very many events happen in such situations, and unique decompositions are only singled out in other situations where there are more complicated quantum states.

On the other hand, a solution readily presents itself when one turns away from states, and towards the unitary process itself. Speaking operationally for a moment, the $\ket{\Phi^+}$ state might have been produced by applying the unitary transformation
\begin{equation}
\texttt{CNOT} := \sum_{i, j=0}^1 \ket{i}_C\ket{j+i}_D\bra{i}_A\bra{j}_B
\end{equation}
to the states $\ket{+}_A:=\frac{1}{\sqrt{2}}(\ket{0}_A+\ket{1}_A)$ and $\ket{0}_B$. Now, it is easily shown that it is impossible for an agent who can only apply phase shifts of the form $V_A =\ket{0}\bra{0}_A + e^{i\phi} \ket{1}\bra{1}_A$ before \texttt{CNOT} is implemented to signal to an agent who measures $D$ after it is implemented. On the other hand, any unitary of the form $V_A = P_A^0 + e^{i\phi} P_A^1$ for $\phi \neq 0$ and projective decomposition $\{P_A^0, P_A^1\} \neq \{\ket{0}\bra{0}_A, \ket{1}\bra{1}_A\}$ \textit{will} allow signalling. 

This suggests that although $\{\ket{0}\bra{0}_A, \ket{1}\bra{1}_A\}$ is not marked out as special by the state $\ket{\Phi^+}$, it \textit{is} marked out as special by the transformation \texttt{CNOT} -- at least relative to the system $D$. With the help of interference influences, we now strip this account of operationalism, and generalize it to arbitrary finite-dimensional unitary transformations.

\begin{definition} \label{def:pref1}
    Given a unitary channel $\cu: A \otimes B \rightarrow C \otimes D$ we say that the projective decomposition $\{P_A^i\}$ is \textit{preferred} by $D$ if and only if the following three conditions are met:
\begin{enumerate}
        \item $\{P_A^i\}$ does not exert an interference influence on any projective decomposition on $D$: $\forall \{P_D^j\}: \{P_A^i\} \not\rightarrow \{P_D^j\} \ $.
    \item If $\{P_A^i\}$  is incompatible with some other projective decomposition $\{Q_A^k\}$, then $\{Q_A^k\}$ exerts an interference influence on at least one projective decomposition on $D$: $\exists i, k: [P_A^i, Q_A^k]\neq0 \implies \{Q_A^k\} \rightarrow \{P_D^j\} $.
    \item Any other decomposition $\{R_A^l\}$ satisfying (1) and (2) is a coarse-graining of $\{P_A^i\}$, in the sense that $\{R_A^l\} \subseteq \texttt{span}(\{P_A^i\})$.
\end{enumerate}
\end{definition}
In short, the preferred $\{P_A^i\}$ is the most fine-grained decomposition such that $D$ is insensitive to relative phase shifts between the $P_A^i$, but \textit{is} sensitive to shifts between projectors that are incompatible with them. So $\{P_A^i\}$ might be thought of as the canonically decoherent projective decomposition from the point of view of $D$.

The notion of preference can be stated more compactly with the help of some algebraic concepts. Given the algebra of linear operators $\cl(\ch)$ on a finite-dimensional Hilbert space $\ch$, the \textit{commutant} $\texttt{comm}(\cx)$ of a subalgebra $\cx \subseteq \cl(\ch)$ is the set of all operators in $\cl(\ch)$ that commute with every operator in $\cx$. The \textit{centre} of $\cx$ is the intersection of $\cx$ with its own commutant, $\texttt{centre}(\cx):= \texttt{comm}(\cx) \cap \cx$.  (In other words, it is the set of operators within $\cx$ that commute with all other operators in $\cx$.) It follows from the lemma in Appendix \ref{app:structure} that $\texttt{centre}(\cx)$ can always be uniquely written as the set of operators obtained from linear complex combinations of projectors from some projective decomposition on the Hilbert space, $\texttt{centre}(\cx)= \texttt{span}(\{P^i\})$. We then have the following theorem, proven in Appendix \ref{app:preference_equivalence}.

\begin{theorem} \label{thm:preference} Consider a unitary channel $\cu: A \otimes B \rightarrow C \otimes D$ and a projective decomposition $\{P_A^i\}$ on $A$. Let $\ca$ denote the algebra of operators of the form $M_A \otimes I_B$, and $\cd$ the algebra of operators of the form $\cu^\dag(I_C \otimes M_D)$.
Then 
\begin{equation}
    D \ \textnormal{prefers} \ \{P_A^i\}  \quad \Longleftrightarrow  \quad \textnormal{\texttt{span}}(\{P_A^i\}) \otimes I_B = \textnormal{\texttt{centre}} (\ca \cap \textnormal{\texttt{comm}}(\cd)). 
\end{equation} 
\end{theorem} 

That is, $D$ prefers the most fine-grained projectors on $A$ that not only commute with all operators on $D$, but also commute with all operators on $A$ that commute with all operators on $D$.

Let us take stock of our current situation. We have seen a sense in which a decomposition on an input to a unitary transformation is preferred by an output of the unitary transformation. But remember, what we really wanted was a preferred \textit{set} of decompositions associated with a given subset of unitarily interacting systems. (And not just any set of decompositions, but one that gives rise to a consistent set of histories for that subset of systems.) To that end, we shall consider a circuit $\mathfrak{C}$ made up of finite-dimensional unitary transformations, and some subset $\mathfrak{B}$ of the systems (i.e.\ the wires). Inspired by \cite{cavalcanti2021view}, we call $\mathfrak{B}$ a \textit{bubble}. We define the preferred set $\mathfrak{P}(\mathfrak{C}, \mathfrak{B})$ of decompositions by taking two decompositions for each system in the bubble; one which is obtained by its interactions with systems in its future; the other, with systems in its past. The following definition makes this more precise.

\begin{figure}
    \centering
    \input{preference_algorithm.tikz}
    \caption{The rule for obtaining the preferred set of decompositions $\mathfrak{P}(\mathfrak{C}, \{A_1, A_2, A_3\})$ for the bubble $\{A_1, A_2, A_3\}$. Here, and throughout the paper, circuits should be read from bottom to top. The colour-coding of wires is only for better readability, and has no formal meaning.}
    \label{fig:preference}
\end{figure}

\begin{definition} \label{def:pref2}
    Given a circuit $\mathfrak{C}$ and a bubble $\mathfrak{B}$, the preferred set of projective decompositions $\mathfrak{P}(\mathfrak{C}, \mathfrak{B})$ is obtained in the following way, as illustrated in Figure \ref{fig:preference}. First, one obtains a broken unitary circuit by making incisions in every wire representing a system $A_k \in \mathfrak{B}$ in the bubble $\mathfrak{B}=\{A_k\}_{k=1}^n$. One then inserts swap gates with an ancilla into each node; or, equivalently, one ``pulls up'' the wires $A_k^{\rm in}$ that go into the new incisions, and pulls down the wires $A_k^{\rm out}$ that come out of them. This results in a single-shot unitary channel $\cu$, whose outputs include the $A_k^{\rm in}$ and inputs include the $A_k^{\rm out}$. We then apply our existing notion of preference to $\cu$. For every input $A_k^{\rm out}$, we take the projective decomposition $\{P_{A_k^{\rm out}}^{e_k'}\}$ preferred by the tensor product $\bigotimes_m A_m^{\rm in}$ of all of the outputs of $\cu$ corresponding to systems in the bubble. Since there is no reason to introduce a temporal asymmetry, we also take for each $A_k^{\rm in}$ the decomposition $\{P_{A_k^{\rm in}}^{e_k}\}$ preferred by $\bigotimes_m A_m^{\rm out}$ given $\cu^\dag$. $\mathfrak{P}(\mathfrak{C}, \mathfrak{B})$ is the set of $2n$ projective decompositions obtained in this way.
\end{definition}

Let us assume from now on, and without loss of generality, that $A_k$ comes higher up than $A_m$ in the circuit whenever $m < k$, so $A_k$ can be thought of as coming later in time\footnote{For ease of expression, we often equate the partial order naturally induced by a unitary circuit with a temporal order. However, the reader should bear in mind that, strictly speaking, the framework here, and the interpretation that we shall develop, is background independent -- we do not explicitly consider the circuits as embedded in spacetime. As we shall discuss later on, we consider it natural to imagine that spacetime itself emerges from quantum causal structure, just as events do.} than $A_m$. Then, by the definition of the unitary channel $\cu$, there is no quantum influence from $A_k^{\rm out}$ to $A_m^{\rm in}$ through $\cu$ for $m \leq k$. It follows from (\ref{eq:quantuminfluence_comm}) and Definition \ref{def:pref1} that the decomposition $\{P_{A_k^{\rm out}}^{e_k'}\}$ that is preferred by $\bigotimes_m A_m^{\rm in}$ is identical to the one preferred by $\bigotimes_{k > m} A_k^{\rm in}$, the ingoing systems in its future. So the preferred outgoing decomposition $\{P_{A_k^{\rm out}}^{e_k'}\}$ on $A_k$ is entirely determined by its interaction with its future, while, similarly, the preferred ingoing $\{P_{A_k^{\rm in}}^{e_k}\}$ is determined entirely its interaction with its past.

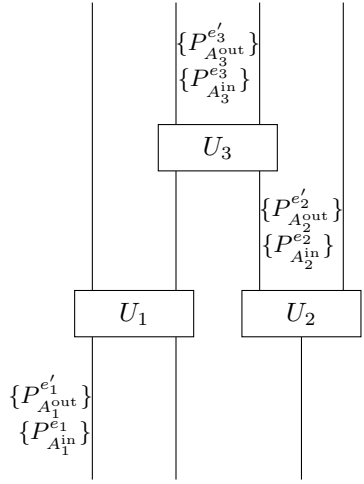
\begin{figure}
    \centering
    \begin{tikzpicture}[scale=1.1]
	\begin{pgfonlayer}{nodelayer}
		\node [style=large map] (0) at (2.5, -0.5) {$U_1$};
		\node [style=large map] (1) at (4.5, 3.5) {$U_3$};
		\node [style=large map] (2) at (6.5, -0.5) {$U_2$};
		\node [style=none] (3) at (3.5, -4.5) {};
		\node [style=none] (4) at (3.5, 7) {};
		\node [style=none] (5) at (1.5, -4.5) {};
		\node [style=none] (6) at (1.5, 7) {};
		\node [style=none] (7) at (6.5, -4.5) {};
		\node [style=none] (8) at (7.5, -0.5) {};
		\node [style=none] (9) at (7.5, 7) {};
		\node [style=none] (10) at (5.5, -0.5) {};
		\node [style=none] (11) at (5.5, 7) {};
		\node [style=none] (12) at (0.6, -3.5) {\small $\{P_{A_1^{\rm in}}^{e_1}\}$};
		\node [style=none] (13) at (6.42, 1) {\small $\{P_{A_2^{\rm in}}^{e_2}\}$};
		\node [style=none] (14) at (4.5, 6) {\small $\{P_{A_3^{\rm out}}^{e_3'}\}$};
		\node [style=none] (15) at (0.5, -2.5) {\small $\{P_{A_1^{\rm out}}^{e_1'}\}$};
		\node [style=none] (16) at (6.5, 2) {\small $\{P_{A_2^{\rm out}}^{e_2'}\}$};
		\node [style=none] (17) at (4.42, 5) {\small $\{P_{A_3^{\rm in}}^{e_3}\}$};
		\node [style=none] (18) at (4.25, -3) {};
	\end{pgfonlayer}
	\begin{pgfonlayer}{edgelayer}
		\draw (4.center) to (3.center);
		\draw (6.center) to (5.center);
		\draw (7.center) to (2);
		\draw (8.center) to (9.center);
		\draw (10.center) to (11.center);
	\end{pgfonlayer}
\end{tikzpicture}
    \caption{\small To determine which interference influences exist between decompositions in $\mathfrak{P}(\mathfrak{C}, \mathfrak{B})$, one has to (1) decorate the original unitary circuit with the projectors, with outgoing decompositions placed just after ingoing ones, and then (2) check whether there is an interference influence from one projective decomposition to another one higher in the circuit given overall unitary transformation in the intervening temporal region that connects them. For example, to check whether $\{P_{A_1^{\rm in}}^{e_1}\} \rightarrow \{P_{A_3^{\rm out}}^{e_3'}\}$ through the circuit above, one applies Definition \ref{def:int} to the unitary transformation $(I \otimes U_3 \otimes I)(U_1 \otimes U_2)$. Or, to check whether $\{P_{A_1^{\rm in}}^{e_1}\} \rightarrow \{P_{A_1^{\rm out}}^{e_1'}\}$, one applies Definition \ref{def:int} to the identity transformation, i.e.\ one simply checks whether the decompositions commute.}
    \label{fig:interference_influence_cicuit}
\end{figure}

Interestingly, the only interference influences that can obtain in the circuit (in the sense described in Figure \ref{fig:interference_influence_cicuit}) between the elements of $\mathfrak{P}(\mathfrak{C}, \mathfrak{B})$ are ones that go from the decompositions associated with the past to decompositions associated with the future. More precisely, Appendix \ref{app:nochains} proves that the only allowed interference influences are of the form
\begin{equation} \label{eq:allowed_int_infl}
    \{P_{A_m^{\rm in}}^{e_m}\} \rightarrow \{P_{A_k^{\rm out}}^{e_k'}\}  \quad \textnormal{where} \ m \leq k.
\end{equation}

The vast majority of sets of projective decompositions do not generate a consistent set of histories. But the causal constraint in equation (\ref{eq:allowed_int_infl}) implies that $\mathfrak{P}(\mathfrak{C}, \mathfrak{B})$ is one of those rare sets that does.

\begin{theorem} \label{thm:linear_prob}
   For any $\mathfrak{P}(\mathfrak{C}, \mathfrak{B})$, and for $\rho=I/d$, (\ref{eqLch_prob}) simplifies to a manifestly linear form. In the Heisenberg picture,
   \begin{equation} \label{eq:simpler_form}
         {\rm Tr}(\tilde{P}_{A_n^{\rm out}}^{e_n'}\tilde{P}_{A_n^{\rm in}}^{e_n}\ldots \tilde{P}_{A_1^{\rm out}}^{e_1'} \tilde{P}_{A_1^{\rm in}}^{e_1} (I/d) \tilde{P}_{A_1^{\rm in}}^{e_1}\tilde{P}_{A_1^{\rm out}}^{e_1'}\ldots \tilde{P}_{A_n^{\rm in}}^{e_n}\tilde{P}_{A_n^{\rm out}}^{e_n'}) = \frac{1}{d}{\rm Tr}(\tilde{P}_{A_1^{\rm in}}^{e_1}\tilde{P}_{A_1^{\rm out}}^{e_1'}\ldots \tilde{P}_{A_n^{\rm in}}^{e_n}\tilde{P}_{A_n^{\rm out}}^{e_n'}).
   \end{equation}
   It follows that $\mathfrak{P}(\mathfrak{C}, \mathfrak{B})$ generates a consistent set of histories.
\end{theorem}
We have finally achieved our goal: a preferred set of consistent histories for every bubble.

In the traditional consistent histories formalism, linearity of probabilities is simply assumed. But Theorem \ref{thm:linear_prob} shows that, when one appeals to causal structure, linearity can be \textit{derived}. Specifically, (\ref{eq:simpler_form}) is obtained from (\ref{eq:allowed_int_infl}) simply by commuting projectors around the trace expression on the left side of (\ref{eq:simpler_form}) (using trace cyclicity in the case of the outgoing projectors) and eliminating then using the idempotency of projectors until one finds the expression on the right. 

The reader might wonder why we have inserted the ``maximally mixed state'' $\rho=I/d$ into the left side of (\ref{eq:simpler_form}). In fact, we would prefer to think of this substitution as ``tracing out the past'': much in same the way that we use the standard trace operation to ignore whatever happens after a certain time, our use of $\rho=I/d$ reflects our decision to ignore whatever comes \textit{before} a certain time.\footnote{Note the duality between the trace operation ${\rm Tr}(\cdot)=\sum_i \bra{i}(\cdot)\ket{i}$ and the identity operator $I = \sum_i \ket{i}\bra{i}$.} The quantum state plays no fundamental role in the interpretation we are laying out, but, as the next section will make clear, it does serve as a useful tool for computing probabilities.

Before we move on, it is worth briefly summarizing the last two sections. Interference influences are noncommutation relations between projective decompositions in the Heisenberg picture (Definition \ref{def:int}), or, equivalently, a  particular sort of dynamical dependence (Theorem \ref{thm:int_equivalence}). Interference influences single out a preferred set $\mathfrak{P}(\mathfrak{C}, \mathfrak{B})$ of $2n$ decompositions relative to a bubble of $n$ unitarily interacting subsystems (Definition \ref{def:pref2}); $\mathfrak{P}(\mathfrak{C}, \mathfrak{B})$ can be thought of as decoherent relative to this bubble. Within a bubble, interference influences only travel from decompositions associated with the past to decompositions associated with the future (\ref{eq:allowed_int_infl}). Remarkably, this causal constraint implies that $\mathfrak{P}(\mathfrak{C}, \mathfrak{B})$ generates a consistent set of histories (Theorem \ref{thm:linear_prob}). All of the core ingredients of the interpretation are now in place. We are ready for axioms.

\section{Theory and models} \label{sec:theory_models}

In this section, we turn all of these ideas into a precise realist interpretation of quantum theory as a description of relational events and their emergence out of causal structure.

In short, if the dynamics of some scenario are described by a unitary circuit $\mathfrak{C}$, we postulate that, for every bubble $\mathfrak{B}$, exactly one history from the consistent set generated by $\mathfrak{P}(\mathfrak{C}, \mathfrak{B})$ is realized. Given a bubble $\mathfrak{B}$, the probability for a given history to be realized is

\begin{equation} \label{eq:bubble_probability_rule}
    p_{\mathfrak{B}}(e_1, e_1', \ldots, e_n, e_n') =\frac{1}{d}{\rm Tr}(\tilde{P}_{A_1^{\rm in}}^{e_1}\tilde{P}_{A_1^{\rm out}}^{e_1'} \ldots \tilde{P}_{A_n^{\rm in}}^{e_n}\tilde{P}_{A_n^{\rm out}}^{e_n'}).
\end{equation}
And that's more or less all there is to it. But let us lay out the interpretation in greater detail with the following axioms.

\begin{enumerate}
    \item The \textsc{dynamics} are given by a finite-dimensional unitary circuit $\mathfrak{C}$. That $\mathfrak{C}$ takes place is taken as a primitive and observer-independent fact about reality.

    \item A \textsc{bubble} is any subset of the systems (i.e.\ individual wires) in $\mathfrak{C}$. A bubble $\mathfrak{B}$ of $n$ systems is associated with the preferred set $\mathfrak{P}(\mathfrak{C}, \mathfrak{B})$ of $2n$ projective decompositions (as described in Definition \ref{def:pref2}).
    
    \item For every bubble of $n$ systems, $2n$ \textsc{events} take place relative to that bubble. Each event is the selection of a unique projector from an element of $\mathfrak{P}(\mathfrak{C}, \mathfrak{B})$.

    \item In a given bubble, the \textsc{probability} of a set of events is given by taking the matrix product of all the corresponding Heisenberg projectors in the order that they appear in the circuit, then tracing and dividing through by the dimension of the Hilbert space (i.e.\ by equation (\ref{eq:bubble_probability_rule})).
\end{enumerate}

In many physical theories, the fundamental object is a state, and the role of the dynamics is merely to constrain this state. But the present theory implies that \textit{dynamics are prior to kinematics}. The most fundamental kinematical object is an event, but the events that happen are determined entirely by the dynamics and the probability rule, and without recourse to any initial condition. The quantum state does not feature in any of the axioms, but nevertheless can be used to compute certain conditional probabilities, as we will soon show.

Events are nonabsolute for two reasons. Firstly, if a projective  decomposition, say $\mathbb{D}_k^{\rm in}$, features in both $\mathfrak{P}(\mathfrak{C}, \mathfrak{B}_1)$ and $\mathfrak{P}(\mathfrak{C}, \mathfrak{B}_2)$, then $P_{A_k^{\rm in}}^{e_k} \in \mathbb{D}_k^{\rm in}$ might get selected relative to $\mathfrak{B}_1$ while a distinct $P_{A_k^{\rm in}}^{\tilde{e}_k} \in \mathbb{D}_k^{\rm in}$ gets selected in $\mathfrak{B}_2$. So $e_k$ happens relative to $\mathfrak{B_1}$, but a different $\tilde{e}_k$ happens relative to $\mathfrak{B}_2$. Secondly, there may be a bubble $\mathfrak{B}_3$ such that $\mathbb{D}_{k}^{\rm in} \not\in \mathfrak{P}(\mathfrak{C}, \mathfrak{B}_3)$, so that none of its projectors get selected relative to $\mathfrak{B}_3$. So our theory is relational, but note also that it isn't ``relations all the way down''. For the explicitly relativized event $e_k^{\mathfrak{B}_1}$ of $P_{A_k^{\rm in}}^{e_k}$ being selected from $\mathbb{D}_k^{\rm in}$ relative to $\mathfrak{B}_1$ is absolute. (Analogously, a spatial interval $\Delta x$ in special relativity is not absolute, but the relativized fact $\Delta x^F$ that the spatial interval is $\Delta x$ relative to the frame $F$ is absolute.)

It is now time to apply this interpretation to some specific physical scenarios. We claim that the interpretation finds itself in a ``Goldilocks zone'': it is able to model any finite-dimensional quantum phenomenon that one would want to model, whilst avoiding models of problematic and unnecessary phenomena. The rest of this section will defend this claim by giving explicit examples and pointing to the general construction in Appendix \ref{app:operational}. \\

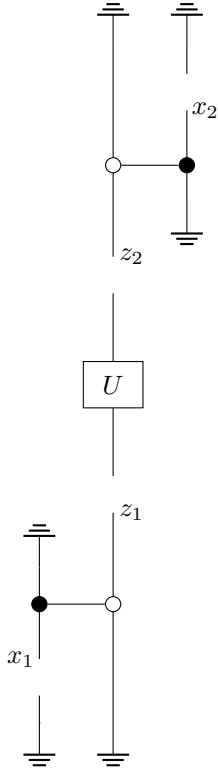
\begin{figure}
    \centering
    \begin{tikzpicture} [scale=0.97]
	\begin{pgfonlayer}{nodelayer}
		\node [style=none] (0) at (0.5, -5.75) {};
		\node [style=none] (1) at (6, 5.5) {};
		\node [style=none] (2) at (6, 0.5) {};
		\node [style=map] (3) at (6, 3) {$U$};
		\node [style=white dot] (4) at (6, -3) {};
		\node [style=none] (5) at (4, -3) {};
		\node [style=black dot] (6) at (4, -3) {};
		\node [style=none] (7) at (4, -6.25) {};
		\node [style=none] (8) at (4, -6.25) {};
		\node [style=none] (9) at (4, -6.25) {};
		\node [style=none] (10) at (4, -5.5) {};
		\node [style=none] (11) at (4, -3) {};
		\node [style=none] (12) at (4, -2) {};
		\node [style=ground] (13) at (4, -6.25) {};
		\node [style=ground] (14) at (6, -6.25) {};
		\node [style=upground] (15) at (4, -2) {};
		\node [style=upground] (16) at (6, 12.25) {};
		\node [style=white dot] (17) at (6, 9) {};
		\node [style=none] (18) at (8, 9) {};
		\node [style=black dot] (19) at (8, 9) {};
		\node [style=none] (20) at (8, 7.75) {};
		\node [style=none] (21) at (8, 9.5) {};
		\node [style=none] (22) at (8, 9.5) {};
		\node [style=none] (23) at (8, 10.5) {};
		\node [style=upground] (24) at (8, 12.25) {};
		\node [style=none] (25) at (6, -0.5) {};
		\node [style=none] (26) at (6, 6.5) {};
		\node [style=none] (27) at (6.5, -0.5) {$z_1$};
		\node [style=none] (28) at (6.5, 6.5) {$z_2$};
		\node [style=ground] (29) at (8, 8) {};
		\node [style=none] (30) at (4, -4.5) {};
		\node [style=none] (31) at (3.5, -4.5) {$x_1$};
		\node [style=none] (32) at (8, 11.5) {};
		\node [style=none] (33) at (8.5, 10.5) {$x_2$};
	\end{pgfonlayer}
	\begin{pgfonlayer}{edgelayer}
		\draw (4) to (6);
		\draw (9.center) to (7.center);
		\draw (8.center) to (10.center);
		\draw (11.center) to (12.center);
		\draw (1.center) to (2.center);
		\draw (17) to (19);
		\draw (22.center) to (23.center);
		\draw (25.center) to (14);
		\draw (26.center) to (16);
		\draw (22.center) to (29);
		\draw (30.center) to (11.center);
		\draw (32.center) to (24);
	\end{pgfonlayer}
\end{tikzpicture}
    \caption{A model of a ``prepare-measure'' scenario, depicting the bubble of four systems.}
    \label{fig:prepare_measure}
\end{figure}

\textbf{Prepare and measure.} To begin with, perhaps the most vanilla sort of experiment in standard quantum theory is one where a qubit is prepared in a state $\ket{\psi}$ and later measured in a different basis, leading to a probability of $|\braket{\phi|\psi}|^2$ for the outcome associated with an element $\ket{\phi}$ of that basis. A model from our theory for this experiment is given in Figure \ref{fig:prepare_measure}. In this circuit, all wires represent qubits; white and black dots represent CNOTs controlled on the white dots; and $U$ is some unitary transformation. Events arise entirely out of dynamics, and are not the consequence of any initial condition. Thus we trace out the inputs as well as the outputs of the circuit. 

Figure \ref{fig:prepare_measure} depicts a particular bubble $\mathfrak{B}_1$ of four systems, corresponding to the four breaks in the wires. The $z_i \in \{0, 1\}$ and $x_j \in \{+, -\}$ label events corresponding to the selection of a unique projector from an element of $\mathfrak{P}(\mathfrak{C}, \mathfrak{B}_1)$. The $z_i$ are selections of projectors onto the $Z$-basis $\{\ket{0}, \ket{1}\}$, while the $x_j$ are selections of projectors onto the $X$-basis $\{\frac{\ket{0} + \ket{1}}{\sqrt{2}}, \frac{\ket{0} - \ket{1}}{\sqrt{2}}\}$. $z_1$ corresponds to an ingoing projective decomposition that arises from the associated system's interaction with the ``preparation device'' in the past and on its left, while $z_2$ corresponds to an outgoing projective decomposition that arises from the associated system's interaction with the ``measurement device'' in its future and on its right.\footnote{C.f.\ the remark after Definition \ref{def:pref2}.} Where wires that go into or come out of breaks are not labelled, this is because the corresponding decompositions in $\mathfrak{P}(\mathfrak{C}, \mathfrak{B}_1)$ are simply $\{I\}$, so that the associated events are trivial.

Although the quantum state does not play any fundamental role in this interpretation, it can be used to compute conditional probabilities for events. For example, if one computes the distribution $p_{\mathfrak{B}_1}(x_1z_1x_2z_2)$ obtained from (\ref{eq:bubble_probability_rule}), one finds that
\begin{equation} \label{eq:transition}       p_{\mathfrak{B}_1}(z_2|z_1) = |\bra{z_2}U\ket{z_1}|^2
\end{equation}
-- which looks rather familiar. We can therefore reasonably say things like ``the system was prepared in the state $\ket{z_1}$, then transformed into $U\ket{z_1}$, and then a measurement returned an outcome corresponding to $\ket{z_2}$ with Born probability''. We note that the state preparation here is stochastic, since $p_{\mathfrak{B}_1}(z_1)=1/2$ for $z_1 \in \{0, 1\}$.

\begin{figure}
    \centering
    \begin{tikzpicture}
	\begin{pgfonlayer}{nodelayer}
		\node [style=none] (0) at (14.5, -5.75) {};
		\node [style=none] (1) at (20, 5.5) {};
		\node [style=none] (2) at (20, 0.5) {};
		\node [style=map] (3) at (20, 3) {$U$};
		\node [style=white dot] (4) at (20, -3) {};
		\node [style=none] (5) at (18, -3) {};
		\node [style=black dot] (6) at (18, -3) {};
		\node [style=white dot] (7) at (18, -6) {};
		\node [style=white dot] (8) at (18, 0) {};
		\node [style=black dot] (9) at (16, 0) {};
		\node [style=black dot] (10) at (16, -6) {};
		\node [style=none] (11) at (18, -6.25) {};
		\node [style=none] (12) at (16, -8) {};
		\node [style=none] (13) at (18, -4) {};
		\node [style=none] (14) at (18, -5) {};
		\node [style=none] (15) at (16, -7) {};
		\node [style=none] (16) at (16, -8) {};
		\node [style=none] (17) at (18, -2) {};
		\node [style=none] (18) at (18, -1) {};
		\node [style=none] (19) at (18, 0.25) {};
		\node [style=none] (20) at (16, -0.25) {};
		\node [style=none] (21) at (16, 1) {};
		\node [style=none] (22) at (16, 2) {};
		\node [style=none] (23) at (16, 2.25) {};
		\node [style=none] (24) at (16, -5.75) {};
		\node [style=none] (25) at (17.5, -5) {$z_3$};
		\node [style=none] (26) at (17.5, -1) {$z_4$};
		\node [style=none] (27) at (15.5, -7) {$x_1$};
		\node [style=none] (28) at (15.5, 1) {$x_2$};
		\node [style=ground] (29) at (18, -6.25) {};
		\node [style=ground] (30) at (20, -6.25) {};
		\node [style=ground] (31) at (16, -8) {};
		\node [style=ground] (32) at (16, -0.25) {};
		\node [style=upground] (33) at (16, -5.75) {};
		\node [style=upground] (34) at (16, 2.25) {};
		\node [style=upground] (35) at (18, 0.25) {};
		\node [style=upground] (36) at (20, 12.25) {};
		\node [style=white dot] (37) at (20, 9) {};
		\node [style=none] (38) at (22, 9) {};
		\node [style=black dot] (39) at (22, 9) {};
		\node [style=white dot] (40) at (22, 6) {};
		\node [style=white dot] (41) at (22, 12) {};
		\node [style=black dot] (42) at (24, 12) {};
		\node [style=black dot] (43) at (24, 6) {};
		\node [style=none] (44) at (22, 5.75) {};
		\node [style=none] (45) at (24, 4) {};
		\node [style=none] (46) at (22, 8) {};
		\node [style=none] (47) at (22, 7) {};
		\node [style=none] (48) at (24, 5) {};
		\node [style=none] (49) at (24, 4) {};
		\node [style=none] (50) at (22, 10) {};
		\node [style=none] (51) at (22, 11) {};
		\node [style=none] (52) at (22, 12.25) {};
		\node [style=none] (53) at (24, 11.75) {};
		\node [style=none] (54) at (24, 13) {};
		\node [style=none] (55) at (24, 14) {};
		\node [style=none] (56) at (24, 14.25) {};
		\node [style=none] (57) at (24, 6.25) {};
		\node [style=none] (58) at (22.5, 7) {$z_5$};
		\node [style=none] (59) at (22.5, 11) {$z_6$};
		\node [style=none] (60) at (24.5, 5) {$x_3$};
		\node [style=none] (61) at (24.5, 13) {$x_4$};
		\node [style=ground] (62) at (22, 5.75) {};
		\node [style=ground] (63) at (24, 4) {};
		\node [style=ground] (64) at (24, 11.75) {};
		\node [style=upground] (65) at (24, 6.25) {};
		\node [style=upground] (66) at (24, 14.25) {};
		\node [style=upground] (67) at (22, 12.25) {};
		\node [style=none] (68) at (20, -0.5) {};
		\node [style=none] (69) at (20, 6.5) {};
		\node [style=none] (70) at (20.5, -0.5) {$z_1$};
		\node [style=none] (71) at (20.5, 6.5) {$z_2$};
	\end{pgfonlayer}
	\begin{pgfonlayer}{edgelayer}
		\draw (4) to (6);
		\draw (16.center) to (12.center);
		\draw (15.center) to (10);
		\draw (14.center) to (11.center);
		\draw (13.center) to (17.center);
		\draw (18.center) to (19.center);
		\draw (8) to (9);
		\draw (23.center) to (22.center);
		\draw (21.center) to (20.center);
		\draw (24.center) to (10);
		\draw (10) to (7);
		\draw (1.center) to (2.center);
		\draw (37) to (39);
		\draw (49.center) to (45.center);
		\draw (48.center) to (43);
		\draw (47.center) to (44.center);
		\draw (46.center) to (50.center);
		\draw (51.center) to (52.center);
		\draw (41) to (42);
		\draw (56.center) to (55.center);
		\draw (54.center) to (53.center);
		\draw (57.center) to (43);
		\draw (43) to (40);
		\draw (68.center) to (30);
		\draw (69.center) to (36);
	\end{pgfonlayer}
\end{tikzpicture}
    \caption{Extended ``prepare-measure'' model, explicitly representing the observable events.}
    \label{fig:prepare_measure_2}
\end{figure}
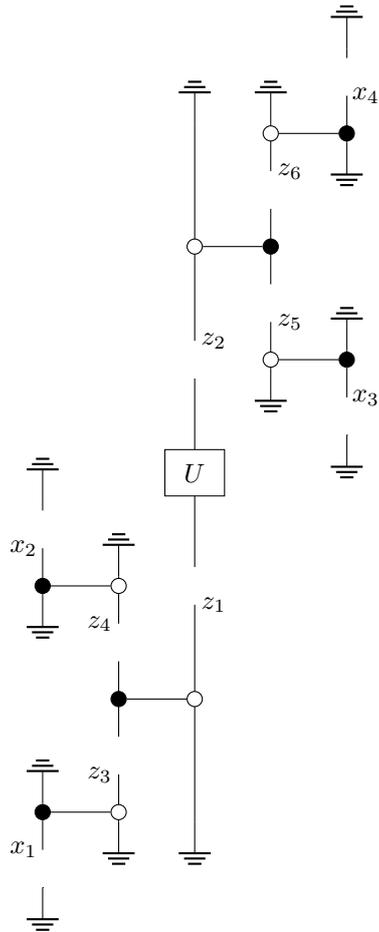
Now, $z_1$ and $z_2$ might not be directly observable; they could, for example, be an electron taking on a particular spin. But in an extended model, $z_1$ can be inferred from directly observable events. In this extended model, we simply have to assume that the preparation and measurement devices also interact with other systems (perhaps the eye of some observer), as depicted in Figure \ref{fig:prepare_measure_2}. When one computes the probability distribution for the 10-system bubble $\mathfrak{B}_2$ associated with Figure \ref{fig:prepare_measure_2}, one finds that $z_1=z_3+z_4$ and $z_2=z_5+z_6$ with certainty. So if $z_3$, $z_4$, $z_5$ and $z_6$ are observed events, then $z_1$ and $z_2$ can be inferred. If we wanted to, then we could extend the model further still so that even these two events could in turn be inferred from other events.


The techniques here can easily be generalized to model any prepare-measure scenario. In fact, as we show Appendix \ref{app:operational}, one can model arbitrary quantum instruments and sequential and parallel combinations thereof. Thus \textit{any model from standard finite-dimensional quantum theory (i.e.\ the sort of theory described in \cite{nielsen2001quantum}) can be reproduced in this interpretation.} \\

\textbf{Wigner's friend.} We can also model scenarios in which the standard quantum theory becomes ambiguous, such as the classic Wigner's friend scenario \cite{wigner1995remarks}. In this scenario, Wigner's friend is in an isolated lab, and measures a particle in a superposition of states $\frac{\ket{0} + \ket{1}}{\sqrt{2}}$. Applying the standard theory from the friend's perspective leads to the friend obtaining a definite outcome and the particle collapsing onto a corresponding state $\ket{0}$ or $\ket{1}$. Applying the same theory from the perspective of Wigner, who sits outside the lab, leads to the apparently contradictory conclusion that the friend ends up entangled with the particle in the state $\ket{\Phi^+}$, as can be checked by Wigner with a subsequent measurement. Standard quantum theory does not tell us which of these two models is correct, nor does it explain their place in some overarching theoretical framework.

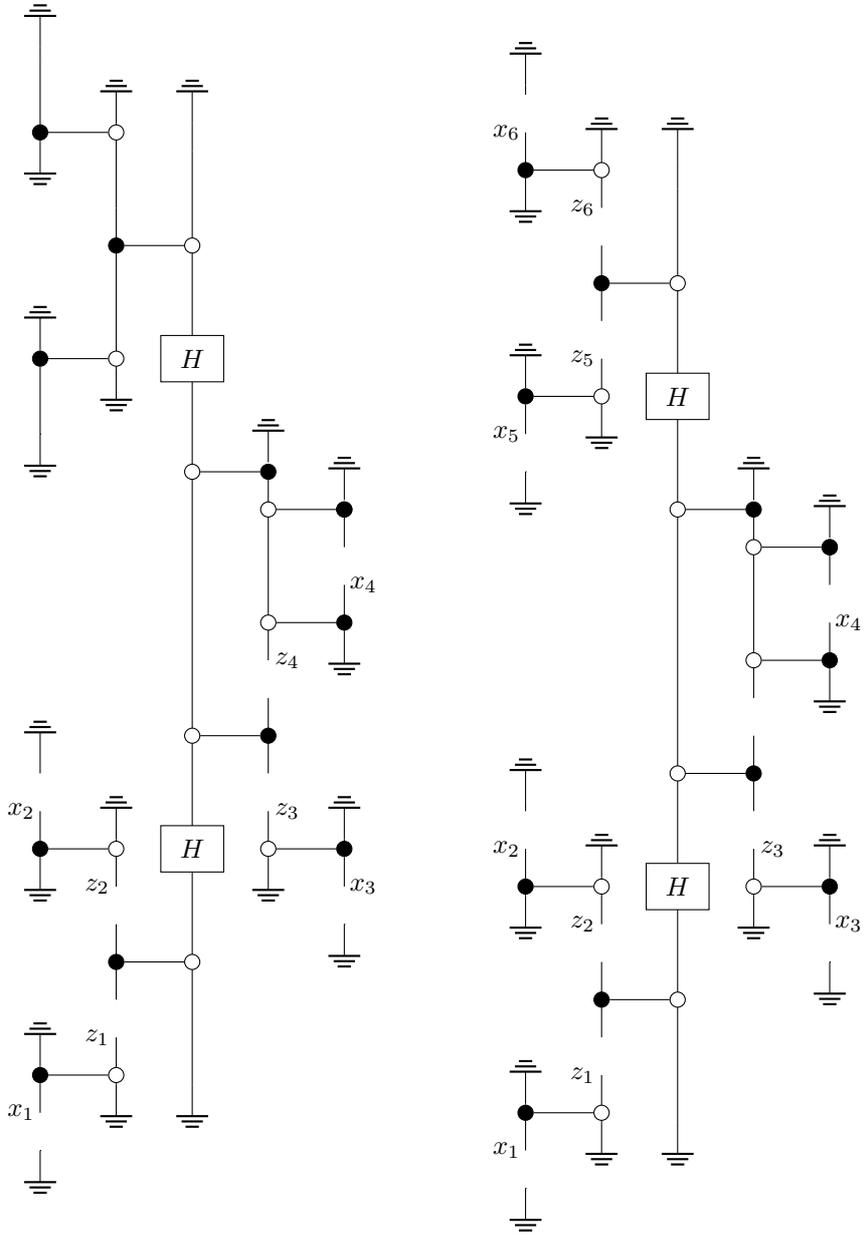
\begin{figure}
    \centering
    \begin{tikzpicture}
	\begin{pgfonlayer}{nodelayer}
		\node [style=none] (0) at (5, 19.5) {};
		\node [style=none] (1) at (5, 19.5) {};
		\node [style=none] (2) at (5, -5.25) {};
		\node [style=none] (3) at (5, 0) {};
		\node [style=map] (4) at (5, 1) {$H$};
		\node [style=white dot] (5) at (5, -2) {};
		\node [style=none] (6) at (3, -2) {};
		\node [style=black dot] (7) at (3, -2) {};
		\node [style=white dot] (8) at (3, -5) {};
		\node [style=white dot] (9) at (3, 1) {};
		\node [style=black dot] (10) at (1, 1) {};
		\node [style=black dot] (11) at (1, -5) {};
		\node [style=none] (12) at (3, -5.25) {};
		\node [style=none] (13) at (1, -7) {};
		\node [style=none] (14) at (3, -3) {};
		\node [style=none] (15) at (3, -4) {};
		\node [style=none] (16) at (1, -6) {};
		\node [style=none] (17) at (1, -7) {};
		\node [style=none] (18) at (3, -1) {};
		\node [style=none] (19) at (3, 0) {};
		\node [style=none] (20) at (3, 1.25) {};
		\node [style=none] (21) at (1, 0.75) {};
		\node [style=none] (22) at (1, 2) {};
		\node [style=none] (23) at (1, 3) {};
		\node [style=none] (24) at (1, 3.25) {};
		\node [style=none] (25) at (1, -4.75) {};
		\node [style=none] (26) at (2.5, -4) {$z_1$};
		\node [style=none] (27) at (2.5, 0) {$z_2$};
		\node [style=none] (28) at (0.5, -6) {$x_1$};
		\node [style=none] (29) at (0.5, 2) {$x_2$};
		\node [style=ground] (30) at (3, -5.25) {};
		\node [style=ground] (31) at (5, -5.25) {};
		\node [style=ground] (32) at (1, -7) {};
		\node [style=ground] (33) at (1, 0.75) {};
		\node [style=upground] (34) at (1, -4.75) {};
		\node [style=upground] (35) at (1, 3.25) {};
		\node [style=upground] (36) at (3, 1.25) {};
		\node [style=white dot] (37) at (5, 4) {};
		\node [style=none] (38) at (7, 4) {};
		\node [style=black dot] (39) at (7, 4) {};
		\node [style=white dot] (40) at (7, 1) {};
		\node [style=white dot] (41) at (7, 7) {};
		\node [style=black dot] (42) at (9, 7) {};
		\node [style=black dot] (43) at (9, 1) {};
		\node [style=none] (44) at (7, 0.75) {};
		\node [style=none] (45) at (9, -1) {};
		\node [style=none] (46) at (7, 3) {};
		\node [style=none] (47) at (7, 2) {};
		\node [style=none] (48) at (9, 0) {};
		\node [style=none] (49) at (9, -1) {};
		\node [style=none] (50) at (7, 5) {};
		\node [style=none] (51) at (7, 6) {};
		\node [style=none] (52) at (7, 7.25) {};
		\node [style=none] (53) at (9, 6.75) {};
		\node [style=none] (54) at (9, 8) {};
		\node [style=none] (55) at (9, 9) {};
		\node [style=none] (56) at (9, 10) {};
		\node [style=none] (57) at (9, 1.25) {};
		\node [style=none] (58) at (7.5, 2) {$z_3$};
		\node [style=none] (59) at (9.5, 0) {$x_3$};
		\node [style=none] (60) at (9.5, 8) {$x_4$};
		\node [style=ground] (61) at (7, 0.75) {};
		\node [style=ground] (62) at (9, 6.75) {};
		\node [style=upground] (63) at (9, 1.25) {};
		\node [style=upground] (64) at (9, 10.25) {};
		\node [style=upground] (65) at (7, 11.25) {};
		\node [style=white dot] (66) at (7, 10) {};
		\node [style=black dot] (67) at (9, 10) {};
		\node [style=none] (68) at (7, 11) {};
		\node [style=white dot] (69) at (5, 11) {};
		\node [style=black dot] (70) at (7, 11) {};
		\node [style=black dot] (71) at (9, 10) {};
		\node [style=white dot] (72) at (5, 17) {};
		\node [style=none] (73) at (3, 17) {};
		\node [style=black dot] (74) at (3, 17) {};
		\node [style=white dot] (75) at (3, 14) {};
		\node [style=white dot] (76) at (3, 20) {};
		\node [style=black dot] (77) at (1, 20) {};
		\node [style=black dot] (78) at (1, 14) {};
		\node [style=none] (79) at (3, 13.75) {};
		\node [style=none] (80) at (1, 12) {};
		\node [style=none] (81) at (3, 16) {};
		\node [style=none] (82) at (3, 16) {};
		\node [style=none] (83) at (1, 12) {};
		\node [style=none] (84) at (1, 12) {};
		\node [style=none] (85) at (3, 18) {};
		\node [style=none] (86) at (3, 18) {};
		\node [style=none] (87) at (3, 20.25) {};
		\node [style=none] (88) at (1, 19.75) {};
		\node [style=none] (89) at (1, 22) {};
		\node [style=none] (90) at (1, 22) {};
		\node [style=none] (91) at (1, 22.25) {};
		\node [style=none] (92) at (1, 14.25) {};
		\node [style=ground] (93) at (3, 13.75) {};
		\node [style=ground] (94) at (1, 12) {};
		\node [style=ground] (95) at (1, 19.75) {};
		\node [style=upground] (96) at (1, 14.25) {};
		\node [style=upground] (97) at (1, 22.25) {};
		\node [style=upground] (98) at (3, 20.25) {};
		\node [style=upground] (99) at (5, 20.25) {};
		\node [style=map] (100) at (5, 14) {$H$};
		\node [style=none] (101) at (7.5, 6) {$z_4$};
		\node [style=ground] (102) at (9, -1) {};
	\end{pgfonlayer}
	\begin{pgfonlayer}{edgelayer}
		\draw (5) to (7);
		\draw (17.center) to (13.center);
		\draw (16.center) to (11);
		\draw (15.center) to (12.center);
		\draw (14.center) to (18.center);
		\draw (19.center) to (20.center);
		\draw (9) to (10);
		\draw (24.center) to (23.center);
		\draw (22.center) to (21.center);
		\draw (25.center) to (11);
		\draw (11) to (8);
		\draw (1.center) to (2.center);
		\draw (37) to (39);
		\draw (49.center) to (45.center);
		\draw (48.center) to (43);
		\draw (47.center) to (44.center);
		\draw (46.center) to (50.center);
		\draw (51.center) to (52.center);
		\draw (41) to (42);
		\draw (56.center) to (55.center);
		\draw (54.center) to (53.center);
		\draw (57.center) to (43);
		\draw (43) to (40);
		\draw (67) to (66);
		\draw (68.center) to (52.center);
		\draw (69) to (70);
		\draw (72) to (74);
		\draw (84.center) to (80.center);
		\draw (83.center) to (78);
		\draw (82.center) to (79.center);
		\draw (81.center) to (85.center);
		\draw (86.center) to (87.center);
		\draw (76) to (77);
		\draw (91.center) to (90.center);
		\draw (89.center) to (88.center);
		\draw (92.center) to (78);
		\draw (78) to (75);
		\draw (1.center) to (99);
	\end{pgfonlayer}
\end{tikzpicture} \  \ \ \ \ \ \ \ \ \ \begin{tikzpicture}
	\begin{pgfonlayer}{nodelayer}
		\node [style=none] (0) at (5, 18.5) {};
		\node [style=none] (1) at (5, 18.5) {};
		\node [style=none] (2) at (5, -6.25) {};
		\node [style=none] (3) at (5, -1) {};
		\node [style=map] (4) at (5, 0) {$H$};
		\node [style=white dot] (5) at (5, -3) {};
		\node [style=none] (6) at (3, -3) {};
		\node [style=black dot] (7) at (3, -3) {};
		\node [style=white dot] (8) at (3, -6) {};
		\node [style=white dot] (9) at (3, 0) {};
		\node [style=black dot] (10) at (1, 0) {};
		\node [style=black dot] (11) at (1, -6) {};
		\node [style=none] (12) at (3, -6.25) {};
		\node [style=none] (13) at (1, -8) {};
		\node [style=none] (14) at (3, -4) {};
		\node [style=none] (15) at (3, -5) {};
		\node [style=none] (16) at (1, -7) {};
		\node [style=none] (17) at (1, -8) {};
		\node [style=none] (18) at (3, -2) {};
		\node [style=none] (19) at (3, -1) {};
		\node [style=none] (20) at (3, 0.25) {};
		\node [style=none] (21) at (1, -0.25) {};
		\node [style=none] (22) at (1, 1) {};
		\node [style=none] (23) at (1, 2) {};
		\node [style=none] (24) at (1, 2.25) {};
		\node [style=none] (25) at (1, -5.75) {};
		\node [style=none] (26) at (2.5, -5) {$z_1$};
		\node [style=none] (27) at (2.5, -1) {$z_2$};
		\node [style=none] (28) at (0.5, -7) {$x_1$};
		\node [style=none] (29) at (0.5, 1) {$x_2$};
		\node [style=ground] (30) at (3, -6.25) {};
		\node [style=ground] (31) at (5, -6.25) {};
		\node [style=ground] (32) at (1, -8) {};
		\node [style=ground] (33) at (1, -0.25) {};
		\node [style=upground] (34) at (1, -5.75) {};
		\node [style=upground] (35) at (1, 2.25) {};
		\node [style=upground] (36) at (3, 0.25) {};
		\node [style=white dot] (37) at (5, 3) {};
		\node [style=none] (38) at (7, 3) {};
		\node [style=black dot] (39) at (7, 3) {};
		\node [style=white dot] (40) at (7, 0) {};
		\node [style=white dot] (41) at (7, 6) {};
		\node [style=black dot] (42) at (9, 6) {};
		\node [style=black dot] (43) at (9, 0) {};
		\node [style=none] (44) at (7, -0.25) {};
		\node [style=none] (45) at (9, -2) {};
		\node [style=none] (46) at (7, 2) {};
		\node [style=none] (47) at (7, 1) {};
		\node [style=none] (48) at (9, -1) {};
		\node [style=none] (49) at (9, -2) {};
		\node [style=none] (50) at (7, 4) {};
		\node [style=none] (51) at (7, 5) {};
		\node [style=none] (52) at (7, 6.25) {};
		\node [style=none] (53) at (9, 5.75) {};
		\node [style=none] (54) at (9, 7) {};
		\node [style=none] (55) at (9, 8) {};
		\node [style=none] (56) at (9, 9) {};
		\node [style=none] (57) at (9, 0.25) {};
		\node [style=none] (58) at (7.5, 1) {$z_3$};
		\node [style=none] (59) at (9.5, -1) {$x_3$};
		\node [style=none] (60) at (9.5, 7) {$x_4$};
		\node [style=ground] (61) at (7, -0.25) {};
		\node [style=ground] (62) at (9, 5.75) {};
		\node [style=upground] (63) at (9, 0.25) {};
		\node [style=upground] (64) at (9, 9.25) {};
		\node [style=upground] (65) at (7, 10.25) {};
		\node [style=white dot] (66) at (7, 9) {};
		\node [style=black dot] (67) at (9, 9) {};
		\node [style=none] (68) at (7, 10) {};
		\node [style=white dot] (69) at (5, 10) {};
		\node [style=black dot] (70) at (7, 10) {};
		\node [style=black dot] (71) at (9, 9) {};
		\node [style=white dot] (72) at (5, 16) {};
		\node [style=none] (73) at (3, 16) {};
		\node [style=black dot] (74) at (3, 16) {};
		\node [style=white dot] (75) at (3, 13) {};
		\node [style=white dot] (76) at (3, 19) {};
		\node [style=black dot] (77) at (1, 19) {};
		\node [style=black dot] (78) at (1, 13) {};
		\node [style=none] (79) at (3, 12.75) {};
		\node [style=none] (80) at (1, 11) {};
		\node [style=none] (81) at (3, 15) {};
		\node [style=none] (82) at (3, 14) {};
		\node [style=none] (83) at (1, 12) {};
		\node [style=none] (84) at (1, 11) {};
		\node [style=none] (85) at (3, 17) {};
		\node [style=none] (86) at (3, 18) {};
		\node [style=none] (87) at (3, 19.25) {};
		\node [style=none] (88) at (1, 18.75) {};
		\node [style=none] (89) at (1, 20) {};
		\node [style=none] (90) at (1, 21) {};
		\node [style=none] (91) at (1, 21.25) {};
		\node [style=none] (92) at (1, 13.25) {};
		\node [style=none] (93) at (2.5, 14) {$z_5$};
		\node [style=none] (94) at (2.5, 18) {$z_6$};
		\node [style=none] (95) at (0.5, 12) {$x_5$};
		\node [style=none] (96) at (0.5, 20) {$x_6$};
		\node [style=ground] (97) at (3, 12.75) {};
		\node [style=ground] (98) at (1, 11) {};
		\node [style=ground] (99) at (1, 18.75) {};
		\node [style=upground] (100) at (1, 13.25) {};
		\node [style=upground] (101) at (1, 21.25) {};
		\node [style=upground] (102) at (3, 19.25) {};
		\node [style=upground] (103) at (5, 19.25) {};
		\node [style=map] (104) at (5, 13) {$H$};
		\node [style=ground] (105) at (9, -2) {};
	\end{pgfonlayer}
	\begin{pgfonlayer}{edgelayer}
		\draw (5) to (7);
		\draw (17.center) to (13.center);
		\draw (16.center) to (11);
		\draw (15.center) to (12.center);
		\draw (14.center) to (18.center);
		\draw (19.center) to (20.center);
		\draw (9) to (10);
		\draw (24.center) to (23.center);
		\draw (22.center) to (21.center);
		\draw (25.center) to (11);
		\draw (11) to (8);
		\draw (1.center) to (2.center);
		\draw (37) to (39);
		\draw (49.center) to (45.center);
		\draw (48.center) to (43);
		\draw (47.center) to (44.center);
		\draw (46.center) to (50.center);
		\draw (51.center) to (52.center);
		\draw (41) to (42);
		\draw (56.center) to (55.center);
		\draw (54.center) to (53.center);
		\draw (57.center) to (43);
		\draw (43) to (40);
		\draw (67) to (66);
		\draw (68.center) to (52.center);
		\draw (69) to (70);
		\draw (72) to (74);
		\draw (84.center) to (80.center);
		\draw (83.center) to (78);
		\draw (82.center) to (79.center);
		\draw (81.center) to (85.center);
		\draw (86.center) to (87.center);
		\draw (76) to (77);
		\draw (91.center) to (90.center);
		\draw (89.center) to (88.center);
		\draw (92.center) to (78);
		\draw (78) to (75);
		\draw (1.center) to (103);
	\end{pgfonlayer}
\end{tikzpicture}
    \caption{A model for the Wigner's friend scenario. The left side depicts the bubble containing state preparations and the friend's measurement, the right depicts the bubble also containing Wigner's measurement.}
    \label{fig:wf}
\end{figure}

In our theory, the two models are associated with different bubbles of the same unitary circuit, each of which constitutes a different and equally real part of the world. Figure \ref{fig:wf} shows a model for the experiment in which we have highlighted the relevant bubbles. The bubble $\mathfrak{B}_1$ on the left includes (systems corresponding to) the preparation of the state, and the friend's outcome. In this bubble, one has a prepare-measure scenario; events are distributed here just as the corresponding events were in the extended prepare-measure experiment described above.

On the other hand, the bubble $\mathfrak{B}_2$ on the right includes all the systems from the one on the left, but it also includes systems corresponding to Wigner's measurement. The events in the lower part of the diagram look almost the same as those in the left picture, with one crucial difference: the friend's outcome $z_4$ has disappeared! The projective decomposition $\{\ket{0}\bra{0}, \ket{1}\bra{1}\}$ on that system is not found in $\mathfrak{P}(\mathfrak{C}, \mathfrak{B}_2)$ because it exerts an interference influence on the decomposition corresponding to Wigner's outcome $z_6$. Instead, the corresponding decomposition in $\mathfrak{P}(\mathfrak{C}, \mathfrak{B}_2)$ is now $\{I\}$. Relative to bubbles that include Wigner's outcome, the friend does not obtain any measurement outcome at all.

We note that the extended Wigner's friend scenarios, introduced in \cite{Frauchiger2018quantum} and commonly used to provide no-go theorems for absoluteness, can be modelled using our theory in a similar way. \\

\textbf{Three-box paradox.} So far, we have shown that the current interpretation can reproduce standard quantum theory, and go beyond it. Consistent histories also goes beyond the standard theory, but in doing so, it opens itself up to some problems that the current interpretation manages to avoid. One example is provided by the ``three-box paradox'', introduced in \cite{aharonov1991complete} and studied from the point of view of consistent histories in \cite{kent1997consistent}.

Let us start with an operational formulation of the ``paradox''\footnote{We'll drop the scare quotes from hereon, but the reader will see that it is debatable whether this is an appropriate label for the phenomenon.} using standard quantum theory. Suppose a particle is prepared at $t_1$ in an equal superposition $\ket{\psi}:=\frac{1}{\sqrt{3}}(\ket{0}+\ket{1}+\ket{2})$ of being in one of three different boxes. Then, at $t_2>t_1$, a PVM $\mathbb{D}_2^i=\{\ket{i}\bra{i}, I - \ket{i}\bra{i}\}$ for either $i=0$ or $i=1$ is performed. This measurement can be thought of as ``checking to see whether or not the particle is in the $\ket{i}$ box''. For the update rule, we assume the projection postulate. Finally, at $t_3>t_2$, we measure an orthonormal basis that includes $\ket{\phi}:=\frac{1}{\sqrt{3}}(\ket{0}+\ket{1}-\ket{2})$. 

Now we ask a question: assuming that the measurement at $t_3$ results in the $\ket{\phi}$ outcome, what was the measurement result at $t_2$? If we checked the $\ket{0}$ box at $t_2$ and found the particle was not there, then standard quantum theory says that the state just after $t_2$ is $(I-\ket{0}\bra{0})\ket{\psi}$ (up to norm), which is orthogonal to $\ket{\phi}$. So if we checked in the $\ket{0}$ box, then we must have found it there, or else we wouldn't have got the $\ket{\phi}$ outcome at $t_3$. But a similar argument shows that if we checked in the $\ket{1}$ box, then we must have found it there. It seems as though our choice of which box to look in determines which box the particle is in!

In fact, standard quantum theory offers a simple, albeit anthropocentric, explanation. As discussed in Section \ref{sec:interpretations_ch}, in the standard theory, measurements are \textit{disturbing}. In the scenario just described, the choice of $\mathbb{D}_2^0$ or $\mathbb{D}_2^1$ determines which sort of projection operator collapses the quantum state at $t_2$. The dynamics in the two different situations are therefore different. Given this difference, there is no reason the same assumptions about what happens at times $t_1$ and $t_3$ should imply the same conclusions about what happens at $t_2$.

The situation is worse for consistent histories. Consider the following  two triplets of projective decompositions:
\begin{equation} \label{eq:consistent_sets}
    (\mathbb{D}_1, \mathbb{D}_2^i, \mathbb{D}_3) \ {\rm for} \ i \in \{0, 1\}.
\end{equation}
where
\begin{equation}
    \begin{split}
        \mathbb{D}_1 &=\{\ket{\psi}\bra{\psi}, I- \ket{\psi}\bra{\psi}\} \\
        \mathbb{D}_2^i &=\{\ket{i}\bra{i}, I - \ket{i}\bra{i}\} \\
        \mathbb{D}_3&=\{\ket{\phi}\bra{\phi}, I- \ket{\phi}\bra{\phi}\}.
    \end{split}
\end{equation}
If we assume that $\rho = I/3$, and that \textit{all time evolution is given by identity channels}, then each of these triplets generates a set of consistent histories as described in Section \ref{sec:interpretations_ch}.\footnote{Consistent historians might prefer to model this scenario with an initial and possibly a final density operator rather than with triplets of projective decompositions and trivial states. Either way, the choice doesn't make any significant difference to the arguments made here. We use triplets of decompositions only because that approach facilitates an easier connection to the interpretation of this paper.} Relative to one consistent set of histories, the events corresponding to $\ket{\psi}\bra{\psi}$ and $\ket{\phi}\bra{\phi}$ imply that the particle was in the $\ket{0}$ box at $t_2$. Relative to the other, the same events imply it was in the $\ket{1}$ box. This time, we cannot explain the difference at $t_2$ by a difference in time evolution, since both consistent sets were generated on the assumption that time evolution was trivial.

Now, the consistent historian does have the option of simply biting the bullet and accepting this phenomenon as just another peculiar feature of the ever-surprising quantum world. There is no logical contradiction if one does so, since the different inferences are made relative to different consistent sets. But, as Adrian Kent argues \cite{kent1997consistent}, the paradox undermines the formalism's claim to be the minimal and natural realist extension of the Copenhagen interpretation, or  ``Copenhagen done right'', since there is no Copenhagen analogue of the paradox. For the purposes of this discussion, the Copenhagen interpretation is essentially equivalent to what we have been calling ``standard quantum theory''. For both theories, three-box paradoxes can always be explained by the influence of the observer on the system.

Unlike consistent histories, the current interpretation can always explain the three-box paradox by appealing to interaction (perhaps it could be called ``consistent histories done right''!). But unlike standard quantum theory, the relevant notion of interaction is not essentially connected with observation. Although the triplets of decompositions in (\ref{eq:consistent_sets})  above form a consistent set of histories, they obviously are not preferred by any bubble, since the dynamics in this situation are trivial. Moreover, note that each triplet in (\ref{eq:consistent_sets}) involves a chain of noncommuting projectors (explicitly, $[\ket{\psi}\bra{\psi}, \ket{i}\bra{i}]\neq 0$ \textit{and} $[\ket{i}\bra{i}, \ket{\phi}\bra{\phi}] \neq 0$). By (\ref{eq:allowed_int_infl}), we can infer that neither triplet of projective decompositions could be all be preferred by \textit{any} bubble.  

This does not mean that the \textit{operational phenomenon} associated with the three-box paradox cannot be modelled using our interpretation -- the construction in Appendix \ref{app:operational} implies that \textit{any} phenomenon from the standard theory can be reproduced, and so, of course, this one can too. But in order to reproduce the operational phenomenon, one will have to explicitly model the measurements as unitary interactions, as we did for the prepare-measure and Wigner's friend scenarios. Differences in the required interactions will explain differences in what can be inferred about $t_2$.

Appendix \ref{app:3box} shows that this story extends to a class of generalized three-box paradoxes. The interested reader can check that the story also generalizes the quantum ``pigeonhole paradox'' of \cite{aharonov2016quantum}. We conjecture that it generalizes even further, to all examples of logical pre- and post-selection paradoxes (as defined in \cite{pusey2015logical}). 
\\

As compared with the bare consistent histories formalism, the distinctive feature of the current interpretation is that a consistent set is only physically significant when singled out relative to a bubble by causal structure. This stipulation does not prevent us from modelling arbitrary operational quantum phenomena. It does not prevent us from modelling (extended) Wigner's friend scenarios. But it does prevent us from modelling an interaction-free version of the three-box paradox. We are in the Goldilocks zone.

\section{Interference influences as explanations of quantum phenomena} \label{sec:classications}

Interference influences are of interest independently of the interpretation presented in this paper. In this section, we will show how they can be used to explain and classify quantum phenomena.

We will define five different ``scenarios'' in terms of the properties of a pair $(\mathfrak{C}, \mathfrak{S})$, where $\mathfrak{C}$ is a unitary circuit and $\mathfrak{S}$ is a set of projective decompositions associated with wires in the circuit. After giving all the definitions, we will show that every one of these scenarios requires a particular set of interference influences. Note that we do not require that $\mathfrak{S}=\mathfrak{P}(\mathfrak{C}, \mathfrak{B})$ for any bubble $\mathfrak{B}$; instead, we simply put in the projective decompositions by hand (although we note that in all five scenarios the decompositions are indeed preferred in appropriate bubbles if the circuit is extended in an appropriate way). The results of this section are therefore valid independently of the interpretation outlined in the previous section; at the same time, they illustrate the significance of the interference influences that lie at the heart of it. 

We note that some of the following definitions of the scenarios will be quite liberal. This is acceptable because we will only seek \textit{necessary}, rather than sufficient conditions for a given scenario to take place. Therefore, our results will remain true if any of the definitions are made logically stronger. Proofs for all of the results of this section are found in Appendix \ref{app:classifications}.

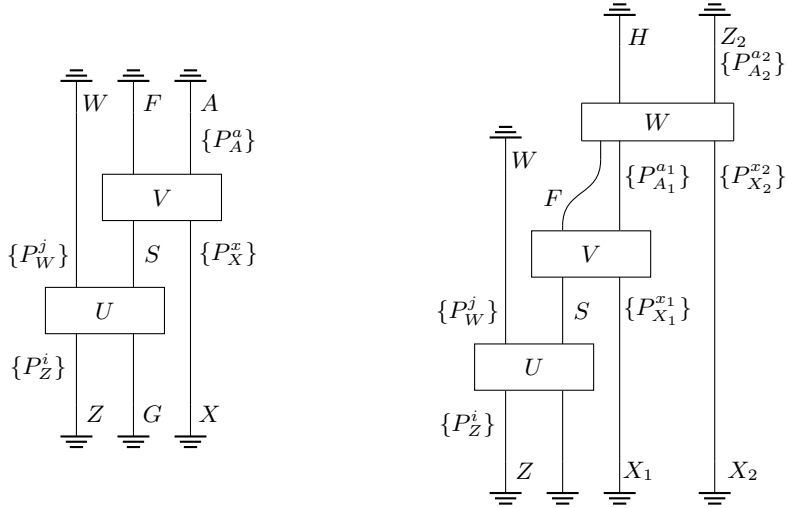
\begin{figure} 
    \centering
    \begin{tikzpicture} \small
	\begin{pgfonlayer}{nodelayer}
		\node [style=large map] (1) at (-0.75, 0) {$U$};
		\node [style=large map] (2) at (0.75, 3) {$V$};
		\node [style=none] (3) at (0, -2.5) {};
		\node [style=none] (5) at (0, 5.25) {};
		\node [style=none] (6) at (-1.5, -2.5) {};
		\node [style=none] (7) at (-1.5, 5.25) {};
		\node [style=none] (8) at (1.5, 5.25) {};
		\node [style=none] (9) at (1.5, -2.5) {};
		\node [style=none] (10) at (-1, -2.75) {$Z$};
		\node [style=none] (12) at (2, -2.75) {$X$};
		\node [style=none] (13) at (0.5, 1.5) {$S$};
        \node [style=none] (28) at (0.5, -2.75) {$G$};
		\node [style=none] (15) at (-1, 5.5) {$W$};
  	\node [style=none] (27) at (0.5, 5.5) {$F$};
		\node [style=none] (16) at (2, 5.5) {$A$};
		\node [style=none] (17) at (-2.5, -1.5) {$\{P_{Z}^i\}$};
		\node [style=none] (18) at (-2.5, 1.5) {$\{P_{W}^j\}$};
		\node [style=none] (19) at (2.5, 1.5) {$\{P_{X}^x\}$};
		\node [style=none] (20) at (2.5, 4.5) {$\{P_A^a\}$};
		\node [style=upground] (21) at (-1.5, 5.25) {};
		\node [style=upground] (22) at (0, 5.25) {};
		\node [style=upground] (23) at (1.5, 5.25) {};
		\node [style=ground] (24) at (1.5, -2.5) {};
		\node [style=ground] (25) at (0, -2.5) {};
		\node [style=ground] (26) at (-1.5, -2.5) {};
	\end{pgfonlayer}
	\begin{pgfonlayer}{edgelayer}
		\draw (5.center) to (3.center);
		\draw (9.center) to (8.center);
		\draw (7.center) to (6.center);
	\end{pgfonlayer}
\end{tikzpicture}    \ \ \ \ \ \ \ \ \ \ \ \ \ \ \ \ \ \begin{tikzpicture} \small
	\begin{pgfonlayer}{nodelayer}
		\node [style=large map] (0) at (2.25, -1.5) {$U$};
		\node [style=large map] (1) at (3.75, 1.5) {$V$};
		\node [style=none] (2) at (3, -4) {};
		\node [style=none] (3) at (3, 2.25) {};
		\node [style=none] (4) at (1.5, -4) {};
		\node [style=none] (5) at (1.5, 3.75) {};
		\node [style=none] (6) at (4.5, 4.5) {};
		\node [style=none] (7) at (4.5, -4) {};
		\node [style=none] (8) at (2, -4.25) {$Z$};
		\node [style=none] (9) at (5, -4.25) {$X_1$};
		\node [style=none] (10) at (3.5, 0) {$S$};
		\node [style=none] (11) at (2, 4) {$W$};
		\node [style=none] (12) at (2.75, 3) {$F$};
		\node [style=none] (13) at (0.5, -3) {$\{P_{Z}^i\}$};
		\node [style=none] (14) at (0.5, 0) {$\{P_{W}^j\}$};
		\node [style=none] (15) at (5.5, 0) {$\{P_{X_1}^{x_1}\}$};
		\node [style=none] (16) at (5.5, 3.5) {$\{P_{A_1}^{a_1}\}$};
		\node [style=upground] (17) at (1.5, 3.75) {};
		\node [style=ground] (18) at (4.5, -4) {};
		\node [style=ground] (19) at (3, -4) {};
		\node [style=ground] (20) at (1.5, -4) {};
		\node [style=none] (21) at (4, 4) {};
		\node [style=none] (22) at (4, 4.5) {};
		\node [style=none] (23) at (3.5, 4.5) {};
		\node [style=none] (24) at (3.5, 5.5) {};
		\node [style=none] (25) at (7.5, 4.5) {};
		\node [style=none] (26) at (7.5, 5.5) {};
		\node [style=none] (27) at (5.5, 5) {$W$};
		\node [style=none] (28) at (4.5, 5.5) {};
		\node [style=none] (29) at (4.5, 7) {};
		\node [style=upground] (30) at (4.5, 7) {};
		\node [style=none] (31) at (7, 4.5) {};
		\node [style=none] (32) at (7, -4) {};
		\node [style=ground] (33) at (7, -4) {};
		\node [style=none] (34) at (7, 5.5) {};
		\node [style=none] (35) at (7, 7) {};
		\node [style=upground] (36) at (7, 7) {};
		\node [style=none] (37) at (8, 3.5) {$\{P_{X_2}^{x_2}\}$};
		\node [style=none] (38) at (8, 6.5) {$\{P_{A_2}^{a_2}\}$};
		\node [style=none] (39) at (7.75, -4.25) {$X_2$};
		\node [style=none] (40) at (7.5, 7.25) {$Z_2$};
		\node [style=none] (41) at (5, 7.25) {$H$};
	\end{pgfonlayer}
	\begin{pgfonlayer}{edgelayer}
		\draw (3.center) to (2.center);
		\draw (7.center) to (6.center);
		\draw (5.center) to (4.center);
		\draw [in=-90, out=90, looseness=1.25] (3.center) to (21.center);
		\draw (22.center) to (21.center);
		\draw (26.center) to (24.center);
		\draw (24.center) to (23.center);
		\draw (23.center) to (25.center);
		\draw (25.center) to (26.center);
		\draw (29.center) to (28.center);
		\draw (32.center) to (31.center);
		\draw (35.center) to (34.center);
	\end{pgfonlayer}
\end{tikzpicture}
    \caption{Complementarity and Wigner's friend scenarios. When a system label is implied by a projective decomposition, we sometimes omit it for better readability.}
    \label{fig:comp_wf}
\end{figure}

To begin with, let us define a \textit{complementarity scenario}. Study the left side of Figure \ref{fig:comp_wf}. $P_Z^i$  and $P_W^j$ correspond to the preparation of a state $\rho^{ij}={\rm Tr}_W(\cu(P_{Z}^i \otimes I_G)(P_W^j \otimes I_S))$ (ignoring normalization). Similarly, $P_X^x$ and $P_A^a$ correspond to a positive operator-valued measurement (POVM) element $\sigma^{ax}:= {\rm Tr}_X(\cv^\dag(I_F \otimes P_{A}^a)(I_S \otimes P_X^x) )$. We say the decompositions $(\{P_Z^i\}, \{P_W^j\}, \{P_X^x\}, \{P_A^a\})$ form a complementarity scenario if and only if $\exists i, j, a, x: [\rho^{ij}, \sigma^{ax}] \neq0$. We then have our first result.

\begin{theorem} \label{thm:complementarity}
    A complementarity scenario requires an interference influence from $\{P_Z^i\}$ to $\{P_A^a\}$.
\end{theorem}

Now consider the right side of Figure \ref{fig:comp_wf}. We say this circuit and set of decompositions forms a \textit{Wigner's friend scenario} just in case it combines two complementarity scenarios; that is, if both $(\{P_Z^i\}, \{P_W^j\}, \{P_{X_1}^{x_1}\}, \{P_{A_1}^{a_1}\})$ and $(\{P_{A_1}^{a_1}\}, \{\}, \{P_{X_2}^{x_2}\}, \{P_{A_2}^{a_2}\})$ form complementarity scenarios.\footnote{In the second case, this means that $\exists a_1, x_2, a_2: [I_F \otimes P_{A_1}^{a_1}, {\rm Tr}_{X_2}(\cw^\dag(I_H \otimes P_{A_2}^{a_2})(I_{FA_1} \otimes P_{X_2}^{x_2}))] \neq 0$.} Unsurprisingly, a Wigner's friend scenario therefore requires two interference influences, combined to form a chain.

\begin{theorem} \label{thm:wf}
    A Wigner's friend scenario requires an interference influence from $\{P_Z^i\}$ to $\{P_{A_1}^{a_1}\}$, and an interference influence from $\{P_{A_1}^{a_1}\}$ to $\{P_{A_2}^{a_2}\}$.
\end{theorem}

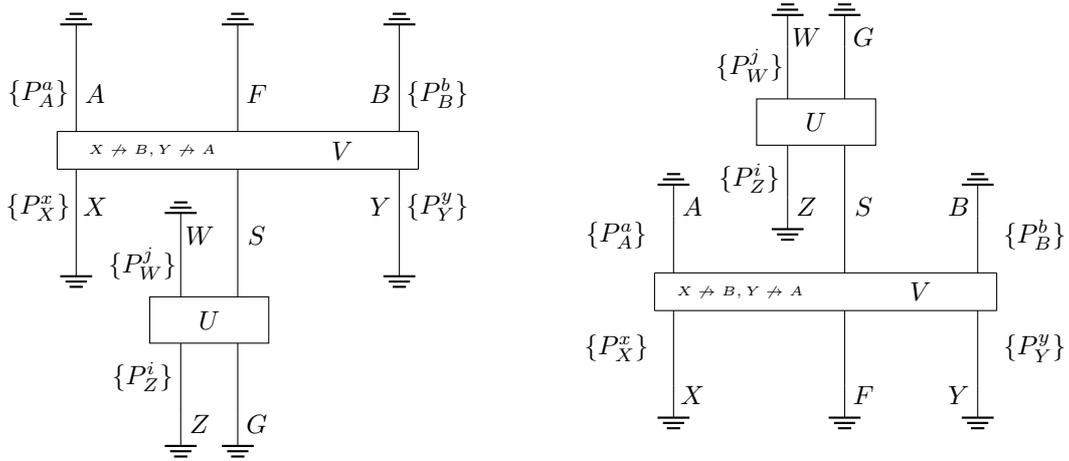
\begin{figure} 
    \centering
    \begin{tikzpicture}
	\begin{pgfonlayer}{nodelayer}
		\node [style=large map] (0) at (5.5, -1.75) {$U$};
		\node [style=none] (1) at (6.25, -4.25) {};
		\node [style=none] (2) at (6.25, 2.25) {};
		\node [style=none] (3) at (4.75, -4.25) {};
		\node [style=none] (4) at (4.75, 0.25) {};
		\node [style=none] (5) at (5.25, -4.5) {$Z$};
		\node [style=none] (6) at (6.75, -4.5) {$G$};
		\node [style=none] (7) at (6.75, 0.5) {$S$};
		\node [style=none] (8) at (5.25, 0.5) {$W$};
		\node [style=none] (9) at (3.75, -3.25) {$\{P_{Z}^i\}$};
		\node [style=none] (10) at (3.75, -0.25) {$\{P_{W}^j\}$};
		\node [style=upground] (11) at (4.75, 0.25) {};
		\node [style=ground] (12) at (6.25, -4.25) {};
		\node [style=ground] (13) at (4.75, -4.25) {};
		\node [style=none] (14) at (1.5, 2.25) {};
		\node [style=none] (15) at (11, 2.25) {};
		\node [style=none] (16) at (1.5, 3.25) {};
		\node [style=none] (17) at (11, 3.25) {};
		\node [style=ground] (18) at (2, 0.25) {};
		\node [style=upground] (19) at (2, 5.25) {};
		\node [style=none] (20) at (2, 3.25) {};
		\node [style=none] (21) at (2, 2.25) {};
		\node [style=none] (22) at (1, 1.25) {$\{P_{X}^x\}$};
		\node [style=none] (23) at (1, 4.25) {$\{P_{A}^a\}$};
		\node [style=none] (24) at (2.5, 1.25) {$X$};
		\node [style=none] (25) at (2.5, 4.25) {$A$};
		\node [style=none] (26) at (6.75, 4.25) {$F$};
		\node [style=ground] (27) at (10.5, 0.25) {};
		\node [style=upground] (28) at (10.5, 5.25) {};
		\node [style=none] (29) at (10.5, 3.25) {};
		\node [style=none] (30) at (10.5, 2.25) {};
		\node [style=none] (31) at (11.5, 1.25) {$\{P_{Y}^y\}$};
		\node [style=none] (32) at (11.5, 4.25) {$\{P_{B}^b\}$};
		\node [style=none] (33) at (10, 1.25) {$Y$};
		\node [style=none] (34) at (10, 4.25) {$B$};
		\node [style=none] (35) at (9, 2.75) {$V$};
		\node [style=none] (36) at (0.5, 0.75) {};
		\node [style=none] (37) at (6.25, 3.25) {};
		\node [style=none] (38) at (6.25, 5.25) {};
		\node [style=none] (39) at (6.25, 5.25) {};
		\node [style=upground] (40) at (6.25, 5.25) {};
		\node [style=none] (41) at (4, 2.75) {\tiny $X \not\rightarrow B, Y \not\rightarrow A$};
		\node [style=none] (42) at (6, 1.75) {};
	\end{pgfonlayer}
	\begin{pgfonlayer}{edgelayer}
		\draw (2.center) to (1.center);
		\draw (4.center) to (3.center);
		\draw (15.center) to (14.center);
		\draw (14.center) to (16.center);
		\draw (16.center) to (17.center);
		\draw (17.center) to (15.center);
		\draw (18) to (21.center);
		\draw (20.center) to (19);
		\draw (27) to (30.center);
		\draw (29.center) to (28);
		\draw (38.center) to (37.center);
	\end{pgfonlayer}
\end{tikzpicture}    \ \ \ \ \ \ \ \ \ \  \begin{tikzpicture}
	\begin{pgfonlayer}{nodelayer}
		\node [style=none] (0) at (5.5, -3.5) {};
		\node [style=none] (1) at (5.5, -1.5) {};
		\node [style=none] (2) at (1, -3.5) {};
		\node [style=none] (3) at (1, -1.5) {};
		\node [style=none] (4) at (1.5, -3.75) {$X$};
		\node [style=none] (5) at (6, -3.75) {$F$};
		\node [style=none] (6) at (6, 1.25) {$S$};
		\node [style=none] (7) at (1.5, 1.25) {$A$};
		\node [style=none] (8) at (-0.5, -2.5) {$\{P_{X}^x\}$};
		\node [style=none] (9) at (-0.5, 0.5) {$\{P_{A}^a\}$};
		\node [style=upground] (10) at (1, 1) {};
		\node [style=ground] (11) at (5.5, -3.5) {};
		\node [style=ground] (12) at (1, -3.5) {};
		\node [style=none] (13) at (9, -3.5) {};
		\node [style=none] (14) at (9, -1.5) {};
		\node [style=none] (15) at (8.5, -3.75) {$Y$};
		\node [style=none] (16) at (8.5, 1.25) {$B$};
		\node [style=upground] (17) at (9, 1) {};
		\node [style=ground] (18) at (9, -3.5) {};
		\node [style=none] (19) at (10.5, -2.5) {$\{P_{Y}^y\}$};
		\node [style=none] (20) at (10.5, 0.5) {$\{P_{B}^b\}$};
		\node [style=none] (21) at (0.5, -1.5) {};
		\node [style=none] (22) at (0.5, -0.5) {};
		\node [style=none] (23) at (9.5, -0.5) {};
		\node [style=none] (24) at (9.5, -1.5) {};
		\node [style=none] (25) at (1, -0.5) {};
		\node [style=none] (26) at (9, -0.5) {};
		\node [style=none] (27) at (5.5, -0.5) {};
		\node [style=none] (28) at (5.5, 5.5) {};
		\node [style=none] (29) at (7.5, -1) {$V$};
		\node [style=none] (30) at (2.75, -1) {\tiny $X \not\rightarrow B, Y \not\rightarrow A$};
		\node [style=large map] (31) at (4.75, 3.5) {$U$};
		\node [style=none] (32) at (4, 1.5) {};
		\node [style=none] (33) at (4, 5.5) {};
		\node [style=none] (34) at (4.5, 1.25) {$Z$};
		\node [style=none] (35) at (4.5, 5.75) {$W$};
		\node [style=upground] (36) at (4, 5.5) {};
		\node [style=ground] (37) at (4, 1.5) {};
		\node [style=upground] (38) at (5.5, 5.5) {};
		\node [style=none] (39) at (6, 5.75) {$G$};
		\node [style=none] (49) at (3, 2) {$\{P_{Z}^i\}$};
		\node [style=none] (50) at (3, 5) {$\{P_{W}^j\}$};
	\end{pgfonlayer}
	\begin{pgfonlayer}{edgelayer}
		\draw (1.center) to (0.center);
		\draw (3.center) to (2.center);
		\draw (14.center) to (13.center);
		\draw (22.center) to (21.center);
		\draw (21.center) to (24.center);
		\draw (24.center) to (23.center);
		\draw (23.center) to (22.center);
		\draw (25.center) to (10);
		\draw (26.center) to (17);
		\draw (28.center) to (27.center);
		\draw (33.center) to (32.center);
	\end{pgfonlayer}
\end{tikzpicture}
    \caption{Bell and PBR scenarios. Some of the unitary transformations are decorated with assumptions about their causal structure, which are designed to ensure these experiments can be undertaken at spacelike separation.}
    \label{fig:bell_ewf}
\end{figure}

Now study the left side of Figure \ref{fig:bell_ewf}. Note that this circuit comes with the causal assumptions that $X \not\rightarrow B$ and $Y \not\rightarrow A$, designed to ensure that the appropriate parts of the circuit can be implemented at spacelike separation without violating relativity theory. The topology of the circuit makes clear that there are no chains of interference influences, so the six decompositions lead to a probability distribution $p(axbyij)$ via (\ref{eq:bubble_probability_rule}). We say the decompositions $(\{P_Z^i\}, \{P_W^j\}, \{P_{X}^{x}\}, \{P_{A}^{a}\}, \{P_{Y}^{y}\}, \{P_{B}^{b}\})$ form a \textit{Bell scenario} if and only if, for some fixed $i$ and $j$, the probabilities $p(axby|ij)$ over ``settings'' $(x, y)$ and ``outcomes'' $(a, b)$ do not admit a local hidden variables model (``local hidden variables models'' are explicitly defined in Appendix \ref{app:classifications}). 

Our next result implies that a Bell scenario requires a ``fork'' of interference influences; that is, it requires that both $\{P_Z^i\} \rightarrow \{P_A^a\}$ and $\{P_Z^i\} \rightarrow \{P_B^b\}$.
But it also requires that this fork is \textit{irreducible}. A fork in the Bell scenario is ``reduced'' by writing each $P_Z^i$ as a sum of products $P_{Z(A)}^mP_{~(B)}^n$ of commuting elements of two projective decompositions $\{P_{Z(A)}^{m}\}$ and $\{P_{Z(B)}^{n}\}$ on $Z$, and likewise writing each $P_W^i$ as a sum of commuting elements $P_{W(A)}^oP_{W(B)}^r$ of two projective decompositions $\{P_{W(A)}^o\}$ and $P_{W(B)}^r\}$ on $W$, where the decompositions are required to satisfy
\begin{equation} \label{eq:bell_reduction}
    \begin{split}
        \{P_{Z(A)}^{m}\} &\not\rightarrow \{P_{W(B)}^{r}\} \\
        \{P_{Z(A)}^{m}\} &\not\rightarrow \{P_B^b\} \\
        \{P_{Z(B)}^{n}\}  &\not\rightarrow \{P_{W(A)}^{o}\} \\
        \{P_{Z(B)}^{n}\}  &\not\rightarrow \{P_A^a\}.
    \end{split}
\end{equation}
An interference fork is called ``irreducible'' if it cannot be reduced.

\begin{theorem} \label{thm:bell}
    A Bell scenario requires an irreducible interference fork made up of interference influences from $\{P_Z^i\}$ to $\{P_A^a\}$ and $\{P_B^b\}$.
\end{theorem}

Our next result concerns the Pusey-Barrett-Rudolph (PBR) theorem \cite{pusey2012reality} concerning the reality of the quantum state. It has been commented that the quantum-theoretical proof of this theorem is somewhat dual to the proof of Bell's theorem (e.g.\ \cite{maudlinPBR2021}). To bring that duality to light, we will define a version of the (bipartite) PBR scenario that is much more general than the one originally discussed in \cite{pusey2012reality}. In this section, we simply define the scenario; Appendix \ref{app:classifications} justifies the definition by showing that it facilitates a proof of a generalized PBR theorem.

To that end, consider the right side of Figure \ref{fig:bell_ewf}. This circuit can be understood as the time reversal of the one on the left, up to some relabellings. The projectors $P_X^x$ and $P_A^a$ correspond to a state prepared on $S$ of the form $\rho^{ax}:={\rm Tr}_{AB}((P_A^a \otimes I_{SB})\cv(P_X^x \otimes I_{FY}))$ (ignoring normalization). Likewise, $P_Y^y$ and $P_B^b$ correspond to $\sigma^{by}:={\rm Tr}_{AB}((I_{AS} \otimes P_B^b)\cv(I_{XF} \otimes P_Y^y))$. It follows from the fact that $X \not\rightarrow B$ and $Y \not\rightarrow A$ through $V$, together with (\ref{eq:quantuminfluence_comm}), that $[\rho^{ax}, \sigma^{by}]=0$. Hence $\rho^{ax}\sigma^{by}$ is also a density operator (up to norm) as long as it is nonzero. The POVM element $\epsilon^{ij}:=\frac{1}{d_z}{\rm Tr}_Z((P_Z^i \otimes I_S)\cu^\dag(P_W^j \otimes I_G))$ corresponds to the projectors $P_Z^i$ and $P_W^j$. For some particular values of the event variables, let us define $\rho:=\rho^{ax}, \rho':=\rho^{a'x'}, \sigma:=\sigma^{by}$, and $\sigma':=\sigma^{b'y'}$. We say the decompositions $(\{P_Z^i\}, \{P_W^j\}, \{P_{X}^{x}\}, \{P_{A}^{a}\}, \{P_{Y}^{y}\}, \{P_{B}^{b}\})$ form a PBR scenario if and only if both of the following conditions are satisfied:
\begin{equation} \label{eq:pbr1}
        {\rm Tr}(\epsilon^{ij}\rho \sigma)  = 0 \
        \lor \ {\rm Tr}(\epsilon^{ij}\rho \sigma') =0 \ 
        \lor \ {\rm Tr}(\epsilon^{ij}\rho' \sigma) =0 \ 
        \lor \ {\rm Tr}(\epsilon^{ij}\rho' \sigma') =0 \quad \forall i, j
\end{equation}

\begin{equation} \label{eq:pbr2}
    \rho \sigma \rho' \sigma' \neq 0.
\end{equation}

We will soon see that a PBR scenario requires a ``collider'' of interference influences; that is, it requires that both $\{P_X^x\} \rightarrow \{P_W^j\}$ and $\{P_Y^y\} \rightarrow \{P_W^j\}$. But just as the fork required for a Bell scenario must be irreducible, so too must the collider in the PBR scenario. The collider is called irreducible if it cannot be reduced by writing the $P_Z^i$ and $P_W^j$ as sums of products $P_{Z(A)}^{m}P_{Z(B)}^{n}$ and $P_{W(A)}^{o}P_{W(B)}^{r}$ of commuting elements of projective decompositions, where this time those decompositions must satisfy
\begin{equation}
    \begin{split}
            \{P_X^x\} &\not\rightarrow  \{P_{W(B)}^{r}\} \\
        \{P_{Z(A)}^{m}\}&\not\rightarrow  \{P_{W(B)}^{r}\} \\
        \{P_Y^y\} &\not\rightarrow  \{P_{W(A)}^{o}\} \\
         \{P_{Z(B)}^{n}\} &\not\rightarrow  \{P_{W(A)}^{o}\}.
    \end{split}
\end{equation}
\begin{theorem} \label{thm:pbr}
    A PBR scenario requires an irreducible interference collider made up of interference influences from $\{P_X^x\}$ and $\{P_Y^y\}$ to $\{P_W^j\}$.
\end{theorem}

\begin{figure}
    \centering
    \begin{tikzpicture} \small 
	\begin{pgfonlayer}{nodelayer}
		\node [style=large map] (0) at (-0.75, -4.5) {$U$};
		\node [style=none] (1) at (0, -7) {};
		\node [style=none] (2) at (0, -0.5) {};
		\node [style=none] (3) at (-1.5, -7) {};
		\node [style=none] (4) at (-1.5, -2.5) {};
		\node [style=none] (5) at (-1, -7.25) {$Z$};
		\node [style=none] (6) at (0.5, -7.25) {$S$};
		\node [style=none] (7) at (0.5, -2.25) {$S$};
		\node [style=none] (8) at (-1, -2.25) {$W$};
		\node [style=none] (9) at (-2.5, -6) {$\{P_{Z}^i\}$};
		\node [style=none] (10) at (-2.5, -3) {$\{P_{W}^j\}$};
		\node [style=upground] (11) at (-1.5, -2.5) {};
		\node [style=ground] (12) at (0, -7) {};
		\node [style=ground] (13) at (-1.5, -7) {};
		\node [style=none] (14) at (-4.75, -0.5) {};
		\node [style=none] (15) at (4.75, -0.5) {};
		\node [style=none] (16) at (-4.75, 0.5) {};
		\node [style=none] (17) at (4.75, 0.5) {};
		\node [style=ground] (18) at (-4.25, -2.5) {};
		\node [style=none] (20) at (-4.25, 0.5) {};
		\node [style=none] (21) at (-4.25, -0.5) {};
		\node [style=none] (22) at (-5.25, -1.5) {  $\{P_{X_1}^{x_1}\}$};
		\node [style=none] (23) at (-5.25, 1.5) {$\{P_{A_1}^{a_1}\}$};
		\node [style=none] (24) at (-3.75, -1.5) {$X_1$};
		\node [style=none] (25) at (-3.75, 1.5) {$A_1$};
		\node [style=ground] (26) at (4.25, -2.5) {};
		\node [style=none] (28) at (4.25, 0.5) {};
		\node [style=none] (29) at (4.25, -0.5) {};
		\node [style=none] (30) at (5.25, -1.5) {$\{P_{Y_1}^{y_1}\}$};
		\node [style=none] (31) at (5.25, 1.5) {$\{P_{B_1}^{b_1}\}$};
		\node [style=none] (32) at (3.75, -1.5) {$Y_1$};
		\node [style=none] (33) at (3.75, 1.5) {$B_1$};
		\node [style=none] (34) at (2.5, 0) {$V_1$};
		\node [style=none] (36) at (0, 0.5) {};
		\node [style=none] (37) at (0, 4) {};
		\node [style=none] (40) at (-7, 4) {};
		\node [style=none] (41) at (7, 4) {};
		\node [style=none] (42) at (7, 5) {};
		\node [style=none] (43) at (-7, 5) {};
		\node [style=ground] (46) at (-6.5, 2) {};
		\node [style=upground] (47) at (-6.5, 7) {};
		\node [style=none] (48) at (-6.5, 5) {};
		\node [style=none] (49) at (-6.5, 4) {};
		\node [style=none] (50) at (-7.5, 3) {$\{P_{X_2}^{x_2}\}$};
		\node [style=none] (51) at (-7.5, 6) {$\{P_{A_2}^{a_2}\}$};
		\node [style=none] (52) at (-6, 3) {$X_2$};
		\node [style=none] (53) at (-6, 6) {$A_2$};
		\node [style=ground] (56) at (6.5, 2) {};
		\node [style=upground] (57) at (6.5, 7) {};
		\node [style=none] (58) at (6.5, 5) {};
		\node [style=none] (59) at (6.5, 4) {};
		\node [style=none] (60) at (7.5, 3) {$\{P_{Y_2}^{y_2}\}$};
		\node [style=none] (61) at (7.5, 6) {$\{P_{B_2}^{b_2}\}$};
		\node [style=none] (62) at (6, 3) {$Y_2$};
		\node [style=none] (63) at (6, 6) {$B_2$};
		\node [style=none] (64) at (-1.5, 0) { \tiny $X_1 \not\rightarrow B_1, \ Y_1 \not\rightarrow A_1$};
		\node [style=none] (65) at (-3.5, 4.5) {\tiny $X_2 \not\rightarrow B_2, \ Y_2 \not\rightarrow A_2$};
		\node [style=none] (66) at (3, 4.5) {$V_2$};
		\node [style=none] (67) at (0, 5) {};
		\node [style=none] (68) at (0, 7) {};
		\node [style=none] (69) at (0, 7) {};
		\node [style=upground] (70) at (0, 7) {};
		\node [style=none] (71) at (-4.25, 2) {};
		\node [style=none] (72) at (-0.5, 3.5) {};
		\node [style=none] (73) at (0.5, 3.5) {};
		\node [style=none] (74) at (-0.5, 4) {};
		\node [style=none] (76) at (4.25, 0.5) {};
		\node [style=none] (78) at (4.25, 2) {};
		\node [style=none] (79) at (0.5, 3.5) {};
		\node [style=none] (80) at (0.5, 4) {};
	\end{pgfonlayer}
	\begin{pgfonlayer}{edgelayer}
		\draw (2.center) to (1.center);
		\draw (4.center) to (3.center);
		\draw (15.center) to (14.center);
		\draw (14.center) to (16.center);
		\draw (16.center) to (17.center);
		\draw (17.center) to (15.center);
		\draw (18) to (21.center);
		\draw (26) to (29.center);
		\draw (37.center) to (36.center);
		\draw (43.center) to (40.center);
		\draw (40.center) to (41.center);
		\draw (41.center) to (42.center);
		\draw (42.center) to (43.center);
		\draw (46) to (49.center);
		\draw (48.center) to (47);
		\draw (56) to (59.center);
		\draw (58.center) to (57);
		\draw (68.center) to (67.center);
		\draw (20.center) to (71.center);
		\draw [in=-90, out=90, looseness=0.75] (71.center) to (72.center);
		\draw (74.center) to (72.center);
		\draw (76.center) to (78.center);
		\draw [in=-90, out=90, looseness=0.75] (78.center) to (79.center);
		\draw (80.center) to (79.center);
	\end{pgfonlayer}
\end{tikzpicture}
    \caption{Local friendliness scenario. }
    \label{fig:pbr}
\end{figure}
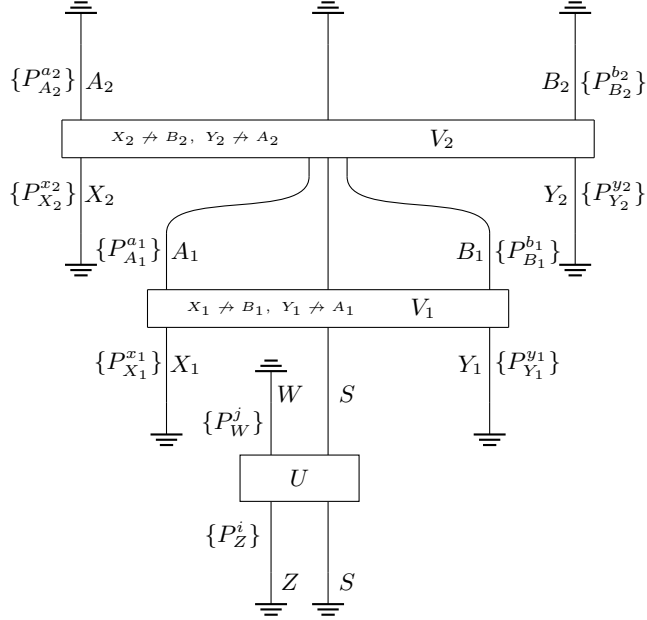

Finally, a \textit{local friendliness scenario} \cite{bong2020strong} combines aspects of a Bell scenario and a Wigner's friend scenario. Specifically, we say the decompositions in Figure \ref{fig:pbr} form a local friendliness scenario if and only if both (a) the decompositions $(\{P_Z^i\},  \{P_W^j\}, \{P_{X_2}^{x_2}\}, \{P_{A_2}^{a_2}\}, \{P_{Y_2}^{y_2}\}, \{P_{B_2}^{b_2}\})$ form a Bell scenario, and (b) the decompositions $(\{P_Z^i\}, \{P_W^j\}, \{P_{X_1}^{x_1}\}, \{P_{A_1}^{a_1}\}, \{P_{X_2}^{x_2}\}, \{P_{A_2}^{a_2}\})$, or the decompositions $(\{P_Z^i\}, \{P_W^j\}, \{P_{Y_1}^{y_1}\}, \{P_{B_1}^{b_1}\}, \{P_{Y_2}^{y_2}\}, \{P_{B_2}^{b_2}\})$, form a Wigner's friend scenario.\footnote{This is a rather permissive definition of a local friendliness scenario, since not all Bell inequality violations are local friendliness inequality violations. But recalling that we are only looking for necessary conditions, and that all local friendliness inequality violations are also Bell inequality violations, this does not create any problems.}
\begin{theorem} \label{thm:lf}
    A local friendliness scenario requires an irreducible interference fork $\{P_{A_2}^{a_2}\} \leftarrow \{P_Z^i\} \rightarrow \{P_{B_2}^{B_2}\}$ and at least one chain. The chain can be of the form $\{P_Z^i\} \rightarrow \{P_{A_1}^{a_1}\} \rightarrow \{P_{A_2}^{a_2}\}$ or $\{P_Z^i\} \rightarrow \{P_{B_1}^{b_1}\} \rightarrow \{P_{B_2}^{b_2}\}$.
\end{theorem}

\begin{figure}
    \centering
    \begin{tikzpicture}
	\begin{pgfonlayer}{nodelayer}
		\node [style=none] (5) at (-8.5, 6.5) {};
		\node [style=none] (6) at (-8.5, 8.5) {};
		\node [style=none] (7) at (-8.5, -4) {};
		\node [style=none] (8) at (-8.5, -2) {};
		\node [style=none] (9) at (-8.5, -1) {};
		\node [style=none] (10) at (-8.5, 1) {};
		\node [style=none] (14) at (-0.5, 6.5) {};
		\node [style=none] (15) at (-2.5, 8.5) {};
		\node [style=none] (16) at (2.5, 8.5) {};
		\node [style=none] (17) at (0.5, 6.5) {};
		\node [style=none] (21) at (-0.5, -4) {};
		\node [style=none] (22) at (-2.5, -2) {};
		\node [style=none] (24) at (0.5, -4) {};
		\node [style=none] (31) at (8.5, 6.5) {};
		\node [style=none] (32) at (10.5, 8.5) {};
		\node [style=none] (33) at (11.5, 8.5) {};
		\node [style=none] (34) at (13.5, 6.5) {};
		\node [style=none] (46) at (-8.5, 5) {complementarity};
		\node [style=none] (47) at (0, 5) {Bell};
		\node [style=none] (48) at (11, 5) {PBR};
		\node [style=none] (49) at (-8.5, -5.5) {Wigner's friend};
		\node [style=none] (50) at (0, -5.5) {local friendliness};
		\node [style=none] (52) at (-8.5, 6) {$\{P_Z^i\}$};
		\node [style=none] (53) at (-8.5, 9) {};
		\node [style=none] (54) at (-8.5, 9) {$\{P_A^a\}$};
		\node [style=none] (55) at (0, 6) {$\{P_Z^i\}$};
		\node [style=none] (56) at (-3, 9) {$\{P_A^a\}$};
		\node [style=none] (57) at (3, 9) {$\{P_B^b\}$};
		\node [style=none] (58) at (8, 6) {$\{P_X^x\}$};
		\node [style=none] (59) at (14, 6) {$\{P_Y^y\}$};
		\node [style=none] (60) at (11, 9) {$\{P_W^j\}$};
		\node [style=none] (61) at (-8.5, -4.5) {$\{P_Z^i\}$};
		\node [style=none] (62) at (-8.5, -1.5) {$\{P_{A_1}^{a_1}\}$};
		\node [style=none] (63) at (-8.5, 1.5) {$\{P_{A_2}^{a_2}\}$};
		\node [style=none] (64) at (0, -4.5) {$\{P_Z^i\}$};
		\node [style=none] (65) at (-2.5, -1) {};
		\node [style=none] (66) at (-2.5, 1) {};
		\node [style=none] (67) at (-2.5, -1.5) {$\{P_{A_1}^{a_1}\}$};
		\node [style=none] (68) at (-2.5, 1.5) {$\{P_{A_2}^{a_2}\}$};
		\node [style=none] (71) at (3, 1.5) {$\{P_{B_2}^{b_2}\}$};
		\node [style=none] (72) at (0, -4) {};
		\node [style=none] (73) at (-2, 1) {};
		\node [style=none] (74) at (2.75, 1) {};
		\node [style=none] (75) at (6.3, -2.5) {(or swapping $A \leftrightarrow B$, $a \leftrightarrow b$)};
	\end{pgfonlayer}
	\begin{pgfonlayer}{edgelayer}
		\draw [style=arrow] (5.center) to (6.center);
		\draw [style=arrow] (7.center) to (8.center);
		\draw [style=arrow] (9.center) to (10.center);
		\draw [style=arrow] (21.center) to (22.center);
		\draw [style=arrow] (65.center) to (66.center);
		\draw [style=red, arrow] (72.center) to (73.center);
		\draw [style=red, arrow] (24.center) to (74.center);
		\draw [style=red, arrow] (14.center) to (15.center);
		\draw [style=red, arrow] (17.center) to (16.center);
		\draw [style=red, arrow] (31.center) to (32.center);
		\draw [style=red, arrow] (34.center) to (33.center);
	\end{pgfonlayer}
\end{tikzpicture}
    \caption{A classification of quantum phenomena according to the structure of interference influences they necessitate. Forks and colliders formed by red arrows are irreducible. A local friendliness scenario requires either the structure displayed explicitly, or else the one obtained by swapping $A \leftrightarrow B$ and $a \leftrightarrow b$, or both.}
    \label{fig:classifications}
\end{figure}
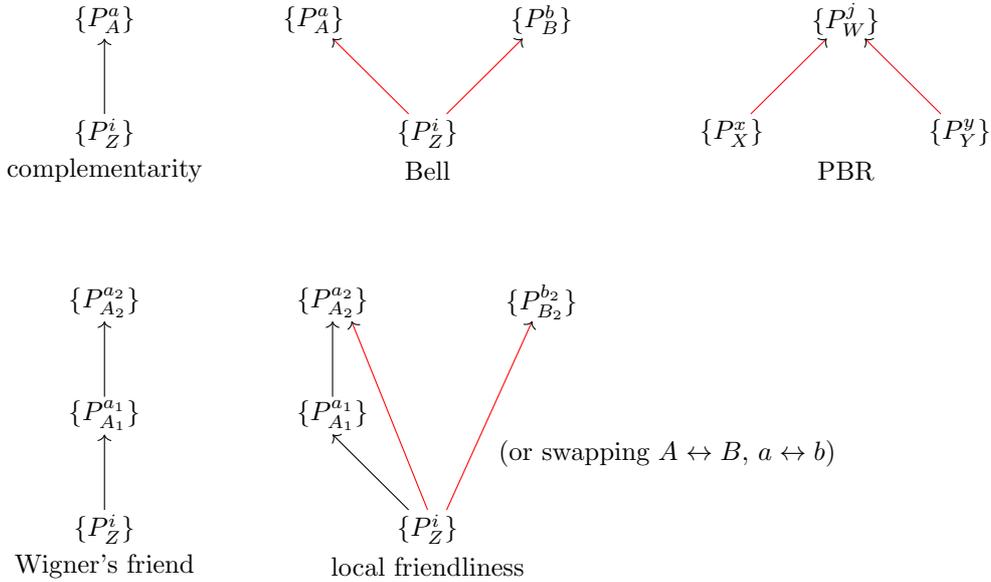

These theorems immediately lead to a classification of quantum phenomena based on causal structure.
\begin{corollary} \label{thm:classifications}
        Each of the five scenarios requires a particular interference causal structure, as depicted in Figure \ref{fig:classifications}. 
\end{corollary}

Corollary \ref{thm:classifications} shows that many of the simplest possible combinations of interference influences lead to various important quantum phenomena. In future work, it is worth exploring whether interesting new phenomena or no-go theorems might be discovered by exploring some simple structures of interference influences that don't already feature in Figure \ref{fig:classifications}. For example, what can be achieved by combining a chain with a collider?

More generally, Corollary \ref{thm:classifications} might form the basis for a new approach to quantum causal modelling, in which it is nonclassicality per se that is to be explained. For example, consider Bell inequality violations. It is true that the quantum common causes of \cite{barrett2020quantum, barrett2021cyclic} and the generalized common causes of \cite{henson2014theory} are necessary conditions for the inequalities to be violated. But both of these sorts of common causes are \textit{also} necessary for correlations that do \textit{not} violate Bell inequalities. So it isn't really the Bell inequality violation in particular that they explain; instead, they explain why there are some correlations rather than no correlations.

On the other hand, any correlations that do not violate Bell inequalities can be recovered without any interference forks at all. So an irreducible interference fork is a necessary condition for \textit{all and only} the correlations that violate Bell inequalities. That is, some correlations requiring an irreducible interference fork is equivalent to those correlations providing a Bell-style proof of nonclassicality. Future work should ask whether the story generalizes, so that \textit{any} proof of nonclassicality is in a similar sense equivalent to a particular structure of interference influences.

\section{On the status of the interpretation} \label{sec:status}

In this penultimate section, we discuss the status of the interpretation laid out in this paper. In particular, we discuss two senses in which the interpretation might be held to provide an accurate description of reality:
\begin{enumerate}
    \item The interpretation, or a modest generalization thereof, can be deployed to accurately describe physics at a fundamental level.
    \item  The interpretation, or a modest generalization thereof, can be deployed to accurately describe physics at an emergent level (and in that respect resembles classical mechanics).
\end{enumerate}

One potential difficulty with both views is epistemological. Emily Adlam \cite{adlam2022does} has described in detail the problem of \textit{intersubjective accessibility} which afflicts certain interpretations (such as RQM without cross-perpsective links). One could argue that the interpretation here faces a similar problem: in short, given two agents, it seems that there should be at least one bubble $\mathfrak{B}_A$ which ``contains'' Alice and not Bob, and another bubble $\mathfrak{B}_B$ which contains Bob but not Alice. But then, assuming there are no correlations between different bubbles, it seems that Alice in $\mathfrak{B}_A$ cannot meaningfully communicate with Bob in $\mathfrak{B}_B$. This might undermine the interpretation's claim to be supported by empirical evidence, particularly if we suppose that Bob is an experimentalist trying to explain to Alice what he saw in the lab, and Alice is a theorist who wishes to decide whether or not to believe in the interpretation. 

Now, there will presumably also be a bubble $\mathfrak{B}_{AB}$ containing both agents. One might attempt to argue that the existence of such ``large'' bubbles in addition to the ``small'' ones is enough to ensure the theory is supported by evidence. Certainly, the existence of big bubbles importantly distinguishes the current interpretation from the ``island universe ontologies'' that Adlam criticizes in \cite{adlam2022does}. But it will take further analysis to determine whether the existence of big bubbles alongside the small ones is really enough to shift the epistemological dial.

If not, then one could attempt to modify the interpretation by positing that histories are only realized relative to ``large'' bubbles. The idea here bears at least a passing resemblance to the extremal action principle of Lagrangian mechanics: just as we are used to saying that particles will follow the paths with the largest (or smallest) action, here we want to say that histories are realized relative to the large bubbles. Of course, the challenge is nailing down a suitable precise definition of ``large''. One candidate definition is that the bubble $\mathfrak{B}$ is large if there does not exist any other bubble $\mathfrak{B}'$ such that any projector that features in $\mathfrak{P}(\mathfrak{C}, \mathfrak{B})$ also features in $\mathfrak{P}(\mathfrak{C}, \mathfrak{B}')$. If the definition of ``large'' is restrictive enough to exclude all but one bubble, then one recovers a sort of absoluteness (but one according to which not all agents in the (extended) Wigner's friend scenario actually obtain an outcome); if not, one obtains a more moderate form of event relativity which may fare better with respect to the problem of intersubjective accessibility. 

Another possible response to the intersubjectivity problem is to postulate that there are in fact correlations between different bubbles. We will return to this idea later on, since it chimes most naturally with view (2).

For now, let us assume the intersubjectivity problem can be resolved, and discuss two other potential difficulties with view (1). One is that so far the interpretation only applies to one of the most simple and empirically limited quantum theories -- the theory of finite-dimensional unitary circuits. Needless to say, such a theory is not usually considered fundamental. While we suspect that no fundamental changes are necessary for an infinite-dimensional generalization of the interpretation, it is less clear whether or not the spirit of the interpretation can survive the transition to \textit{continuous} dynamics, where there is no fundamental division of the dynamics into discrete transformations (i.e.\ into boxes in a circuit). Another issue with viewing the interpretation as fundamental physics is that a single unitary transformation can always be decomposed into many different circuits, and the choice of a \textit{particular} circuit representation is often regarded as somewhat arbitrary. But in the present interpretation, different circuits lead to different bubbles, and thus to different physical situations. To summarize, the interpretation in its current form appears to rely significantly on (1) discreteness of the dynamics, and (2) a preferred circuit representation of the dynamics, both of which might be regarded as surprising attributes of a fundamental theory of physics.

But perhaps they should not be so regarded. Dynamics are indeed continuous in quantum field theory and general relativity. But it is commonly speculated that this continuity must ultimately give way to a more fundamental discreteness in some theory of quantum gravity. Moreover, there is some evidence that a discrete theory of quantum gravity could successfully recover the continuous structures of our current theories as approximations. For example, some free quantum field theories approximate discrete quantum cellular automata \cite{Bisio_2015, D_Ariano_2016}, and some general relativistic spacetimes approximate discrete partial orders (as studied in the context of causal set theory; see \cite{Surya_2019} for a review). 

Moreover, it is striking that a quantum cellular automaton admits a canonical representation as a particular unitary circuit, and its quantum causal structure gives rise to a discrete partial order which one could aim to argue is approximated by a general relativistic spacetime. This raises the speculative possibility of a fundamental theory of physics based on the interpretation of quantum theory proposed in this paper, in which both events and spacetime emerge from a discrete quantum causal structure.

Readers who are unconvinced by such a possibility might consider view (2). One other motivation for doing so is to uncover more elegant structures at a deeper level than the ones we have been discussing in this paper. For example, the current interpretation has not posited any correlations between events that are relative to one bubble and events that are relative to another. Of course, we know from no-go theorems that certain correlations are forbidden 
-- namely, the sort of correlations that would make the absoluteness assumption (effectively) true -- but that doesn't mean that there are \textit{no} correlations whatsoever. And the intersubjectivity problem arguably provides an independent motivation for positing such correlations, since they might facilitate meaningful communication across bubbles. So perhaps there are correlations between bubbles, and perhaps identifying them will allow one to glimpse a deeper level of reality, admitting a simpler and more beautiful description. 

We mention another couple of possible attitudes towards the formalism before closing. Everettians -- and particularly Everettians of the Deutsch-Hayden \cite{deutsch2000information} persuasion -- might wish to employ the notion of preference to describe the branching of the multiverse without relying on the state vector. Indeed, in a precise sense, one can see preference as generalizing the Heisenberg-picture relative states defined in \cite{kuypers2021everettian}.\footnote{On this point, we thank Charles Alexandre Bédard and
Nicetu Tibau Vidal for enlightening discussions.}  Finally, one could argue that the formalism plays primarily an epistemic role, doing more to describe our beliefs and inferences than to describe nature itself.

\section{Discussion} \label{sec:discussion}

Bell's theorem pushed us towards quantum influences; Wigner's friend, towards event relativity. Here, we have argued that a deeper understanding of quantum theory is to be sought in the marriage of these two ideas. At the core of this argument is Theorem \ref{thm:linear_prob}, which shows that quantum causal structure singles out a unique set of consistent histories relative to every subset of a set of unitarily interacting subsystems. 

This facilitates an interpretation of quantum theory according to which all dynamics are unitary, and yet the universe is represented not by a unitarily evolving quantum state, but by emergent events. Every interpretation of quantum theory must advocate at least one radical conceptual shift, and the current one advocates the following: dynamics do not merely \textit{constrain} the state of reality; they \textit{create} it.

This interpretation rejects the absoluteness of events, but it does contain absolutes: just like in the Everrett interpretation, events are only nonabsolute \textit{until you relativize them} to a suitable reference (for the Everettians, branches; for us, bubbles). But, unlike the Everett interpretation, this one is genuinely stochastic, meaning that there is no difficulty in arguing that the theory would be falsified by non-Born-rule frequencies.  

The interpretation makes precise the idea from RQM that ``events arise from interaction'' (with the important caveat that, on the current interpretation, the interaction is among general sets, rather than pairs, of systems). In doing so, it also provides a way of vastly reducing the number of incompatible sets of events that one finds in consistent histories, and thus it provides a way of addressing Dowker and Kent's criticism \cite{dowker1996consistent}, to which we shall soon return. The interference influences on which the interpretation is based facilitate causal explanations of specifically nonclassical features of quantum correlations in a number of scenarios, and it seems possible that these piecemeal results could be developed into a significant extension of the quantum causal modelling framework. As the previous section discusses, the interpretation does not come without its difficulties. Nevertheless, we consider it a promising route towards an ever deeper understanding of the quantum world. \\

There are a number of directions for future work. It is worth exploring which projective decompositions are preferred in much more general scenarios. A particular area of interest is the emergence of approximate classicality. Typically, this topic has been approached with an (implicit or explicit) assumption that the quantum state plays a crucial role in shaping the ontology. Hence there is often much emphasis on the block-diagonality of a density matrix and its approximately classical evolution through time. One of the core ideas of this work is that decoherence can instead be understood in purely causal terms as an absence of interference influences.
This provides us with a way of studying the emergence of classicality that doesn't suffer the pitfalls associated with realism about the quantum state (e.g.\, the existence of Everettian branches that violate the Born rule).


And so a natural next step after this paper is to determine whether or not this causal conception of decoherence permits a satisfying explanation of emergent classicality. If unitary circuits are generated from realistic Hamiltonians, then do the $\mathfrak{P}(\mathfrak{C}, \mathfrak{B})$ for an appropriate class of bubbles contain projectors that approximately localize particles in position and momentum, and are the corresponding events distributed in a way that approximates classical equations of motion? If so, then the problems raised by Dowker and Kent \cite{dowker1996consistent}  are arguably resolved. That is, relative to the right sort of bubble, one can then unambiguously predict that the sun will rise again.

To that end, it would be worthwhile developing an infinite-dimensional generalization of the interpretation. It is also worth exploring in further detail the connection between the emergence of events as we have described it and time symmetry, touched on in the remarks after Definition \ref{def:pref2}.

Another question for future work is whether our \textsc{dynamics} assumption might be replaced with a ``\textsc{causation}'' one. For on our account, the preferred set of projective decompositions $\mathfrak{P}(\mathfrak{C}, \mathfrak{B})$ is derived only from the interference influences through the circuit -- the full dynamical structure is not required. Therefore, it is conceivable that the interpretation could be modified to only posit a set of interference influences, rather than a full dynamical structure. Whether or not this is possible depends on whether the following conjecture is true: \textit{any pair of unitary circuits with exactly the same interference influences lead to exactly the same probabilities via (\ref{eq:bubble_probability_rule})}. If so, then perhaps we have found a way of making good on a suggestion by Spekkens \cite{spekkens2015paradigm}, that the traditional dualistic paradigm of kinematics and dynamics should be replaced by a more unified paradigm based on causal structure.

Most ambitiously, it is also worth exploring the connection with quantum gravity (touched on in the previous section). There, it is often speculated that spacetime emerges from causation. In our interpretation, which is background-independent, \textit{events} emerge from causation. Is there a more unified story, waiting to be told?

\section*{Acknowledgements}

This project has been going on for a long time, and the list of people to thank is correspondingly long. We are very grateful to
Emily Adlam,
Marina Maciel Ansanelli,
Charles Alexandre Bédard,
{\v{C}}aslav Brukner,
Eric Cavalcanti,
Giulio Chiribella,
Daniel Gore,
Caroline L. Jones,
Adrian Kent,
Hl\'er Kristj\'ansson,
Robin Lorenz,
Tein van der Lught,
Markus M{\" u}ller,
Simon Saunders,
David Schmid,
Rob Spekkens,
Chris Timpson,
Augustin Vanrietvelde,
Nicetu Tibau Vidal, 
Matt Wilson, and 
Y\`il\`e Yīng
for a number of useful discussions. 
Special thanks go to Daniel Gore and Matthew Wilson for reading and commenting on drafts of the manuscript. We would also like to thank the audiences and organizers of talks given by JB at the 2023 QISS workshop in Oxford and the seminars by NO at the Perimeter Institute \cite{ormrod2023quantum} and the Institute for Quantum Optics and Quantum  Information in Vienna.  

This research was funded in part by the Engineering and Physical Sciences
Research Council (EPSRC), and was supported by
the John Templeton Foundation through the ID\#
62312 grant, as part of the ‘The Quantum Information Structure of Spacetime’ Project (QISS). The
opinions expressed in this publication are those of the
authors and do not necessarily reflect the views of
the John Templeton Foundation. For the purpose of
Open Access, the authors have applied a CC BY public copyright licence to any Author Accepted Manuscript (AAM) version arising from this submission.

\bibliography{refs}

\appendix

\section{A structure lemma for operator algebras} \label{app:structure}

Here, we state a fairly well-known result about the structure of algebras of operators on finite-dimensional Hilbert spaces. It follows from Theorem III.1.1 of \cite{davidson1996c}.

\begin{lemma} \cite{davidson1996c}. \label{thm:structure_fundamental}
    For any subalgebra $\cx \subseteq \cl(\ch)$ of the algebra $\cl(\ch)$ of operators on a finite-dimensional Hilbert space $\ch$, there exists some decomposition of $\ch$ of the form $\ch = \bigoplus_i \ch_L^i \otimes \ch_R^i$ such that 
    \begin{equation}
        M \in \cx \quad \Longleftrightarrow \quad \exists \{M_L^i\}: \ \sum_{i} M_{L^i} \otimes \pi_{R^i}
    \end{equation}
    where each $M_{L^i}$ is an operator on $\ch_L$ that has null support outside $\ch_L^i$, and each $\pi_{R^i}$ is the operator on $\ch_R$ that projects onto $\ch_R^i$.
\end{lemma}

\section{Proof of Theorem \ref{thm:int_equivalence}} \label{app:interference_equivalence}

On the one hand, suppose $[\tilde{P}_A^i, \tilde{P}_D^j] = 0 \ \ \forall i, j$. Then, for any $j$ and $V_{\vec{\phi}}$ of the form $V_{\vec{\phi}} = \sum_i e^{i\phi_i}P_A^i \otimes I_B$,
\begin{equation}
    \begin{split}
        {\rm Tr}((I_C \otimes P_D^j)UV_{\vec{\phi}}(\cdot)V^\dag_{\vec{\phi}} U^\dag)&={\rm Tr}(U\tilde{P}_D^jV_{\vec{\phi}}(\cdot) V_{\vec{\phi}}^\dag U^\dag) \\
        &={\rm Tr}(\tilde{P}_D^jV_{\vec{\phi}}(\cdot) V_{\vec{\phi}}^\dag) \\
        &= {\rm Tr}(\tilde{P}_D^j V_{\vec{\phi}}^\dag V_{\vec{\phi}}(\cdot)) \\
        &= {\rm Tr}(\tilde{P}_D^j(\cdot)) \\
        &={\rm Tr}((I_C \otimes P_D^j)U(\cdot) U^\dag).
    \end{split}
\end{equation}

On the other hand, that ${\rm Tr} \big((I_C \otimes P_D^j)\cu(V_{\vec{\phi}}(\cdot)V_{\vec{\phi}}^\dag)\big)={\rm Tr} \big((I_C \otimes P_D^j)\cu(\cdot)\big)  \ \ \forall j, V_{\vec{\phi}}$. Since ${\rm Tr}(M(\cdot))= {\rm Tr}(N(\cdot))$ implies that $M=N$ for any operators $M$ and $N$, it follows that $V_{\vec{\phi}}^\dag U^\dag  (I_C \otimes P_D^j) U V_{\vec{\phi}} = U^\dag (I_C \otimes P_D^j) U$. Therefore, $[V_{\vec{\phi}}, \tilde{P}_D^j]=0$. Define $V_{\vec{0}}:= I_A \otimes I_B$ and $V_{\vec{\phi_i}}:=(I_A - 2 P_A^i) \otimes I_B$. For any $i$ and $j$, one can deduce that $[\tilde{P}_A^i, \tilde{P}_D^j]=\frac{1}{2}[V_{\vec{0}}-V_{\vec{\phi_i}}, \tilde{P}_D^j]=0$. This proves the theorem.

We note that one can also generalize the proof to show that $\{P_A^i\} \rightarrow  \{P_D^j\}$ is equivalent to the possibility of signalling via a protocol in which the sender performs a transformation on $A$ with Kraus operators that are complex linear combinations of the $P_A^i$ and the receiver does a measurement with Kraus operators that are complex linear combinations of the $P_D^j$. Alternatively, one can see this fact as corollary of Theorem 3.2 of \cite{ormrod2023causal}. 

\section{Proof of Theorem \ref{thm:preference}} \label{app:preference_equivalence}

We'll show that each of the three conditions that define preference is equivalent to an inclusion relation between operator algebras, and that the conjunction of these three inclusion relations is equivalent to $\texttt{span}(\{P_A^i\}) \otimes I_B = \texttt{centre} (\ca \cap \texttt{comm}(\cd))$.

Lemma \ref{thm:structure_fundamental} implies that $\cd$ is spanned by its projectors. Therefore, commuting with every projector in $\cd$ means commuting with $\cd$ itself. And so, condition (1) of preference is equivalent to the inclusion relation
\begin{equation}
    \texttt{span}(\{P_A^i\}) \otimes I_B \subseteq \ca \cap \texttt{comm}(\cd).
\end{equation}

Therefore, condition (2) is equivalent to the statement that $\texttt{span}(\{Q_A^k\}) \otimes I_B \subseteq \ca \cap \texttt{comm}(\cd)$ implies $\texttt{span}(\{Q_A^k\}) \otimes I_B \subseteq \texttt{comm}(\texttt{span}(\{P_A^i\}) \otimes I_B)$.
Since the algebra $\ca \cap \texttt{comm}(\cd)$ is spanned by its projectors, this is equivalent to this statement that $\ca \cap \texttt{comm}(\cd) \subseteq\texttt{comm}(\texttt{span}(\{P_A^i\}) \otimes I_B)$, which is in turn equivalent to the inclusion relation
\begin{equation}
    \texttt{span}(\{P_A^i\}) \otimes I_B \subseteq \texttt{comm}(\ca \cap \texttt{comm}(\cd)).
\end{equation}

It follows that the conjunction of (1) and (2) is equivalent to the statement that $\texttt{span}(\{P_A^i\}) \otimes I_B\subseteq \texttt{centre}(\ca \cap \texttt{comm}(\cd))$. Hence, condition (3) is saying that $\texttt{span}(\{R_A^l\}) \otimes I_B \subseteq \texttt{centre}(\ca \cap \texttt{comm}(\cd))$ implies $ \texttt{span}(\{R_A^l\}) \otimes I_B \subseteq \texttt{span}(\{P_A^i\}) \otimes I_B \subseteq \ca \cap \texttt{comm}(\cd) \otimes I_B$.
And hence, it is equivalent to 
\begin{equation}
    \texttt{centre}(\ca \cap \texttt{comm}(\cd) \subseteq \texttt{span}(\{P_A^i\}) \otimes I_B.
\end{equation}

Therefore, the conjunction of all three conditions for preference is equivalent to $\texttt{span}(\{P_A^i\} \otimes I_B= \texttt{centre}(\ca \cap \texttt{comm}(\cd)$.

\section{Proof of equation (\ref{eq:allowed_int_infl})} \label{app:nochains}

In this section, we prove equation (\ref{eq:allowed_int_infl}), i.e.\ we show that the only permitted interference influences between projective decompositions within a single $\mathfrak{P}(\mathfrak{C}, \mathfrak{B})$ go from an ingoing decomposition to an outgoing one, where the latter is either associated with the same system as the former, or else another system higher up in the circuit.

To begin with, consider a unitary circuit and the simple case of a bubble with just three systems, $\mathfrak{B}=\{A, B, C\}$. The circuit can always thought of as a combination of four unitary transformations (which we think of here as unitary operators between Hilbert spaces)
\begin{equation}
    \begin{split}
        U_1&: G \rightarrow A \otimes \overline{A} \\
        U_2&: A \otimes \overline{A} \rightarrow B \otimes \overline{B} \\
        U_3&: B \otimes \overline{B} \rightarrow C \otimes \overline{C} \\
        U_4&: C \otimes \overline{C} \rightarrow F,
    \end{split}
\end{equation}
where we have assumed, without loss of generality, that $A<B<C$ is compatible with the temporal order induced by the circuit.
$\mathfrak{P}(\mathfrak{C}, \mathfrak{B})$ contains six projective decompositions, which we label $\mathbb{D}_{A/B/C}^{\rm in / out}$. They are derived by considering a unitary operator of the type $U:  A^{\rm out} \otimes B^{\rm out}\otimes C^{\rm out}   \otimes G \rightarrow A^{\rm in} \otimes B^{\rm in}\otimes C^{\rm in}   \otimes F$, corresponding to the bottom-right of Figure \ref{fig:preference}. As described in Figure \ref{fig:interference_influence_cicuit}, we say there is an interference influence between a pair of these decompositions if, when we embed them back into the original circuit, they fail to commute in the Heisenberg picture.

We'll start by showing that there can be no interference influence from $\mathbb{D}_A^{\rm out}$ to $\mathbb{D}_C^{\rm out}$. To this end, we note that, by the definition of $U$, the unitary operator for the whole circuit $V:=U_4U_3U_2U_1$ can be written
\begin{equation}
    {\rm Tr}_{A^{\rm out}  B^{\rm out} C^{\rm out}}((I_{A^{\rm in}  B^{\rm in} C^{\rm in} \rightarrow A^{\rm out} B^{\rm out} C^{\rm out}} \otimes I_F)U)=V,
\end{equation}
where $I_{A^{\rm in}  B^{\rm in} C^{\rm in} \rightarrow A^{\rm out} B^{\rm out} C^{\rm out}}$ is the identity operator from the tensor product of the ingoing Hilbert spaces to the tensor product of the outgoing Hilbert spaces. (Diagrammatically, one can think of this operation as bending around each $X^{\rm in}$ wire that comes out of $U$, and then inserting it into the corresponding $X^{\rm out}$ that goes into $U$.) For an arbitrary $P \in \mathbb{D}_{A}^{\rm out}$,
\begin{equation} \label{eq:looping_in}
    \begin{split}
            {\rm Tr}_{A^{\rm out}  B^{\rm out} C^{\rm out}}((I_{A^{\rm in}  B^{\rm in} C^{\rm in} \rightarrow A^{\rm out}  B^{\rm out} C^{\rm out}} \otimes I_F)U(P \otimes I_{B^{\rm out}C^{\rm out}G})) &= U_4U_3U_2(P \otimes I_{\overline{A}})U_1 \\
            &= \tilde{P} V,
    \end{split}
\end{equation}
where $\tilde{P} :=U_4U_3U_2(P \otimes I_{\overline{A}})U_2^\dag U_3^\dag U_4^\dag$. Similarly, for any $Q \in \mathbb{D}_C^{\rm out}$,
\begin{equation} 
    \begin{split}
            {\rm Tr}_{A^{\rm out}  B^{\rm out} C^{\rm out}}((I_{A^{\rm in}  B^{\rm in} C^{\rm in} \rightarrow A^{\rm out}  B^{\rm out} C^{\rm out}} \otimes I_F)U(Q \otimes I_{A^{\rm out}B^{\rm out}G})) &= U_4(Q \otimes I_{\overline{C}})U_3U_2U_1 \\
            &= \tilde{Q} V,
    \end{split}
\end{equation}
where $\tilde{Q}:=U_4QU_4^\dag$. Now recall that $\mathbb{D}_A^{\rm out}$ is preferred by $A^{\rm in} \otimes B^{\rm in}\otimes C^{\rm in}$ given $U$, implying that $U(P \otimes I_{B^{\rm out}C^{\rm out}P})U^\dag= I_{A^{\rm in}  B^{\rm in} C^{\rm in}} \otimes M_F$ for some $M_F$. Thus we can write
\begin{equation} \label{eq:pulling_through}
    \begin{split}
         {\rm Tr}_{A^{\rm out}  B^{\rm out} C^{\rm out}}&((I_{A^{\rm in}  B^{\rm in} C^{\rm in} \rightarrow A^{\rm out}  B^{\rm out} C^{\rm out}} \otimes I_F)U(P \otimes I_{B^{\rm out}C^{\rm out}G})) \\
          &={\rm Tr}_{A^{\rm out}  B^{\rm out} C^{\rm out}}((I_{A^{\rm in}  B^{\rm in} C^{\rm in} \rightarrow A^{\rm out}  B^{\rm out} C^{\rm out}} \otimes M_F)U) \\
          &=M_FV.
    \end{split}
\end{equation}
Comparing (\ref{eq:looping_in}) and (\ref{eq:pulling_through}), we conclude that $M_F=\tilde{P}.$ Similarly, we know that $U(Q \otimes I_{A^{\rm out}B^{\rm out}G})U^\dag = I_{A^{\rm in}  B^{\rm in} C^{\rm in}} \otimes N_F$ for some $N_F$, and can show that $N_F=\tilde{Q}$. Since unitary transformations preserve commutation relations, we can then argue that 
\begin{equation}
    [P \otimes I_{B^{\rm out}C^{\rm out}G}, Q \otimes I_{A^{\rm out}B^{\rm out}G}]=0 \quad \implies \quad  [M_F, N_F]=0 \quad  \implies \quad [\tilde{P}, \tilde{Q}]=0.
\end{equation}
It follows that $\mathbb{D}_A^{\rm out} \not\rightarrow \mathbb{D}_C^{\rm out}$.

Since the relevant notions are symmetric in time, we can give a closely analogous argument that $\mathbb{D}_A^{\rm in} \not\rightarrow \mathbb{D}_C^{\rm in}$.

Now, we show that $\mathbb{D}_A^{\rm out} \not\rightarrow \mathbb{D}_C^{\rm in}$. On the one hand, for any $P \in \mathbb{D}_A^{\rm out}$ and $R \in \mathbb{D}_C^{\rm in}$, 
\begin{equation}
    \begin{split}
            {\rm Tr}_{A^{\rm out}  B^{\rm out} C^{\rm out}}&((I_{A^{\rm in}  B^{\rm in} C^{\rm in} \rightarrow A^{\rm out}  B^{\rm out} C^{\rm out}}  \otimes I_F)(R \otimes I_{A^{\rm in}B^{\rm in}F})U(P \otimes I_{B^{\rm out}C^{\rm out}G})) \\
            &= U_4(R \otimes I_{\overline{C}})U_3U_2(P \otimes I_{\overline{A}})U_1 \\
            &= \tilde{R} \tilde{P} V
    \end{split}
\end{equation}
where $\tilde{P} :=U_4U_3U_2(P \otimes I_{\overline{A}})U_2^\dag U_3^\dag U_4^\dag$, $\tilde{R}:=U_4RU_4^\dag$. On the other hand,
\begin{equation}
    \begin{split}
            {\rm Tr}_{A^{\rm out}  B^{\rm out} C^{\rm out}}&((I_{A^{\rm in}  B^{\rm in} C^{\rm in} \rightarrow A^{\rm out}  B^{\rm out} C^{\rm out}}  \otimes I_F)(R \otimes I_{A^{\rm in}B^{\rm in}F})U(P \otimes I_{B^{\rm out}C^{\rm out}G})) \\
             &={\rm Tr}_{A^{\rm out}  B^{\rm out} C^{\rm out}}((I_{A^{\rm in}  B^{\rm in} C^{\rm in} \rightarrow A^{\rm out}  B^{\rm out} C^{\rm out}}  \otimes I_F)(R \otimes I_{A^{\rm in}B^{\rm in}} \otimes M_F)U) \\
             &=M_F{\rm Tr}_{A^{\rm out}  B^{\rm out} C^{\rm out}}((I_{A^{\rm in}  B^{\rm in} C^{\rm in} \rightarrow A^{\rm out}  B^{\rm out} C^{\rm out}}  \otimes I_F)(R \otimes I_{A^{\rm in}B^{\rm in}F})
             U) \\
            &= \tilde{P} \tilde{R} V,
    \end{split}
\end{equation}
where $M_F$ is defined as above. It follows that $[\tilde{P}, \tilde{R}]=0$, so $\mathbb{D}_A^{\rm out} \not\rightarrow \mathbb{D}_C^{\rm in}$.

What we have shown so far is that for the circuit and bubble we have been considering, the only possible interference influence from $A$ to $C$ has the form $\mathbb{D}_A^{\rm in} \rightarrow \mathbb{D}_C^{\rm out}$. Now consider a unitary circuit and a bubble $(A_1, \ldots, A_n)$, where the subscript corresponds to the temporal order. We can write the circuit in the form
\begin{equation}
    \begin{split}
       & V_1: G \rightarrow A_1 \otimes \overline{A}_1 \\
       & V_i: A_{i-1} \otimes \overline{A}_{i-1} \rightarrow A_i \otimes \overline{A}_i \ 
{\rm for} \ i \in \{2, \ldots n\} \\
    &V_{n+1}: A_n \otimes \overline{A}_n \rightarrow F.
    \end{split}
\end{equation}
The arguments above straightforwardly generalize to show that for any $j \geq i$, the only possible interference influence from (a decomposition associated with) $A_i$ to (a decomposition associated with $A_j$) is $\mathbb{D}_{A_i}^{\rm in} \rightarrow \mathbb{D}_{A_j}^{\rm out}$. This proves the theorem.

\section{Reproducing standard quantum theory} \label{app:operational}

This appendix serves as an instruction manual for reconstructing an arbitrary circuit from standard finite-dimensional quantum theory using the interpretation from this paper. 

The most general sort of deterministic transformation one can perform in standard finite-dimensional quantum theory is a \textit{quantum channel}. This is a completely positive map $\cc: A \rightarrow B$ from linear operators $\ch_A$ to linear operators on $\ch_B$ that preserves the trace of the operators. The most general (possibly non-deterministic) sort of transformation is a \textit{quantum instrument}. This is a set $\{\cc_i\}_i$ of completely positive maps $\cc_i: A \rightarrow B$ from operators on $\ch_A$ to operators on $\ch_B$ whose sum $\sum_i \cc_i$ is a quantum channel. We will explain how one can construct any quantum instrument, as well as any circuit formed by composing quantum instruments in sequence and in parallel. Let us start by showing that any quantum instrument can be thought of as an orthonormal basis measurement of the ancillary output of a unitary that acts on a larger system.

\begin{lemma} \label{lemma:instrument}
For any quantum instrument $\{\cc_i\}_i$ of type $\cc_i: A \rightarrow B$ there exists a unitary channel $\cu: A \otimes X \rightarrow B \otimes Y \otimes Z$ and a state $\ket{\psi} \in \ch_X$ such that 
\begin{equation}
    \cc_i(\cdot) = {\rm Tr}_{YZ} \big((I_B \otimes \ket{i}\bra{i}_Y \otimes I_Z)\cu( (\cdot) \otimes \ket{\psi}\bra{\psi}) \big).
\end{equation}
\end{lemma}

To prove Lemma \ref{lemma:instrument}, define the channel $\cd: A \rightarrow B \otimes Y$ as $\cd(\cdot) := \sum_i \cc_i(\cdot) \otimes \ket{i}\bra{i}_Y$. By the Stinespring dilation theorem, there exists a unitary channel $\cu: A \otimes X \rightarrow B \otimes Y \otimes Z$ and a state $\ket{\psi}_X$ such that $\cd(\cdot) = {\rm Tr}_Z \cu( (\cdot) \otimes \ket{\psi}\bra{\psi}_X)$. The lemma immediately follows. 

The next lemma shows how a unitary can be chosen so that projectors onto a given basis end up being preferred by an appropriate system.

\begin{lemma} \label{lemma:comp_pref}
The projective decomposition $\{\ket{i}\bra{i}_A\}$ on $A$ is preferred by $D$ given the unitary $V := \sum_{i, j=0}^{d-1} \ket{i}_C\bra{i}_A \otimes \ket{j+i}_D\bra{j}_B$ (where the addition is modulo $d$).
\end{lemma}

For the proof, first note that  $\{\ket{i}\bra{i}_A\}$ is preferred by $D$ if and only if the following condition holds.
\begin{equation}
   M_A \in \texttt{span} (\{\ket{i}\bra{i}_A\}) \quad \Longleftrightarrow  \quad [V(M_A \otimes I_B)V^\dag, I_C \otimes M_D]=0  \ \forall M_D
\end{equation}
The $\Rightarrow$ direction is obvious. For $\Leftarrow$, suppose that $[V(M_A \otimes I_B)V^\dag, I_C \otimes D]=0$ for all $M_D$. It follows that $ V(M_A \otimes I_B)V^\dag = \tilde{M}_C \otimes I_D$ for some $\tilde{M}_C$. Using the definition of $V$, this can be rewritten as $\sum \bra{i}M_A\ket{i’}  \ket{i}\bra{i'}_C \otimes (\sum_j \ket{j+i}\bra{j+i’}_D) = \tilde{M}_C \otimes I_D$ for some $\tilde{M}_C$. Applying $(\bra{k}_C \otimes \bra{k}_D)(\cdot)(\ket{k'}_C \otimes \ket{k'}_D)$ to both sides leaves us with $\bra{k}M_A\ket{k’}=0$ for all $k \neq k'$. It follows that $ M_A \in \texttt{span} (\{\ket{i}\bra{i}_A\})$. 

\begin{figure}
    \centering
    \begin{tikzpicture}
	\begin{pgfonlayer}{nodelayer}
		\node [style=none] (0) at (0.75, 0.5) {};
		\node [style=none] (1) at (0.75, -0.5) {};
		\node [style=none] (2) at (5, -0.5) {};
		\node [style=none] (3) at (5, 0.5) {};
		\node [style=none] (4) at (2.75, 0) {$U$};
		\node [style=none] (5) at (4.5, -0.5) {};
		\node [style=none] (6) at (4.5, 0.5) {};
		\node [style=none] (7) at (1.25, -0.5) {};
		\node [style=none] (8) at (1.25, 0.5) {};
		\node [style=none] (9) at (4.5, -1.5) {};
		\node [style=none] (10) at (4.5, -2.5) {};
		\node [style=none] (11) at (4.25, -2.5) {};
		\node [style=none] (12) at (4.25, -1.5) {};
		\node [style=large map] (13) at (5.5, -6.25) {$V^\dag$};
		\node [style=none] (14) at (4.5, -8.25) {};
		\node [style=none] (15) at (6.5, -7.5) {};
		\node [style=none] (16) at (6.5, -5.25) {};
		\node [style=none] (17) at (6.5, -7.5) {};
		\node [style=none] (18) at (6.5, -8.5) {};
		\node [style=none] (19) at (6.5, -8.5) {};
		\node [style=none] (20) at (4.5, 2.5) {};
		\node [style=none] (21) at (4.5, 1.5) {};
		\node [style=large map] (22) at (5.5, 4) {$V$};
		\node [style=none] (23) at (4.5, 0.5) {};
		\node [style=none] (24) at (6.5, 6.25) {};
		\node [style=none] (25) at (6.5, 6.25) {};
		\node [style=none] (26) at (6.5, 6.25) {};
		\node [style=none] (27) at (6.5, 5.25) {};
		\node [style=none] (28) at (6.5, 3) {};
		\node [style=none] (29) at (4.5, 6) {};
		\node [style=none] (30) at (1.25, -2) {};
		\node [style=none] (31) at (1.25, 2) {};
		\node [style=none] (32) at (0.5, -1.75) {$A$};
		\node [style=map] (33) at (4.5, -4) {$W$};
		\node [style=none] (34) at (3.75, -2) {$X_3$};
		\node [style=none] (35) at (3.75, -7.75) {$X_1$};
		\node [style=none] (36) at (3.75, -5.25) {$X_2$};
		\node [style=none] (37) at (0.5, 1.75) {$B$};
		\node [style=none] (38) at (2, 1.75) {$Y$};
		\node [style=none] (39) at (3.75, 2) {$Z_1$};
		\node [style=ground] (40) at (1.25, -2) {};
		\node [style=upground] (41) at (1.25, 2) {};
		\node [style=none] (42) at (2.75, 0.5) {};
		\node [style=none] (43) at (2.75, 2) {};
		\node [style=upground] (44) at (2.75, 2) {};
		\node [style=upground] (45) at (4.5, 6) {};
		\node [style=upground] (46) at (6.5, 6.25) {};
		\node [style=ground] (47) at (6.5, 3) {};
		\node [style=ground] (48) at (6.5, -8.5) {};
		\node [style=upground] (49) at (6.5, -5.25) {};
		\node [style=none] (50) at (4.5, -8.25) {};
		\node [style=none] (51) at (4.5, -8.25) {};
		\node [style=ground] (52) at (4.5, -8.25) {};
		\node [style=none] (53) at (7, -8.75) {$e$};
		\node [style=none] (54) at (7, -7.25) {$e'$};
		\node [style=none] (55) at (5, -2.75) {$f$};
		\node [style=none] (56) at (5, -1.25) {$f'$};
		\node [style=none] (57) at (5, 1.25) {$g$};
		\node [style=none] (58) at (5, 2.75) {$g'$};
		\node [style=none] (59) at (7.25, 5) {$h$};
		\node [style=none] (60) at (7.25, 6.5) {$h'$};
		\node [style=none] (61) at (3.75, 5.5) {$Z_2$};
	\end{pgfonlayer}
	\begin{pgfonlayer}{edgelayer}
		\draw (1.center) to (2.center);
		\draw (2.center) to (3.center);
		\draw (3.center) to (0.center);
		\draw (0.center) to (1.center);
		\draw (9.center) to (5.center);
		\draw (14.center) to (10.center);
		\draw (16.center) to (15.center);
		\draw (19.center) to (18.center);
		\draw (23.center) to (21.center);
		\draw (25.center) to (24.center);
		\draw (28.center) to (27.center);
		\draw (29.center) to (20.center);
		\draw (31.center) to (8.center);
		\draw (7.center) to (30.center);
		\draw (43.center) to (42.center);
	\end{pgfonlayer}
\end{tikzpicture}
    \caption{A model of a quantum instrument. Six events take place relative to the bubble of four systems indicated by the broken wires. Only two events, $f$ and $g'$, are important for reproducing the quantum instrument. Specifically, conditioning on $f=0$ amounts to assuming that the measurement device has been successfully prepared, and $g'$ can then be understood as the outcome of the instrument.}
    \label{fig:instrument_construction}
\end{figure}
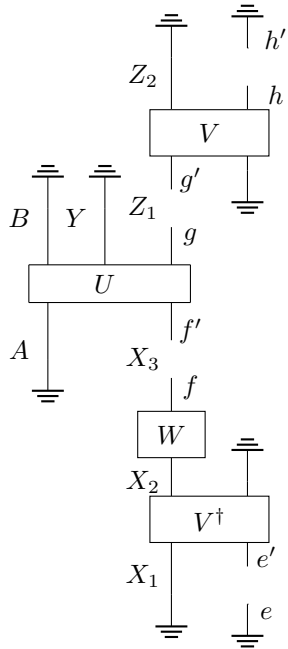

The method for reproducing an arbitrary quantum instrument is given in Figure \ref{fig:instrument_construction}. One uses the unitary channel $\cu$ whose existence is guaranteed by Lemma \ref{lemma:instrument}. Beforehand, one acts on $X$ and ancillas with a unitary of the form $V^\dag$ (where the $X$ output of this unitary is analogous to the $A$ from the definition of $V$, and the $X$ input is analogous to the $C$) and then with another unitary $W$ with the property that $W \ket{0}_{X_2}=\ket{\psi}_{X_3}$, where $\ket{\psi}$ is the state from Lemma \ref{lemma:instrument}. After $\cu$ one performs a unitary of the form $V$ on $Z$ and ancillas (where the $Z$ input is analogous to the $A$ from $V$'s definition, and the $Z$ output is analogous to $C$). One considers the bubble $\mathfrak{B}$ corresponding to the four cuts in the wires in the figure. $\mathfrak{P}(\mathfrak{C}, \mathfrak{B})$ contains $\mathbb{D}_{X_3}^{\rm in}= \{W\ket{i}\bra{i}_{X_2}W^\dag\}$ and $\mathbb{D}_Y^{\rm out} = \{\ket{i}\bra{i}_Y\}$. When $f=0$, the projector $\ket{\psi}\bra{\psi}_{X_3} \in \mathbb{D}_{X_3}^{\rm in}$ is selected, meaning that the instrument is successfully implemented. The event $g'$ can then be identified with the outcome of the instrument, and is associated with the completely positive map $\cc_{g'}$.

To model an experiment in which the instrument $\{\cc^1_i\}$ of type $\cc_i^1:  A \rightarrow B$ is followed by another instrument  $\{\cc^2_i\}$ of type $\cc_i^2:  B \rightarrow C$, one simply performs the construction from Figure \ref{fig:instrument_construction} for each instrument (using separate unitaries and ancillary systems in each case) and then plugs the $B$ output of the first construction into the $B$ input of the second construction. One then considers a bubble of 8 systems, formally the union $\mathfrak{B} = \mathfrak{B}_1 \cup \mathfrak{B}_2$ of the two bubbles from each of the individual constructions. Relative to this new bubble, the probability of getting the outcome $g_2$ of the second instrument given the outcome $g_1$ of the first instrument, assuming that each instrument is successfully implemented (i.e.\ that $f_1=f_2=0$) is indeed what one would expect from standard quantum theory:
\begin{equation}
   p_\mathfrak{B}(g'_2|f_1=f_2=0, g'_1) = {\rm Tr} \big(\cc_{g_2}^2(\cc_{g_1}^1(I/d)) \big).
\end{equation}
For a concrete example, suppose the first instrument was a preparation of a particular density operator, so $\{\cc^1_i\}_i=\{\rho\}$ (in this case, the input system of the first instrument is trivial, $\ch_A \cong \mathbb{C}$). Then $g_1'$ only takes one possible value and can be ignored. The expression above becomes $p_\mathfrak{B}(g'_2|f_1=f_2=0) = {\rm Tr} \big(\cc_{g_2}^2(\rho) \big)$. 

Composing instruments in parallel simply involves taking the tensor product of two constructions. Again, one recovers the probability formulae one expects from the standard theory.

\section{The three-box paradox} \label{app:3box}

This appendix articulates a generalized sense in which a three-box paradox could conceivably arise in the current interpretation, and then shows that it actually never does. 

Suppose that, given a unitary circuit, the triplets of decompositions $(\mathbb{D}_1, \mathbb{D}_2^i, \mathbb{D}_3)$ are contained in $\mathfrak{P}(\mathfrak{C}, \mathfrak{B}_i)$ respectively for $i \in \{0, 1\}$, and assume they are given in an order compatible with ``temporal'' order induced by the circuit. It follows that there exists a pair of probability distributions  $p_{\mathfrak{B}_0}$ and $p_{\mathfrak{B}_1}$, each over three events, and each corresponding to one of the triplets.

Now further suppose that all the Heisenberg projectors in $\mathbb{D}_2^0$ commute with all those in $\mathbb{D}_2^1$ (this is indeed the case for the three-box paradox described in Section \ref{app:3box}). Then consider the set of decompositions $(\mathbb{D}_1, \mathbb{D}_2^0, \mathbb{D}_2^1, \mathbb{D}_3)$. This four-element set might not be contained in the preferred set of decompositions for any bubble. However, since there are no chains of interference influences, it will still satisfy (\ref{eq:simpler_form}) with $\rho=I/3$. It is therefore mathematically possible to write down a probability distribution $p_2$ over all four events using (\ref{eq:bubble_probability_rule}).

Now suppose, for contradiction, that the decompositions $(\mathbb{D}_1, \mathbb{D}_2^i, \mathbb{D}_3)$ form a (generalized) three-box paradox. That is, assume that $p_{\mathfrak{B}_0}(e_{2,0}|e_1, e_3)=p_{\mathfrak{B}_1}(e_{2,1}|e_1, e_3)=1$, where the fixed events $e_{2, 0}$ and $e_{2, 1}$ correspond to orthogonal projectors. The form of (\ref{eq:bubble_probability_rule}) implies that the probability distribution $p_2$ for $(\mathbb{D}_1, \mathbb{D}_2^0, \mathbb{D}_2^1, \mathbb{D}_3)$ embeds $p_{\mathfrak{B}_1}$ and $p_{\mathfrak{B}_2}$ as its marginals, so that:
\begin{equation}
    \begin{split}
       \sum_{\tilde{e}_{2, 1}}p_2(e_{2,0}\tilde{e}_{2, 1}|e_1, e_3) &= p_{\mathfrak{B}_0}(e_{2,0}|e_1, e_3) =1 \\
        \sum_{\tilde{e}_{2, 0}}p_2(\tilde{e}_{2,0}e_{2, 1}|e_1, e_3) &= p_{\mathfrak{B}_1}(e_{2,1}|e_1, e_3) =1.
    \end{split}
\end{equation}
But since $e_{2, 0}$ and $e_{2, 1}$ correspond to orthogonal projectors, the form of (\ref{eq:bubble_probability_rule}) also implies that $p_2(e_{2, 0}e_{2, 1}|e_1, e_3)=0$. This is in contradiction with a rule that is always satisfied by probability distributions, namely that if $q(\alpha|\gamma)=q(\beta|\gamma)=1$ then $q(\alpha \beta|\gamma)=1$. In summary, the assumption that each $(\mathbb{D}_1, \mathbb{D}_2^i, \mathbb{D}_3)$ not only forms a consistent set of histories, but also lacks chains of interference influences (\ref{eq:simpler_form}), allows us to embed each marginal probability distribution in a single ``global'' distribution, blocking a three-box paradox.

\section{Classifications} \label{app:classifications}

In this final appendix, we prove Theorems \ref{thm:classifications} -- \ref{thm:lf}, and elaborate on some of the scenarios. 

First though, let us state a useful lemma, which follows from Lemma \ref{thm:structure_fundamental}.

\begin{lemma} \label{lemma:structure}
    Consider a Hilbert space $\ch_X \otimes \ch_Y \otimes \ch_Z$, an algebra $\ca$ of operators that all have the form $M=M_{XY} \otimes I_Z$, and an algebra $\cb$ of operators that all have the form $N=I_X \otimes N_{YZ}$. If $\ca \subseteq \texttt{comm}(\cb)$, then $\ch_Y$ admits a decomposition $\bigoplus_i \ch_{Y_L^i} \otimes \ch_{Y_R^i}$ such that any $M \in \ca$ and $N \in \cb$ can be written in the forms $M=\sum_iM_{X{Y_L^i}} \otimes \pi_{Y_R^iZ}$ and $N=\sum_i \pi_{X{Y_L^i}} \otimes N_{Y_R^iZ}$ respectively, where the $\pi_{Y_R^iZ}$ are projectors onto $\ch_{Y_R^i} \otimes \ch_Z$ and similarly the $\pi_{X{Y_L^i}}$ project onto $\ch_X \otimes \ch_{Y_L^i}$.
\end{lemma}

\textbf{Proof of Theorems \ref{thm:complementarity} and \ref{thm:wf}.} Suppose that the left side of Figure \ref{fig:comp_wf} forms a complementarity scenario. Then assume, for contradiction, $\{P_{Z}^i\} \not\rightarrow \{P_{A}^a\}$. By the definition of an interference influence, 
\begin{equation}
    [\cu(P_{Z}^i \otimes I_G) \otimes I_{X}, I_{W} \otimes \cv^\dag(I_F \otimes P_{A}^a)]=0.
\end{equation}
But the commutator $[\rho^{ij}, \sigma^{kl}]$ can be rewritten as
\begin{equation}
   [\rho^{ij}, \sigma^{kl}]= {\rm Tr}_{WX} \big((P_{W}^j \otimes I_S \otimes P_{X}^x)[\cu(P_{Z}^i \otimes I_G) \otimes I_{X}, I_{W} \otimes \cv^\dag(I_F \otimes P_{A}^a)] \big) =0.
\end{equation}
Therefore, we do not have a complementarity scenario. 

This proves Theorem \ref{thm:complementarity}. Due to the definition of a Wigner's friend scenario as a pair of complementarity scenarios, Theorem \ref{thm:wf} then follows immediately. \\

\textbf{Proof of Theorems \ref{thm:bell} and \ref{thm:lf}.} Before giving the proof for the Bell scenario, let us explicitly define the phrase ``local hidden variables model'' that appeared in its definition. The probability distribution $p(axby|ij)$ for a fixed $i$ and $j$ admits a local hidden variables model if and only if it is mathematically possible to express it as the marginal $p(axby|ij)= \sum_{\lambda} p(axby\lambda|ij)$ of a probability distribution $p(axby\lambda|ij)$, where two constraints called ``Bell locality'' and ``statistical independence'' are satisfied. Bell locality says that the left and right variables are screened off from one another by $\lambda$, in the sense that $p(axby|\lambda ij)=p(ax|\lambda ij)p(by|\lambda ij)$. Statistical independence says that the settings are uncorrelated from $\lambda$, i.e.\ $p(xy\lambda| ij)=p(xy|ij)p(\lambda| ij)$.

Now for the proof. If there is no irreducible interference fork in the circuit on the left of Figure \ref{fig:bell_ewf}, then there exist projective decompositions $\{P_{Z(A)}^{m}\}$ and $ \{P_{Z(B)}^{n}\}$ on $Z$ and $ \{P_{W(A)}^{o}\}$ and $ \{P_{W(B)}^{r}\}$ on $W$ satisfying
\begin{equation} \label{eq:no_int_1}
    \begin{split}
    & [P_{Z(A)}^{m}, P_{Z(B)}^{n}]= 0 \quad \forall  mn \\
    & [P_{W(A)}^{o}, P_{W(B)}^{r}]=0 \quad \forall op \\
      &  \{P_{Z(A)}^{m}\} \not\rightarrow \{P_B^b\} \\
       & \{P_{Z(A)}^{m}\} \not\rightarrow \{P_{W(B)}^{r}\} \\
        & \{P_{Z(B)}^{n}\} \not\rightarrow \{P_A^a\} \\
        &\{P_{Z(B)}^{n}\} \not\rightarrow \{P_{W(A)}^{o}\},
    \end{split}
\end{equation}
and such that each $P_Z^i$ can be written as a sum of products $P_{Z(A)}^{m}P_{Z(B)}^{n}$, and each $P_W^j$ can be written as a sum of products $P_{W(A)}^{o}P_{W(B)}^{r}$. Formally, this last statement means that the set $M \times N$ of possible joint values $(m, n)$ can be partitioned into disjoint subsets $(M \times N)_i$ such that $P_Z^i=\sum_{(m, n) \in (M \times N)_i} P_{Z(A)}^{m}P_{Z(B)}^{n}$, and similarly we can write $P_W^j=\sum_{(o, r) \in (O \times R)_j} P_{W(A)}^{o}P_{W(B)}^{r}$. Note then that every joint value $(m, n)$ is associated with exactly one value of $i$, and likewise each $(o, r)$ with one value of $j$. We denote these values $i_{mn}$ and $j_{or}$ respectively.

Since $X \not\rightarrow B$ and $Y \not\rightarrow A$ through $V$, it follows from (\ref{eq:fine_graining}) that in particular
\begin{equation} \label{eq:no_int_2}
\begin{split}
    \{P_X^x\} &\not\rightarrow \{P_B^b\} \\
    \{P_Y^y\} &\not\rightarrow \{P_A^a\}.
\end{split}
\end{equation}

Let us use tildes to denote projectors that have been embedded into the Hilbert space for the whole circuit fragment and transformed into the time slice after $U$ but before $V$. So, for example, $\tilde{P}_A^a:=I_{W} \otimes \cv^\dag(P_A^a \otimes I_{FB})$, and $\tilde{P}_{Z(A)}^{m} := I_X \otimes \cu(P_{Z(A)}^{m} \otimes I_G) \otimes I_Y$. Now, consider the operator algebras
\begin{equation}
    \begin{split}
        \ca :&= \texttt{aspan} \big(\{\tilde{P}_{Z(A)}^{m}\} \cup \{\tilde{P}_X^x\} \cup  \{\tilde{P}_{W(A)}^{o}\} \cup \{\tilde{P}_A^a\} \big) \\
        \cb :&= \texttt{aspan} \big(\{\tilde{P}_{Z(B)}^{n}\} \cup \{\tilde{P}_Y^y\} \cup \{\tilde{P}_{W(B)}^{r}\} \cup \{\tilde{P}_B^b\} \big),
    \end{split}
\end{equation}
where $\texttt{aspan}(s)$ is the algebra of operators obtained by taking matrix products and convex linear combinations of operators in the set $s$. (\ref{eq:no_int_1}), (\ref{eq:no_int_2}), and temporal order of the circuit imply that $\ca \subseteq \texttt{comm}(\cb)$. Define the composite system $C$ by $\ch_C = \ch_{W} \otimes \ch_S$. Lemma \ref{lemma:structure} implies that there exists a decomposition $\ch_C = \bigoplus_k \ch_{C_L^k} \otimes \ch_{C_R^k}$ such that any $M \in \ca$ and $N \in \cb$ can be written 
\begin{equation} \label{eq:l3app}
    \begin{split}
        M=\sum_k M_{XC_L^k} \otimes \pi_{C_R^kY} \\
        N=\sum_k \pi_{XC_L^k} \otimes N_{C_R^kY}.
    \end{split}
\end{equation}

The probability distribution $p(axbyij)$ is given by
\begin{equation}
    p(axbyij)= \frac{1}{d} {\rm Tr}(\tilde{P}_Z^i \tilde{P}_X^x \tilde{P}_Y^y\tilde{P}_W^j\tilde{P}_A^a\tilde{P}_B^b). 
\end{equation}
Defining the projectors $\tilde{P}_C^k:=\pi_{XC_L^k} \otimes \pi_{C_R^kY}$ this can be rewritten as $p(axbyij)=\sum_{mnopk}q(axbyijmnork)$, where the probability distribution $q(axbyijmnork)$ is defined by
\begin{equation} \label{eq:eminem}
\begin{split}
       q(axbyijmnork)  := \frac{1}{d} {\rm Tr}(\tilde{P}_{Z(A)}^{m} \tilde{P}_{Z(B)}^{n}  \tilde{P}_X^x \tilde{P}_Y^y \tilde{P}_C^k   \tilde{P}_{W(A)}^{o} \tilde{P}_{W(B)}^{r} \tilde{P}_A^a\tilde{P}_B^b) \delta_{i, i_{mn}} \delta_{j, j_{or}}.  
\end{split}
\end{equation}
{(One can verify this is a valid probably by an argument similar to the proof of Theorem \ref{thm:linear_prob}.)} Now we show that $p(axby|ij)$ admits a local hidden variables model for $\lambda := (m, n, o, r, k)$. For the proof of Bell locality, we start by commuting around some projectors in (\ref{eq:eminem}).
\begin{equation}
    \begin{split}
        q(axbyijmnork) &= \frac{1}{d} {\rm Tr}(\tilde{P}_C^k  
        P_{Z(A)}^{m}  \tilde{P}_X^x \tilde{P}_{W(A)}^{o} \tilde{P}_A^a
        P_{Z(B)}^{n} \tilde{P}_Y^y \tilde{P}_{W(B)}^{r} \tilde{P}_B^b) \delta_{i, i_{mn}} \delta_{j, j_{or}} \\
        &=\frac{1}{d}{\rm Tr}(M_{XC_L^k}^{mxoa} \otimes  N_{C_R^kY}^{nyrb})\delta_{i, i_{mn}} \delta_{j, j_{or}} \\
        &= \frac{1}{d}{\rm Tr}(M_{XC_L^k}^{mxoa}) {\rm Tr}( N_{C_R^kY}^{nyrb}) \delta_{i, i_{mn}} 
 \delta_{j, j_{or}}\\
        & \implies q(axby|ijmnork)=q(ax|ijmnork)q(by|ijmnork).
    \end{split}
\end{equation}
Here, the second line uses (\ref{eq:l3app}) and the facts that $P_{Z(A)}^{m}  \tilde{P}_X^x \tilde{P}_{W(A)}^{o} \tilde{P}_A^a \in \ca$ and $P_{Z(B)}^{n} \tilde{P}_Y^y \tilde{P}_{W(B)}^{r} \tilde{P}_B^b \in \cb$. The final step uses the fact that if a probability distribution $p(\alpha \beta \gamma)$ can be written as $p(\alpha \beta \gamma)=f(\alpha \gamma)g(\beta \gamma)$ for some functions $f$ and $g$, then $p(\alpha \beta |\gamma)= p(\alpha|\gamma)p(\beta|\gamma)$, using the substitutions $\alpha=ax, \beta=by$ and $\lambda=ijmnork$. For statistical independence,
\begin{equation}
    \begin{split}
         q(xyijmnork) &= \frac{1}{d} {\rm Tr}(\tilde{P}_C^k  
        P_{Z(A)}^{m}  \tilde{P}_{W(A)}^{o} 
        P_{Z(B)}^{n}\tilde{P}_{W(B)}^{r} \tilde{P}_X^x \tilde{P}_Y^y ) \delta_{i, i_{mn}} \delta_{j, j_{or}} \\
        &= \frac{1}{d} {\rm Tr}(O^{ijmnork}_{C}) \delta_{i, i_{mn}} \delta_{j, j_{or}} {\rm Tr}(P_X^x \otimes P_Y^y)  \\
        & \implies q(xymnork|ij)=q(xy|ij)q(mnork|ij).
    \end{split}
\end{equation}
The second equality makes use of the fact that the first five traced-over operators in the first line act trivially on $X$ and $Y$. The final inference exploits the aforementioned fact about distributions of the form $p(\alpha \beta \gamma)=f(\alpha \gamma)g(\beta \gamma)$,  this time using the substitutions $\alpha=xy$, $\beta=mnork$ and $\gamma=ij$. This proves Theorem \ref{thm:bell}

Theorem \ref{thm:lf} follows immediately from Theorems \ref{thm:wf} and \ref{thm:bell}: one needs both a chain for the Wigner's friend scenario required by the definition of the local friendliness scenario, and an irreducible interference fork for the Bell inequality violations. \\

\textbf{Explaining the (generalized) PBR scenario.} Since our PBR scenario is in some respects a generalization of the one from \cite{pusey2012reality}, it is worth explaining its relationship with the PBR theorem.

We start with a recap of the PBR scenario as described in \cite{pusey2012reality}. Suppose that when a quantum state $\ket{0}$ is prepared, that really means that some state $\lambda$ in a more detailed, but as yet undiscovered, theory is prepared, with a probability density of $\mu_0(\lambda)$. And suppose that when $\ket{+}$ is prepared, that really means that some state $\lambda$ is prepared with probability density $\mu_+(\lambda)$. And suppose that the two distributions overlap. That is, suppose that the intersection of the supports of $\mu_0$ and $\mu_+$ is attributed a nonzero probability by both measures. If the $\lambda$'s are regarded as complete physical descriptions of the system, then the state space of quantum theory can then be regarded as merely \textit{epistemic}, because two different quantum states can correspond to exactly the same physical states of affairs.

However, \cite{pusey2012reality} shows that the distributions cannot overlap given some natural assumptions. First up, one has to assume that the probability for obtaining the outcome corresponding to $\ket{\phi_i}$ when performing basis measurement on a quantum state $\ket{\psi}$ is given by a weighted sum of probabilities for that outcome given some fixed $\lambda$:
\begin{equation} \label{eq:prob_ass}
    p(i|\psi) = \int d\lambda \mu(i|\lambda)\mu_\psi(\lambda).
\end{equation}
We call this the \textit{probability assumption.} It is one of the defining assumptions of the ontological models framework \cite{harrigan2010einstein}, on which the PBR theorem is based. Secondly, one needs to assume that if $A$ is prepared in $\ket{\alpha}$ for $\alpha \in \{0, +\}$, and likewise $B$ is prepared in $\ket{\beta}$ for $\beta \in \{0, +\}$, then the description of $A \otimes B$ in the underlying theory is
\begin{equation} \label{eq:pi}
    \mu_{\alpha \beta}(\lambda_A, \lambda_B)=\mu_\alpha(\lambda_A) \mu_\beta(\lambda_B).
\end{equation}
This is called \textit{preparation independence}. Assuming the predictions of quantum theory are correct, it follows that the Born probability $p(i|\alpha \beta)=|\bra{\Phi_i}\ket{\Psi_{\alpha, \beta}}|^2$ for getting the outcome for $\bra{\Phi_i}$ when performing a measurement of a basis on the state $\ket{\Psi_{\alpha, \beta}}:=\ket{\alpha}\ket{\beta}$ is given by
\begin{equation} \label{eq:prob}
    p(i|\alpha \beta)= \int d \lambda_A d\lambda_B \mu(i|\lambda_A, \lambda_B) \mu_\alpha(\lambda_A) \mu_\beta(\lambda_B)
\end{equation}
A little thought shows that, given these assumptions, if we can find a basis with the property that
\begin{equation} \label{eq:pbr_property}
    \forall i  \ \exists \alpha, \beta: \ \braket{\Phi_i|\Psi_{\alpha, \beta}}=0,
\end{equation}
i.e.\ with the property that \textit{every} possible outcome of the measurement rules out one of the $\ket{\Psi_{\alpha, \beta}}$, then $\mu_0$ and $\mu_+$ cannot overlap. 

\cite{pusey2012reality} shows that such a basis does indeed exist. It then generalizes the argument beyond a bipartite scenario to show that there can be no overlap between any pair of distinct pure states, again given (generalizations of) the last two assumptions. \\

We now move on to our version of the PBR scenario. Our version sticks to the bipartite case. However, it is more general in the sense that (i) we consider mixed, rather than pure, states, (ii) we consider possibly different pairs of states on each of the two subsystems, and (iii) we do not assume that the two subsystems are isomorphic, or that they should be understood as tensor factors of the overall Hilbert space. Instead, we let the two subsystems correspond to the left and right parts of a decomposition of the overall Hilbert space of the form
\begin{equation} \label{eq:structure}
    \ch_S = \bigoplus_i \ch_{S_L^i} \otimes \ch_{S_R^i}.
\end{equation}
The ``left'' subsystem $L$ is associated with density operators of the form
\begin{equation}
    \rho= \bigoplus_i p_i \rho_{S_L^i} \otimes I_{S_R^i}
\end{equation}
where each $\rho_{S_L^i}$ is itself a density operator on the Hilbert space $\ch_{S_L^i}$, and the $p_i$ form a probability distribution. Similarly, ``right'' subsystem $R$ is associated with density operators of the form
\begin{equation}
    \sigma = \bigoplus_i q_i I_{S_L^i} \otimes \sigma_{S_R^i}
\end{equation}
where the $\sigma_{S_R^i}$ are density operators and the $q_i$ form a probability distribution. 

As the proof of Theorem \ref{thm:pbr} will show, the $\rho$, $\rho'$, $\sigma$ and $\sigma'$ that appear in the definition of the PBR scenario do indeed have this form. Roughly speaking, a PBR scenario as we define it facilitates a no-go theorem for the claim that the distributions for the two states on $L$ overlap, and the distributions for the two states on $R$ overlap, and, moreover, the overlap corresponds to the same subspace $\ch_{S_L^i} \otimes  \ch_{S_R^i}$ of $\ch_S$.

Let us make this more precise. We would like to associate $\rho^{(')}$ with some probability density function $\mu_{\rho^{(')}}$ over the states $\lambda_L \in \Lambda_L$, and $\sigma^{(')}$ with some probability density function $\mu_{\sigma^{(')}}$ over the states $\lambda_R \in \Lambda_R$. Now, let us assume that in general, if a density operator $\tau$ can be written as a probabilistic combination of different density operators $\tau=\sum r_k \tau_k$, then the probability distribution associated with $\tau$ is a corresponding probability distribution over the different $\tau_k$. That is,
\begin{equation} \label{eq:pbr_conv}
   \tau=\sum r_k \tau_k \quad \implies \quad  \mu_{\tau} = \sum_k r_k \mu_{\tau_k}.
\end{equation}
If we imagine that we subject $\rho^{(')}$ to a PVM $\{\pi_S^i\}$ made up of projectors onto the different orthogonal subspaces in 
(\ref{eq:structure}), this implies that the state space $\Lambda_L$ must decompose as $\Lambda_L = \cup_i \Lambda_L^i$, where each $\Lambda_L^i \subseteq \Lambda_L$ is made up of states that return the $i$th outcome of the POVM with certainty, and thus $\Lambda_L^i \cap \Lambda_L^j=\emptyset$ for $i \neq j$. Similarly, we can decompose $\Lambda_R = \cup_i \Lambda_R^i$ into disjoint regions for each outcome of the POVM. Then we can write the probability functions for $\rho$, $\rho'$, $\sigma$, and $\sigma'$ as mixtures of probability functions for different values of $i$. For example, $\mu_{\rho}= \sum_i p_i \mu_{\rho_{S_L^i}}$, where the support of each $\mu_{\rho_{S_L^i}}$ is contained in the corresponding $\Lambda_L^i$.

As in \cite{pusey2012reality}, we shall make use of the probability assumption (\ref{eq:prob_ass}). But in this setting, it no longer makes sense to impose the same preparation independence assumption we did before. For if the probability distribution associated to, say, $\rho\sigma$ was the product of probability distributions associated to $\rho$ and $\sigma$ individually, then the resulting distribution might have nontrivial support on a region $\Lambda^i_L \times  \Lambda_R^j$ where $i \neq j$, even though the two regions in the product expression correspond to contradictory outcomes of the PVM $\{\pi_S^i\}$. Instead, we need a generalized preparation independence assumption. Since $\rho\sigma$ has the form
\begin{equation}
    \frac{\sum_i p_i q_i \rho^{}_{S_L^i} \otimes \sigma^{}_{S_R^i}}{\sum_j p_j q_j}
\end{equation}
upon renormalization, we shall assume that the associated probability distribution is defined by
\begin{equation} \label{eq:gen_pi}
    \mu_{\rho\sigma}(\lambda_L, \lambda_R)=\frac{\sum_i p_iq_i \mu_{\rho_{S_L^i}}(\lambda_L)\mu_{\sigma_{S_R^i}}(\lambda_R)}{\sum_j p_j q_j }.
\end{equation}
Thus the two subsystems are not prepared independently, in so far as the preparation is not successful if $\lambda_L \in \Lambda_L^i$ and $\lambda_R \in \Lambda_R^j$ for some $i \neq j$. However, \textit{given the assumption} that both subsystems are successfully prepared in matching subspaces $\Lambda_L^i$ and $\Lambda_R^i$ for some fixed $i$, they are independent. (We make analogous assumptions for all $\rho^{(')}\sigma^{(')}$). 

Let us define $\co_L^i$ as the part of the overlap of the distributions $\mu_\rho$ and $\mu_{\rho'}$ that is contained in $\Lambda_L^i$, and $\co_R^i$ similarly. That is, 
\begin{equation}
\begin{split}
        \co_L^i &:= \texttt{supp}(\mu_{\rho}) \cap \texttt{supp}(\mu_{\rho'}) \cap \Lambda_L^i \\
        \co_R^i &:= \texttt{supp}(\mu_{\sigma}) \cap \texttt{supp}(\mu_{\sigma'}) \cap \Lambda_R^i.
\end{split}
\end{equation}
Recall that we are aiming for a no-go theorem for the claim that the distributions corresponding to $\rho$ and $\rho'$ have a nontrivial overlap \textit{and} the distributions corresponding to $\sigma$ and $\sigma'$ have a nontrivial overlap \textit{on the same subspace}. That is, 
\begin{equation} \label{eq:pbr_claim}
    \exists i: \quad p_i\mu_{\rho}(\co_L^i) \neq 0, \ p_i'\mu_{\rho'} \ (\co_L^i) \neq 0, \  q_i\mu_{\sigma}(\co_R^i) \neq 0, \  q_i'\mu_{\sigma'}(\co_R^i) \neq 0. 
\end{equation}
If (\ref{eq:pbr_claim}) holds in conjunction with generalized preparation independence (\ref{eq:gen_pi}), then we can infer that there is a nontrivial overlap of all four distributions $\mu_{\rho^{(')} \sigma^{(')}}$. 
{That is, (\ref{eq:gen_pi}) and (\ref{eq:pbr_claim}) together imply that} 
\begin{equation} \label{eq:pbr_jointclaim}
    \exists \cs:  \cs \subseteq  \co_{LR}, \ \mu_{\rho^{(')}\sigma^{(')}}(\cs) \neq 0,
\end{equation}
where for readability we have defined 
\begin{equation}
    \co_{LR} :=\texttt{supp}(\mu_{\rho\sigma}) \cap \texttt{supp}(\mu_{\rho\sigma'}) \cap \texttt{supp}(\mu_{\rho'\sigma}) \cap \texttt{supp}(\mu_{\rho'\sigma'}),
\end{equation}
and specifically $\cs = \co_L^i \times \co_R^i$. 

If we also assume the existence of the POVM from equation (\ref{eq:pbr1}) and we make the probability assumption (\ref{eq:prob_ass}), then (\ref{eq:pbr_jointclaim}) implies that the probability for any outcome $(i, j)$ given that the state $\lambda \in \cs$ is $\mu(ij|\lambda)=0$. But this also contradicts the probability assumption, since then $\sum_{ij}\mu(ij|\lambda) \neq 1$ and so $\mu(ij|\lambda)$ cannot be a (conditional) probability distribution. So (\ref{eq:pbr1}), (\ref{eq:prob_ass}), and (\ref{eq:pbr_jointclaim}) imply a contradiction. But ultimately, (\ref{eq:pbr_jointclaim}) was derived from generalized preparation independence (\ref{eq:gen_pi}) and the existence of the overlap as described in equation (\ref{eq:pbr_claim}). Therefore, given the probability assumption and generalized preparation independence, the existence of a POVM with the property (\ref{eq:pbr1}) implies that the overlap from (\ref{eq:pbr_claim}) does not exist. Therefore, if one can show that such a POVM exists, one can rule out epistemic interpretations of the quantum state that assume both the probability rule and generalized preparation independence.

It remains to justify the assumption of equation (\ref{eq:pbr2}), namely, that the product $\rho\sigma\rho\sigma'$ is nonzero. This follows from the other assumptions we have already made in setting out this generalized PBR no-go theorem. The probability assumption implies that orthogonal density operators are associated with nonoverlapping probability densities. Therefore, the existence of the overlap from (\ref{eq:pbr_claim}) implies that, for some $i$, $\rho_{S_L^i}\rho_{S_L^i}' \neq 0$ and $\sigma_{S_R^i}\sigma_{S_R^i}' \neq 0$. But then the structure of the operators with respect to  (\ref{eq:structure}) implies that $\rho \sigma \rho' \sigma' \neq 0$. Thus whenever one can derive a no-go theorem along the lines we can describe, (\ref{eq:pbr2}) holds -- this is why it is acceptable to make it part of the definition of a PBR scenario.

We have just sketched out how a PBR-style no-go theorem can be derived in the bipartite case when the subsystems are associated with density operators belonging to commuting operator algebras. The proof below will show, amongst other things, that one does indeed find such subsystems whenever one has a PBR scenario as defined above. \\

\textbf{Proof of Theorem \ref{thm:pbr}.} As in the Bell scenario, suppose that, for all $i$ and $j$, $P_Z^i$ and $P_W^j$ can be written as sums of products of commuting elements of projective decompositions on $Z$ and $W$ respectively. That is, assume that $P_Z^i = \sum_{(m, n) \in (M \times N)_i} 
P_{Z(A)}^{m}P_{Z(B)}^{n}$, and $P_W^j=\sum_{(o, r) \in (O \times R)_j} P_{W(A)}^{o}P_{W(B)}^{r}$, where the $(M \times N)_i$ are nonoverlapping sets of joint valuations of the indices $m$ and $n$, and similarly $(O \times R)_j$. Assume that $\{P_{Z(A)}^{m}\} \not\rightarrow \{P_{W(B)}^{r}\}$, $\{P_X^x\} \not\rightarrow \{P_{W(B)}^{r}\}$, $\{P_{Z(B)}^{n}\} \not\rightarrow \{P_{W(A)}^{o}\}$, and $\{P_Y^y\} \not\rightarrow \{P_{W(A)}^{o}\}$. Again define the algebras $\ca := \texttt{aspan} \big(\{\tilde{P}_{Z(A)}^{m}\} \cup \{\tilde{P}_X^x\} \cup  \{\tilde{P}_{W(A)}^{o}\} \cup \{\tilde{P}_A^a\} \big)$ and $\cb := \texttt{aspan} \big(\{\tilde{P}_{Z(B)}^{n}\} \cup \{\tilde{P}_Y^y\} \cup \{\tilde{P}_{W(B)}^{r}\} \cup \{\tilde{P}_B^b\} \big)$, where tildes now denote Heisenberg projectors on the system $Z \otimes A \otimes S \otimes B$.  Deduce that $\ca \subseteq \texttt{comm}(\cb)$, and thus from Lemma \ref{lemma:structure} that $\ch_D:=\ch_Z \otimes \ch_S$ admits a decomposition
$\ch_D = \bigoplus_k \ch_{D_L^k} \otimes \ch_{D_R^k}$, such that any $M \in \ca$ and $N \in \cb$ can be written 
\begin{equation} \label{eq:l3app2}
    \begin{split}
        M=\sum_k M_{AD_L^k} \otimes \pi_{D_R^kB} \\
        N=\sum_k \pi_{AD_L^k} \otimes N_{D_R^kB}.
    \end{split}
\end{equation}

$\rho$ and $\rho'$ are defined on $D_L$, while $\sigma$ and $\sigma'$ are defined on $D_R$.
Since $[\rho, \sigma]=[\rho', \sigma']=0$, $\rho\sigma$ and $\rho' \sigma'$ are both positive operators. For any pair of positive operators $O$ and $O'$, we have that $OO' = 0 \Leftrightarrow {\rm Tr}(OO')=0$. Thus (\ref{eq:pbr2}) implies that ${\rm Tr}(\rho \sigma \rho' \sigma') = \sum_{ij} {\rm Tr}(\epsilon^{ij}\rho \sigma \rho' \sigma')  \neq 0$. Since $\rho \sigma \rho' \sigma'$ is also a positive operator it follows that
\begin{equation} \label{eq:pbrexists}
    \exists i, j: \ {\rm Tr}(\epsilon^{ij}\rho \sigma \rho' \sigma')  \neq 0.
\end{equation}
This will allow us to show that (given the lack of an irreducible interference collider) if (\ref{eq:pbr2}) holds then (\ref{eq:pbr1}) does not. To that end, we note that
\begin{equation}
    \begin{split}
        {\rm Tr}(\epsilon^{ij}\rho \sigma \rho' \sigma') &= \frac{1}{d_Zd_Ad_B} {\rm Tr}(\tilde{P}_Z^i\tilde{P}_W^j(I_{AB} \otimes I_Z \otimes  \rho \sigma \rho' \sigma')) \\
        &= \frac{1}{d_Zbd_Ad_B}\sum_{mnor} {\rm Tr}\big( \tilde{P}_{Z(A)}^{m}\tilde{P}_{W(A)}^{o} \tilde{P}_{Z(B)}^{n}\tilde{P}_{W(B)}^{r}(I_{AB} \otimes I_{Z} \otimes  \rho)(I_{AB} \otimes I_{Z}\otimes  \rho') \\
        & \ \ \ \ \ \ \ \ \ \ \ \ \ \ \ \ \ \ \ \ \ \ \ \ \ \ \ \ \times (I_{AB} \otimes I_{Z} \otimes  \sigma)(I_{AB} \otimes I_{Z} \otimes  \sigma') \big) \delta_{i, i_{mn}\delta_{j, j_{or}}}.
    \end{split}
\end{equation}
The operator $I_{Z} \otimes \rho$ can be rewritten in the form
\begin{equation}
    \begin{split}
        I_{Z} \otimes \rho &= {\rm Tr}_{AB}(\tilde{P}_A^a\tilde{P}_X^x) \\
        &=\sum_k {\rm Tr}_{AB}(\rho_{AD_L^k} \otimes \pi_{D_R^kB}) \\
        &= \sum_k \rho_{D_L^k} \otimes \pi_{D_R^k},
    \end{split}
\end{equation}
where in the second line we used the fact that $\tilde{P}_A^a\tilde{P}_X^x \in \ca$. Using this and similar expressions, along with the facts that $\tilde{P}_{Z(A)}^{m}\tilde{P}_{W(A)}^{o} \in \ca$ and $\tilde{P}_{Z(B)}^{n}\tilde{P}_{W(B)}^{r} \in \cb$, we can write
\begin{equation}
     {\rm Tr}(\epsilon^{ij}\rho \sigma \rho' \sigma') = \frac{1}{d_Z} \sum_{mnork} {\rm Tr}(M_{D_L^k}^{imjo}\rho_{D_L^k} \rho_{D_L^k}') {\rm Tr}(N_{D_R^k}^{injp}\sigma_{D_R^k} \sigma_{D_R^k}').
\end{equation}
Together with (\ref{eq:pbrexists}), this implies that
\begin{equation}
    \exists ijmnork: \ {\rm Tr}(M_{D_L^k}^{imjo}\rho_{D_L^k} \rho_{D_L^k}') \neq 0 \ {\rm and} \ {\rm Tr}(N_{D_R^k}^{injr}\sigma_{D_R^k}\sigma_{D_R^k}') \neq 0.
\end{equation}
It follows that 
\begin{equation}
  \exists ijmnork: \   M_{D_L^k}^{imjo}\rho_{D_L^k} \neq 0, \ M_{D_L^k}^{imjo}\rho_{D_L^k}' \neq 0, \ N_{D_R^k}^{injp}\sigma_{D_R^k} \neq 0, \ N_{D_R^k}^{injp}\sigma_{D_R^k}' \neq 0.
\end{equation}
We can then infer that, for example, ${\rm Tr}(M_{D_L^k}^{imjo}\rho_{D_L^k}){\rm Tr}(N_{D_R^k}^{injp}\sigma_{D_R^k}) \neq 0$, and so
\begin{equation}
    {\rm Tr}(\epsilon^{ij}\rho\sigma)= \frac{1}{d_Z} \sum_{mnork} {\rm Tr}(M_{D_L^k}^{imjo}\rho_{D_L^k}) {\rm Tr}(N_{D_R^k}^{injp}\sigma_{D_R^k}) \neq 0
\end{equation}
for this $i$ and $j$ (since each summed-over term is greater than or equal to zero). Similarly, we can show that ${\rm Tr}(\epsilon^{ij}\rho\sigma')$, ${\rm Tr}(\epsilon^{ij}\rho'\sigma)$, and ${\rm Tr}(\epsilon^{ij}\rho'\sigma')$ are all nonzero for the same $i$ and $j$, providing us with a counterexample to (\ref{eq:pbr1}). To summarize the proof, the lack of an irreducible interference collider means that (\ref{eq:pbr2}) implies that (\ref{eq:pbr1}) fails.

\end{document}